\documentclass[preprint,amsmath,amssymb,aps,a4]{revtex4-2}
\usepackage{graphicx}
\usepackage{bm}% bold math
\usepackage{txfonts}
\usepackage{hyperref}
\date{\today}
\newcommand{\be}{\begin{eqnarray}}
\newcommand{\ee}{\end{eqnarray}}

\newcommand{\bfp}{{\bf p}_{\perp}}

\newcommand{\Dp}{{\bf \Delta}_{\perp}}

\usepackage{xcolor}
\usepackage{rotating}
\begin{document}
%
%\preprint{APS/123-QED}
%
\title{Twist-4 proton GTMDs in the light-front quark-diquark model}
\author{Shubham Sharma}
\email{s.sharma.hep@gmail.com}
\affiliation{Department of Physics, Dr. B. R. Ambedkar National Institute of Technology, Jalandhar 144008, India}
\author{Harleen Dahiya}
\email{dahiyah@nitj.ac.in}
\affiliation{Department of Physics, Dr. B. R. Ambedkar National Institute of Technology, Jalandhar 144008, India}

\date{\today}% It is always \today, today,
\begin{abstract}
By integrating across the light-cone energy, totally unintegrated, off-diagonal quark-quark generalized parton correlation functions (GPCFs) can yield generalized transverse-momentum dependent parton distributions (GTMDs). We have obtained the twist-4 GTMDs for the case of proton in the light-front quark-diquark model (LFQDM).  We have decoded the quark-quark GTMD correlator for proton and solved the parametrization equations for Dirac matrix structure at twist-4. We have derived the explicit equations of GTMDs for the diquark being scalar or vector leading to the results for the struck quark being a $u$ or $d$ quark. Being a multidimensional function, the dependence is studied with two variables at a time, keeping the others fixed. Transverse-momentum dependent form factors (TMFFs) have been obtained from GTMDs via integration over the longitudinal momentum fraction $x$ of the quark. We have debated over the possibility to specify the initial and final state polarizations of the smacked quark and the nucleon in twist-4 GTMDs as well as the Wigner distributions. Additionally, from twist-4 GTMDs, we have deduced the twist-4 T-even TMDs and found its synchronization with the already published results.
\par
 \vspace{0.1cm}
\end{abstract}
%====================================================
%
\maketitle
%
%\tableofcontents
%
\section{Introduction\label{secintro}}
%							$Refined Below$
Quantum chromodynamics (QCD) describes the anatomy and dynamics of the hadron in terms of its constituent quark and gluon degrees of freedom. To comprehend the hadron structure in terms of partons, various distribution categories exist. Parton distribution functions (PDFs), which are derived from parton densities and carry information about detecting the parton with the longitudinal momentum fraction $x$ inside the hadron \cite{Gluck:1994uf,Collins:1981uw,9803445}, are the simplest. Since they represent an one-dimensional (1-D) structure, their significance is limited. Transverse momentum-dependent parton distributions (TMDs) \cite{Radici14,Collins81uk,Collins:1981uw,Mulders95,Sivers:1989cc,Kotzinian94,Boer:1997nt} and generalized parton distributions (GPDs) \cite{Mueller:1998fv,Goeke:2001tz,Diehl03,Ji04,Belitsky05,Boffi07,Ji96,Brodsky06,Radyushkin97,Burkardt00,Diehl02,DC05,Ji97,Hagler03,Kanazawa14,Rajan16} demonstrate the multidimensional nature of the hadron. TMDs provide a three-dimensional (3-D) depiction of a hadron in the momentum space by providing transverse momentum information ${ p_\perp}$ in addition to what we obtain from the PDFs \cite{Collins:2003fm,Collins:2007ph,Collins:1999dz,Hautmann:2007uw}. This information can be obtained experimentally through the semi-inclusive deep inelastic scattering (SIDIS) and the Drell-Yan (DY) processes. TMDs have also encrypted data pertaining to spin-orbit correlations and the angular momentum of the nucleon \cite{Bacchetta:2006tn,D'Alesio:2007jt,Burkardt:2008jw,Barone:2010zz,Aidala:2012mv,Collins:1981uk,Ji:2004wu,Collins:2004nx,Cahn:1978se,Konig:1982uk,Chiappetta:1986yg,Collins:1984kg,Sivers:1989cc,Efremov:1992pe,Collins:1992kk,Collins:1993kq,Kotzinian:1994dv,Mulders:1995dh,Boer:1997nt,Boer:1997mf,Boer:1999mm,Bacchetta:1999kz,Brodsky:2002cx,Collins:2002kn,Belitsky:2002sm,Burkardt:2002ks,Pobylitsa:2003ty,Goeke:2005hb,Bacchetta:2006tn,Cherednikov:2007tw,Brodsky:2006hj,Avakian:2007xa,Miller:2007ae,Arnold:2008kf,Brodsky:2010vs,lattice-TMD}. GPDs encode information about the longitudinal momentum $x$ and transverse position $b_\perp$ of the parton and were experimentally introduced in the ambience of deeply virtual Compton scattering (DVCS) \cite{Mueller:1998fv,Ji97,Radyushkin:1996nd,Goeke:2001tz,Diehl03,Belitsky05,Boffi07}. In addition, double parton distribution functions (DPDFs) have attracted attention recently because they provide crucial information for comprehending the hadron's 3-D structure \cite{Kasemets:2017vyh,Rinaldi:2018slz,Diehl:2011yj}. DPDFs are essential for describing processes involving simultaneous interaction of two partons in high-energy collisions, such as those occurring in multi-parton scattering events at hadron colliders like the Large Hadron Collider (LHC) at CERN. 
\begin{figure*}
	\centering
	\begin{minipage}[c]{0.98\textwidth}
		\includegraphics[width=17cm]{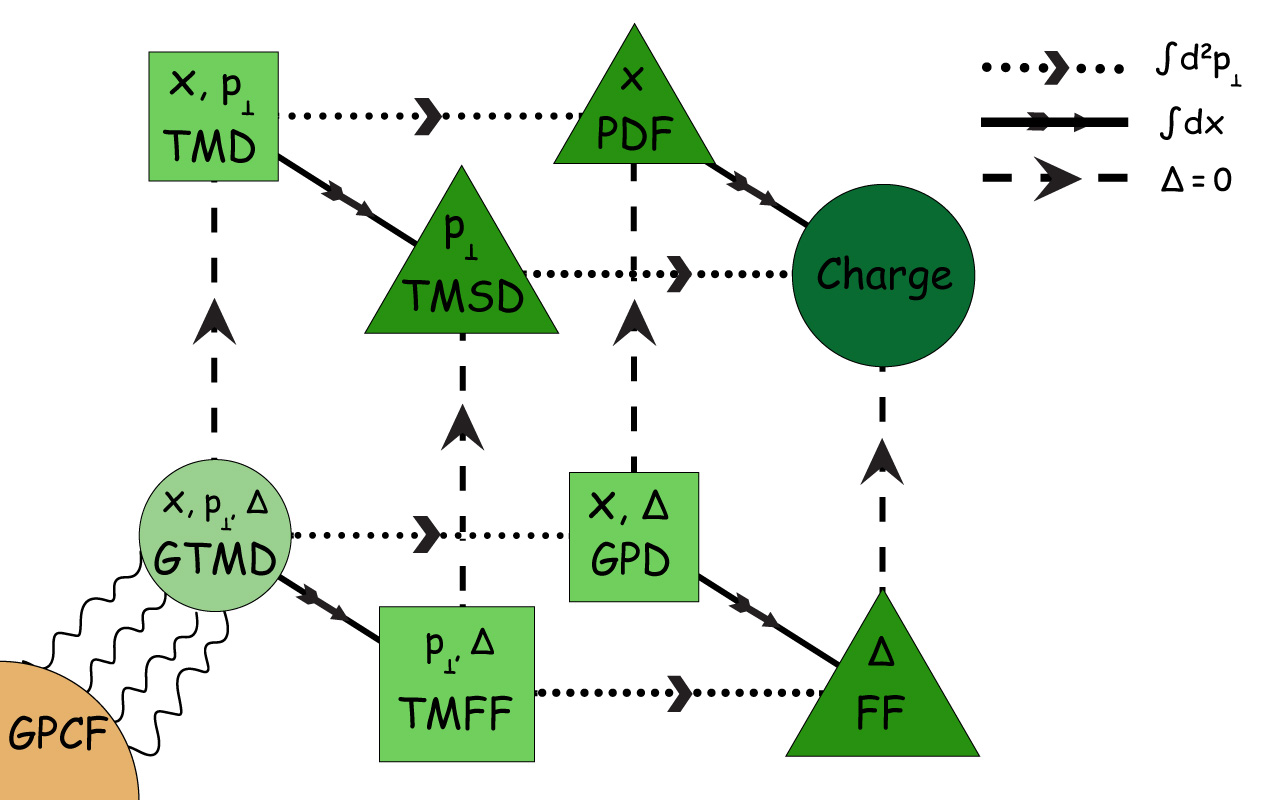}
		\hspace{0.05cm}\\
	\end{minipage}
	\caption{\label{figtree} (Color online) Illustration of the family tree of generalized parton correlation functions (GPCFs). Different arrows correspond to different limits on the GTMDs. Dotted line represents the integration over the quark's transverse momentum $\bfp$. Solid line signifies the integration over the longitudinal momentum fraction $x$. Dashed line is used to represent the case of no momentum transfer.
	}
\end{figure*}
\par
The most general correlator with the highest quantity of information on the parton inside the parent hardon is the fully unintegrated, off-diagonal quark-quark correlator, also known as the generalized parton correlation functions (GPCFs) \cite{Lorce13,Meissner:2009ww}. The GPCFs have dependence on the four-momentum quark and the transfer of momentum to the hadron. Generalized transverse momentum-dependent distributions (GTMDs) arise when GPCFs are integrated over the light cone component of the quark momentum \cite{Meissner:2008ay,Meissner:2009ww,Echevarria:2016mrc,Lorce13}. GTMDs can also be called ``mother distributions'' due to their connection to GPDs and TMDs within certain kinematic limits \cite{Lorce13}. It is essential to mention that the link between GTMDs and GPD is not preserved under preserved under perturbative QCD (pQCD) evolution. GPDs can be evolved using Dokshitzer-Gribov-Lipatov-Altarelli-Parisi (DGLAP) and Efremov-Radyushkin-Brodsky-Lepage (ERBL) equations, whereas for GTMDs Collins-Soper-Sterman (CSS) equations are used. The situation is very similar to the relation among TMDs and PDFs where they evolve respectively by CSS and DGLAP equations\cite{Echevarria:2022ztg}.
Wigner distributions (WDs) are quantum mechanical analogs of the classical phase space distribution and are obtained by Fourier transform of the GTMDs with respect to the momentum transfer $ \Delta_\perp$~\cite{Ji:2003ak,Belitsky:2003nz,Lorce11,Wigner32}. Being a quantum distribution, it is constrained by the principle of uncertainty. As a consequence, it cannot be probabilistically interpreted across the entire phase space and is not positively definitive. The complete family tree of GPCFs along with other distribution functions like form factors (FFs), transverse-momentum dependent form factors (TMFFs) and transverse-momentum dependent spin
densities (TMSDs) have been shown in Fig. \ref{figtree}.
 \par
GTMDs have been introduced in Refs. \cite{Meissner:2008ay,Meissner:2009ww,Echevarria:2016mrc,Lorce13}. The parameterization of quark GTMDs upto twist-4 for spin-$0$ and spin-$\frac{1}{2}$ target has been given in Ref.~\cite{Meissner:2008ay} and Ref. \cite{Meissner:2009ww} respectively. While studying the high-energy diffractive processes such as Higgs or vector meson generation in QCD's $k_T$-factorization, gluon GTMDs appear quite naturally \cite{Martin:1999wb,Khoze:2000cy,Martin:2001ms}. In Ref.~\cite{Lorce13}, a complete classification of gluon GTMDs along with the independent confirmation of the counting of quark GTMDs has been given. Twist-2 quark GTMDs have been computed in the light-front constituent quark model (LFCQM)~\cite{Lorce:2011dv,Lorce11}, light-cone quark model \cite{Ma:2018ysi}, chiral quark soliton model \cite{Lorce:2011dv,Lorce11}, spectator model~\cite{Meissner:2008ay,Meissner:2009ww} and  the light-cone constituent quark model \cite{Lorce:2011dv}. Model calculations of GTMDs in the color-glass condensate has been performed in Ref. \cite{Hagiwara:2016kam}. In Ref. \cite{zhou:2016rnt}, elliptical gluon GTMDs for a nucleus have been presented.
Ref. \cite{Lorce:2011ni, Hatta:2011ku, Lorce:2011ni} demonstrates
the relation of GTMDs with the elusive quark, spin-orbit correlations and gluon orbital angular momentum. Evolution of the GTMDs have been studied in \cite{Echevarria:2016mrc}. Considering the final state interaction, twist-2 time-reversal odd (T-odd) GTMDs and single spin asymmetries have been investigated \cite{Brodsky:2002cx}. In Ref. \cite{Echevarria:2022ztg}, sufficient linkage between double Drell-Yan process (DDY) and GTMDs calculations have been claimed. \par
%================
%PARA 4
%================'
A plethora of fascinating nucleon characteristics, especially the shape of the wave function has been predicted by the light-front AdS/QCD \cite{BT}. It has compatibility with the Drell-Yan-West relationship \cite{DY70,West70} and the counting rule for quarks \cite{Maji:2016yqo}. In the light-front quark-diquark model (LFQDM), the proton's structure has been assumed to be made up of an active quark and a diquark spectator with a particular mass \cite{Chakrabarti:2019wjx}. From the AdS/QCD predictions, the light-front wave functions (LFWFs) have been derived containing contributions from the scalar ($S=0$) and axial vector ($S=1$) diquarks with a $SU(4)$ spin-flavor structure \cite{Maji:2016yqo}. LFQDM is sufficiently capable to evolve the PDFs from 
its model scale $\mu^2=0.09 ~\mathrm{GeV}^2$ to any desired scale upto $\mu^2=10^4~ ~\mathrm{GeV}^2$. As demonstrated by other models,  GPDs possess a diffraction pattern like structure in the longitudinal position space \cite{Mondal:2015uha}.\par

In the LFQDM, presentation of the transverse configuration of the proton has been done successfully \cite{Maji:2017bcz}. The parallels of SIDIS spin asymmetries in the LFQDM with HERMES and COMPASS findings have been shown in Ref. \cite{Gurjar:2022rcl}. In the LFQDM, twist-2 T-even TMDs have been studied where the inequalities customary of diquark models have been shown to be fulfilled \cite{Maji:2017bcz}. Despite the fact that $x$-$p_\perp$ factorization does not occur in the LFQDM, unlike other phenomenological models for the TMD $f_1^\nu(x,{p_\perp})$, numerical analysis of TMDs agrees with the phenomenological ansatz. The GPD-TMD relations and the orbital angular momentum of quarks has been calculated and compared to the outcomes of other models with identical characteristics \cite{Gurjar:2021dyv}. Gravitational form factors (GFFs), $A(Q^2)$ and $B(Q^2)$, derived in the LFQDM are in line with the lattice QCD. The qualitative behavior of the D-term form factor matches with data gathered from DVCS experiments at JLab, the lattice QCD, and the predictions of several other models \cite{Chakrabarti:2020kdc}. The distributions of pressure and shear force are also compatible with the outcomes of other models \cite{Chakrabarti:2020kdc}. Recently, twist-3 \cite{sstwist3} and twist-4 \cite{sstwist4} T-even TMDs have been derived in the LFQDM. The calculations of average square momenta and average square transverse momenta for these higher twist TMDs \cite{sstwist3,sstwist4} have been found to be in close agreement with those of the LFCQM and the bag model. Within this model, the relations of higher twist TMDs with twist-2 TMDs have been obtained and some of them are similar to as in other quark models. In Ref. \cite{sstwist3}, flavor combination of the PDF $e(x)$ has also been compared with the recent CLAS data. Twist-2 GTMDs for non zero skewness have been obained in the LFQDM \cite{majigtmd}. It has been shown that the corresponding WDs in longitudinal space exhibit diffraction pattern just like the wave in optics. 
\par
%_____________________________________________________________________________
% tp11 end
%====================================================
According to our knowledge, the determination of GTMDs in any model at
subleading twist (twist-3) and twist-4 has not been performed yet. Although, the investigation on higher twist TMDs have been done for hadrons \cite{Avakian:2010br,Jakob:1997wg,Lorce:2014hxa,
Pasquini:2018oyz,Kundu:2001pk,Mukherjee:2010iw}. 
Twist-4 distributions have been the subject of discussion \cite{Lorce:2014hxa,PhysRevLett.67.552,SIGNAL1997415,PhysRevD.95.074017,ELLIS19821,ELLIS198329,QIU1991105,QIU1991137,PhysRevD.83.054010,liu21}.
Specifically, in the frame of LFQDM,  both twist-3 \cite{sstwist3} and twist-4 \cite{sstwist4} T-even TMDs have been derived. Taking these points into consideration, it would be interesting to go to higher twist GTMDs calculation in the LFQDM.
\par
%_____________________________________________________________________________
% PARA 5
%====================================================
The objective of this study is to examine the twist-4 GTMDs for the case of proton using the LFQDM. By integrating across the light-cone energy, the totally unintegrated, off-diagonal quark-quark GPCFs give rise to GTMDs. In the LFQDM, we have obtained the twist-4 GTMDs for the case of proton. Specifically, we have decoded the fully unintegrated quark-quark GTMD correlator for the twist-4 Dirac matrix structure and hence obtained the explicit equations for twist-4 GTMDs of proton by comparing it with the parameterization equations. We have derived the explicit equations of GTMDs for both the $u$ and $d$ struck quark possibilities from the scalar and vector diquark parts. Being a multidimensional function, the dependence is studied with two variables at a time, while keeping the others fixed or integrated. We have debated over the possibility to specify the initial and final state polarizations of the smacked quark and the nucleon in the twist-4 GTMDs as well as the WDs. Additionally, from the twist-4 GTMDs we have derived the results of twist-4 T-even TMDs and have been compared with the available results.
\par
We have arranged our work as follows: Essential details of the LFQDM along with its input parameters and other constants have been discussed in Sec.\ref{secmodel}. In Sec.\ref{seccor}, the details of twist-4 quark-quark GTMD correlator along with the corresponding parameterization equations has been presented.
The explicit results of the twist-4 GTMDs have been presented in Sec.\ref{secresults}. Interpretation of GTMDs with the help of 3D-Plots has been done in Sec.\ref{secdiscussion}. At the end, the results have been concluded in Sec.\ref{seccon}.
%_____________________________________________________________________________
%DONE LAST PARA
%====================================================
%% _____________________________________________________________________________

\section{Light-Front Quark-Diquark Model (LFQDM) \label{secmodel}}
%_____________________________________________________________________________
% tq1 start
%====================================================
In the LFQDM \cite{Maji:2016yqo}, proton is described as an accumulation of active quark and a diquark observer with a certain mass \cite{Chakrabarti:2019wjx}. The spin-flavor $SU(4)$ structure is possessed by a proton and its description includes the blend of isoscalar-scalar diquark singlet $|u~ S^0\rangle$, isoscalar-vector diquark $|u~ A^0\rangle$ and isovector-vector diquark $|d~ A^1\rangle$ states as \cite{Jakob:1997wg,Bacchetta:2008af}
\begin{equation}
	|P; \pm\rangle = C_S|u~ S^0\rangle^\pm + C_V|u~ A^0\rangle^\pm + C_{VV}|d~ A^1\rangle^\pm. \label{PS_state}
\end{equation}
Here, $S$ and $A=V,VV$ are employed to designate scalar and vector diquarks respectively. Their respective isospins are signified by the $0$ or $1$ superscripts on them. In Ref. \cite{Maji:2016yqo}, the coefficients $C_{i}$ of scalar and vector diquark states have been determined and are given as
\begin{equation}
	\begin{aligned}
		C_{S}^{2} &=1.3872, \\
		C_{V}^{2} &=0.6128, \\
		C_{V V}^{2} &=1.
	\end{aligned}
	\label{Eq3d1}
\end{equation}
The fraction of longitudinal momentum acquired by the struck quark is expressed as $x=p^+/P^+$ where the smacked quark's ($p$) and diquark ($P_X$) momentum can be expressed as
\begin{eqnarray}
	p &&\equiv \bigg(xP^+, p^-,\bfp \bigg)\,,\label{qu} \\
	P_X &&\equiv \bigg((1-x)P^+,P^-_X,-\bfp\bigg). \label{diq}
\end{eqnarray}
The expansion of Fock-state in the two particle case for $J^z =\pm1/2$ for the scalar $|\nu~ S\rangle^\pm $ and vector diquark $|\nu~ A \rangle^\pm$ in the situation of two particles can be expressed as \cite{majiref25}
\begin{eqnarray}
|\nu~ S\rangle^\pm & =& \int \frac{dx~ d^2\bfp}{2(2\pi)^3\sqrt{x(1-x)}} \Bigg[ \psi^{\pm(\nu)}_{+}(x,\bfp)\bigg|+\frac{1}{2},~s; xP^+,\bfp\bigg\rangle \nonumber \\
&+& \psi^{\pm(\nu)}_{-}(x,\bfp) \bigg|-\frac{1}{2},~s; xP^+,\bfp\bigg\rangle\Bigg],\label{fockSD}\\
|\nu~ A \rangle^\pm & =& \int \frac{dx~ d^2\bfp}{2(2\pi)^3\sqrt{x(1-x)}} \Bigg[ \psi^{\pm(\nu)}_{++}(x,\bfp)\bigg|+\frac{1}{2},~+1; xP^+,\bfp\bigg\rangle \nonumber\\
&+& \psi^{\pm(\nu)}_{-+}(x,\bfp)\bigg|-\frac{1}{2},~+1; xP^+,\bfp\bigg\rangle +\psi^{\pm(\nu)}_{+0}(x,\bfp)\bigg|+\frac{1}{2},~0; xP^+,\bfp\bigg\rangle \nonumber \\
&+& \psi^{\pm(\nu)}_{-0}(x,\bfp)\bigg|-\frac{1}{2},~0; xP^+,\bfp\bigg\rangle + \psi^{\pm(\nu)}_{+-}(x,\bfp)\bigg|+\frac{1}{2},~-1; xP^+,\bfp\bigg\rangle \nonumber\\
&+& \psi^{\pm(\nu)}_{--}(x,\bfp)\bigg|-\frac{1}{2},~-1; xP^+,\bfp\bigg\rangle  \Bigg].\label{fockVD}
\end{eqnarray}
Here, the flavor index $\nu ~=u$ (for the case of scalar) and   $\nu ~=u,d$ (for the case of vector)). We have  $|\lambda^q,~\lambda^{Sp};  xP^+,\bfp\rangle$ representing the two particle state with quark helicity of $\lambda^q=\pm\frac{1}{2}$ and spectator diquark helicity of $\lambda^{Sp}$. The helicity of spectator for scalar diquark is $\lambda^{Sp}=\lambda^{S}=0$ (singlet) and that for vector diquark is $\lambda^{Sp}=\lambda^{D}=\pm 1,0$ (triplet).
With the possibility of the diquarks being a scalar or a vector, the LFWFs have been listed when $J^z=\pm1/2$ in Table \ref{tab_LFWF} \cite{Maji:2017bcz}.
\begin{table}[h]
	\centering % used for centering table
	\begin{tabular}{ |p{1.5cm}|p{1.4cm}|p{1.2cm}|p{1.8cm} p{4.0cm}|p{1.8cm} p{4.0cm}|  }
		%  \hline
		%  \multicolumn{8}{|c|}{Model Parameters corresponding to up \& down quarks } \\
		\hline
		&~~$\lambda^q$~~&~~$\lambda^{Sp}$~~&\multicolumn{2}{c|}{LFWFs for $J^z=+1/2$} & \multicolumn{2}{c|}{LFWFs for $J^z=-1/2$}\\
		\hline
		~~$\rm{Scalar}$&~~$+1/2$~~&~~$~~0$~~&~~$\psi^{+(\nu)}_{+}(x,\bfp)$~&~~$=~N_S~ \varphi^{(\nu)}_{1}$~~&~~$\psi^{-(\nu)}_{+}(x,\bfp)$~&~~$=~N_S \bigg(\frac{p^1-ip^2}{xM}\bigg)~ \varphi^{(\nu)}_{2}$~~  \\
		&~~$-1/2$~~&~~$~~0$~~&~~$\psi^{+(\nu)}_{-}(x,\bfp)$~&~~$=~-N_S\bigg(\frac{p^1+ip^2}{xM} \bigg)~ \varphi^{(\nu)}_{2}$~~&~~$\psi^{-(\nu)}_{-}(x,\bfp)$~&~~$=~N_S~ \varphi^{(\nu)}_{1}$~~   \\
%		~~S No.~~&~~$\lambda_q$~~&~~$\lambda_D$~~&\multicolumn{2}{c|}{LFWFs for $J^z=+1/2$} & \multicolumn{2}{c|}{LFWFs for $J^z=-1/2$}\\
		\hline
		&~~$+1/2$~~&~~$+1$~~&~~$\psi^{+(\nu)}_{+~+}(x,\bfp)$~&~~$=~~N^{(\nu)}_1 \sqrt{\frac{2}{3}} \bigg(\frac{p^1-ip^2}{xM}\bigg)~  \varphi^{(\nu)}_{2}$~~&~~$\psi^{-(\nu)}_{+~+}(x,\bfp)$~&~~$=~~0$~~  \\
		~~~~&~~$-1/2$~~&~~$+1$~~&~~$\psi^{+(\nu)}_{-~+}(x,\bfp)$~&~~$=~~N^{(\nu)}_1 \sqrt{\frac{2}{3}}~ \varphi^{(\nu)}_{1}$~~&~~$\psi^{-(\nu)}_{-~+}(x,\bfp)$~&~~$=~~0$~~   \\
		~~$\rm{Vector}$&~~$+1/2$~~&~~$~~0$~~&~~$\psi^{+(\nu)}_{+~0}(x,\bfp)$~&~~$=~~-N^{(\nu)}_0 \sqrt{\frac{1}{3}}~  \varphi^{(\nu)}_{1}$~~&~~$\psi^{-(\nu)}_{+~0}(x,\bfp)$~&~~$=~~N^{(\nu)}_0 \sqrt{\frac{1}{3}} \bigg( \frac{p^1-ip^2}{xM} \bigg)~  \varphi^{(\nu)}_{2}$~~   \\
		&~~$-1/2$~~&~~$~~0$~~&~~$\psi^{+(\nu)}_{-~0}(x,\bfp)$~&~~$=~~N^{(\nu)}_0 \sqrt{\frac{1}{3}} \bigg(\frac{p^1+ip^2}{xM} \bigg)~ \varphi^{(\nu)}_{2}$~~&~~$\psi^{-(\nu)}_{-~0}(x,\bfp)$~&~~$=~~N^{(\nu)}_0\sqrt{\frac{1}{3}}~  \varphi^{(\nu)}_{1}$~~   \\
	&~~$+1/2$~~&~~$-1$~~&~~$\psi^{+(\nu)}_{+~-}(x,\bfp)$~&~~$=~~0$~~&~~$\psi^{-(\nu)}_{+~-}(x,\bfp)$~&~~$=~~- N^{(\nu)}_1 \sqrt{\frac{2}{3}}~  \varphi^{(\nu)}_{1}$~~   \\
		&~~$-1/2$~~&~~$-1$~~&~~$\psi^{+(\nu)}_{-~-}(x,\bfp)$~&~~$=~~0$~~&~~$\psi^{-(\nu)}_{-~-}(x,\bfp)$~&~~$=~~N^{(\nu)}_1 \sqrt{\frac{2}{3}} \bigg(\frac{p^1+ip^2}{xM}\bigg)~  \varphi^{(\nu)}_{2}$~~   \\
		\hline
	\end{tabular}
	\caption{The LFWFs for both diquark cases when $J^z=\pm1/2$, for various values of helicities of smacked quark $\lambda^q$ and the spectator diquark $\lambda^{Sp}$. The normalization constants are $N_S$, $N^{(\nu)}_0$ and $N^{(\nu)}_1$.}
	\label{tab_LFWF} % is used to refer this table in the text
\end{table}
The general form of LFWFs $\varphi^{(\nu)}_{i}=\varphi^{(\nu)}_{i}(x,\bfp)$ in Table \ref{tab_LFWF} has been obtained from the prediction of soft-wall AdS/QCD, and the establishment of parameters $a^\nu_i,~b^\nu_i$ and $\delta^\nu$ have been followed from Ref. \cite{Maji:2016yqo}. We have
\begin{eqnarray}
\varphi_i^{(\nu)}(x,\bfp)=\frac{4\pi}{\kappa}\sqrt{\frac{\log(1/x)}{1-x}}x^{a_i^\nu}(1-x)^{b_i^\nu}\exp\Bigg[-\delta^\nu\frac{\bfp^2}{2\kappa^2}\frac{\log(1/x)}{(1-x)^2}\bigg].
\label{LFWF_phi}
\end{eqnarray}
These LFWFs are enhanced version of the soft-wall AdS/QCD prediction \cite{Gutsche:2013zia}. For $a_i^\nu=b_i^\nu=0$  and $\delta^\nu=1.0$, the wave functions $\varphi_i^\nu ~(i=1,2)$ collapse to AdS/QCD. By using the Dirac and Pauli form factor data, the fitting of parameters $a_i^{\nu}$ and $b_i^{\nu}$ has been performed at the model scale $\mu_0=0.313{\ \rm GeV}$  \cite{Maji:2016yqo,majiref16,majiref17}. The value of parameter $\delta^{\nu}$ has been adopted for LFQDM \cite{Maji:2016yqo} from AdS/QCD \cite{deTeramond:2011aml}. Apart from these, the normalization constants $N_{i}^{2}$ in Table {\ref{tab_LFWF}} are derived from Ref. \cite{Maji:2016yqo}. For the sake of completeness, the model parameters for both struck quark flavors have been listed in Table \ref{tab_par}. In accordance with Ref. \cite{Chakrabarti:2019wjx}, we have assumed the proton mass ($M$) and the constituent quark mass ($m$) to be $0.938~\mathrm{GeV}$ and $0.055~\mathrm{GeV}$, sequentially. The value $0.4~\mathrm{GeV}$ \cite{Chakrabarti:2013dda,Chakrabarti:2013gra} has been assigned to the AdS/QCD scale parameter $\kappa$ appearing in  Eq. (\ref{LFWF_phi}).
\begin{table}[h]
	\centering % used for centering table
	\begin{tabular}{ |c|c|c|c|c|c|c|c| }
		%  \hline
		%  \multicolumn{8}{|c|}{Model Parameters corresponding to up \& down quarks } \\
		\hline
		~~$\nu$~~&~~$a_1^{\nu}$~~&~~$b_1^{\nu}$~~&~~$a_2^{\nu}$~~&~~$b_2^{\nu}$~~&~~$N_{S}$~~&~~$N_0^{\nu}$~~&~~$N_1^{\nu}$~~   \\
		\hline
		~~$u$~~&~~$0.280\pm 0.001$~~&~~$0.1716 \pm 0.0051$~~&~~$0.84 \pm 0.02$~~&~~$0.2284 \pm 0.0035$~~&~~$2.0191$~~&~~$3.2050$~~&~~$0.9895$~~  \\
		~~$d$~~&~~$0.5850 \pm 0.0003$~~&~~$0.7000 \pm 0.0002$~~&~~$0.9434^{+0.0017}_{-0.0013}$~~&~~$0.64^{+0.0082}_{-0.0022}$~~&~~$0$~~&~~$5.9423$~~&~~$1.1616$~~    \\
		\hline
	\end{tabular}
	\caption{Values of model parameters and normalization constants $N_{i}^{2}$.}
	\label{tab_par} % is used to refer this table in the text
\end{table}

%%%%%%%%%%%%%%%%%%%%%%%%%%%%%%%%%%%%%%%%%%%%%%%
\section{Twist-4 GTMDs in LFQDM}\label{seccor}
%%%%%%%%%%%%%%%%%%%%%%%%%%%%%%%%%%%%%%%%%%%%%%%
We have presented a detailed analysis of the twist-4 GTMDs in the LFQDM in this section. The fully unintegrated quark-quark correlator for a spin-$\frac{1}{2}$ hadron at the fixed light-cone time $ z^+=0$, defined following Ref. \cite{Meissner:2009ww}, is given as
\begin{eqnarray} 
W^{\nu [\Gamma]}_{[\Lambda^{N_i}\Lambda^{N_f}]}(x, p_{\perp},\Delta_{\perp},\theta)=\frac{1}{2}\int \frac{dz^-}{(2\pi)} \frac{d^2z_T}{(2\pi)^2} e^{ip.z} 
\langle P^{f}; \Lambda^{N_f} |\bar{\psi} (-z/2)\Gamma \mathcal{W}_{[-z/2,z/2]} \psi (z/2) |P^{i};\Lambda^{N_i}\rangle \bigg|_{z^+=0}\,,
\label{corr}
\end{eqnarray} 
where $|P^{i};\Lambda^{N_i}\rangle $ and $|P^{f}; \Lambda^{N_f}\rangle$ are the initial and final states of the proton with helicities $\Lambda^{N_i}$ and $\Lambda^{N_f}$, respectively and $\psi (\bar{\psi})$ is the quark field operator. The GTMD corelator depends on the group of variables $\{x,\xi, p_{\perp},\Delta_{\perp},\theta\}$ \cite{Meissner:2009ww}. We have considered the case of zero skewness i.e., $\xi=- \Delta^+/2P^+=0$. Also, we have written the dependence on $\bfp \cdot \Dp$ in terms of angle ($\theta$) between the plane $\bfp$ and $\Dp$. Hence, throughout the article we have written GTMDs correlator $W^{\nu [\Gamma]}_{[\Lambda^{N_i}\Lambda^{N_f}]}(x,\xi, p_{\perp},\Delta_{\perp},\bfp \cdot \Dp)$ as $W^{\nu [\Gamma]}_{[\Lambda^{N_i}\Lambda^{N_f}]}(x,p_{\perp},\Delta_{\perp},\theta)$ or compactly as $W^{\nu [\Gamma]}_{[\Lambda^{N_i}\Lambda^{N_f}]}$, where $\Gamma$ represents the twist-4 Dirac $\gamma$-matrices, i.e., $\Gamma=\{\gamma^-,\, \gamma^-\gamma^5,\, i\sigma^{j-} \gamma^5\}$.
The Wilson line, $\mathcal{W}_{[-z/2,z/2]}$, ensures that the resulting bilocal quark operator has $SU(3)$ color gauge invariance and we have considered it to be $1$ for convenience. Here, we follow the convention $z^\pm=(z^0 \pm z^3)$ and the kinematics are given by  
\begin{eqnarray}
P &\equiv& \bigg(P^+,\frac{M^2+\Dp^2/4}{P^+},\textbf{0}_\perp\bigg)\,,\\
\Delta &\equiv& \bigg(0, 0,\Dp \bigg)\,.
%\Delta &\equiv& (0,0,\Dp)\,,
\end{eqnarray}
In the symmetric frame, proton's average momentum is stated as $P= \frac{1}{2} (P^{f}+P^{i})$, while momentum transfer is represented as $\Delta=(P^{f}-P^{i})$. The initial and final four momenta of the proton are then correspondingly expressed as
\begin{eqnarray}
P^{i} &\equiv& \bigg(P^+,\frac{M^2+\Dp^2/4}{P^+},-\Dp/2\bigg)\,,\label{Pp}\\
P^{f} &\equiv& \bigg(P^+,\frac{M^2+\Dp^2/4}{P^+},\Dp/2\bigg)\,. \label{Ppp}
\end{eqnarray}
The struck quark's ($p$) and diquark's ($P_X$) momenta are given by  Eq.~(\ref{qu}) and  Eq.~(\ref{diq}) respectively. $t= \Delta^2=-\Dp^2$ represents the square of the total momentum transfer. By substituting  Eq.~(\ref{fockSD}) and Eq.~(\ref{fockVD}) in Eq.~(\ref{corr}) by the use of Eq.~(\ref{PS_state}), one can express the GTMD correlator for the scalar and vector diquark parts in the form of overlaps of the LFWFs provided in Table {\ref{tab_LFWF}} as
%The correlator $W^{\nu [\Gamma]}_{[\Lambda^{N_f}\Lambda^{N_i}]}$ defined in Eq.~(\ref{corr})  can be expressed in terms of overlaps of the LFWFs given in  Table {\ref{tab_LFWF}}.
%We obtain for the scalar diquark
\begin{eqnarray} 
W^{ [\Gamma](S)}_{[\Lambda^{N_i}\Lambda^{N_f}]}(x, p_{\perp},\Delta_{\perp},\theta)&=&\frac{C_{S}^{2}}{16\pi^3} \sum_{\lambda^{q_i}} \sum_{\lambda^{q_f}} \psi^{\Lambda^{N_f}\dagger}_{\lambda^{q_f}}(x,\bfp+(1-x^\prime)\frac{\Dp}{2})\psi^{\Lambda^{N_i}}_{\lambda^{q_i}}(x,\bfp-(1-x^\prime)\frac{\Dp}{2}) \nonumber\\  &&\frac{u^{\dagger}_{\lambda^{q_f}}(x P^{+},\bfp+\frac{\Dp}{2})\gamma^{0} \Gamma u_{\lambda^{q_i}}(x P^{+},\bfp-\frac{\Dp}{2})}{2 x P^{+}}\,, \label{cors} \\
W^{ [\Gamma](A)}_{[\Lambda^{N_i}\Lambda^{N_f}]}(x, p_{\perp},\Delta_{\perp},\theta)&=&\frac{C_{A}^{2}}{16\pi^3} \sum_{\lambda^{q_i}} \sum_{\lambda^{q_f}} \sum_{\lambda^{D}} \psi^{\Lambda^{N_f}\dagger}_{\lambda^{q_f} \lambda^D}(x,\bfp+(1-x^\prime)\frac{\Dp}{2})\psi^{\Lambda^{N_i}}_{\lambda^{q_i}\lambda^D}(x,\bfp-(1-x^\prime)\frac{\Dp}{2}) \nonumber\\  &&\frac{u^{\dagger}_{\lambda^{q_f}}(x P^{+},\bfp+\frac{\Dp}{2})\gamma^{0} \Gamma u_{\lambda^{q_i}}(x P^{+},\bfp-\frac{\Dp}{2})}{2 x P^{+}}\,, \label{corv} 
\end{eqnarray} 
where, $C_A=C_V, C_{VV}$ for $u$ and $d$ quarks sequentially. $u^{\dagger}_{\lambda^{q_f}}(x P^{+},\bfp+\frac{\Dp}{2})\gamma^{0} \Gamma u_{\lambda^{q_i}}(x P^{+},\bfp-\frac{\Dp}{2})$ represents the spinor product corresponding to the Dirac matrices at twist-4 $\Gamma=\gamma^-,~\gamma^-\gamma^5$, and $i\sigma^{j-}\gamma^5$. Explicit Dirac spinors are given in Ref.
\cite{Harindranath:1996hq,Brodsky:1997de}. Here, $\lambda^{q_i}$ and $\lambda^{q_f}$ symbolize the quark helicity in the initial and the final state consequently. For the vector diquark, additional summation over diquark helicity $\lambda^{D}$ exists.\par
By using the parts representing the scalar and the vector diquark, the correlator for $u$ and $d$ quarks in the LFQDM model is written as
\begin{eqnarray} 
W^{u[\Gamma]}_{[\Lambda^{N_i}\Lambda^{N_f}]}(x, p_{\perp},\Delta_{\perp},\theta) &=&  ~W^{u[\Gamma](S)}_{[\Lambda^{N_i}\Lambda^{N_f}]}(x, p_{\perp},\Delta_{\perp},\theta) + ~W^{u[\Gamma](V)}_{[\Lambda^{N_i}\Lambda^{N_f}]}(x, p_{\perp},\Delta_{\perp},\theta)\,,\\
W^{d[\Gamma]}_{[\Lambda^{N_i}\Lambda^{N_f}]}(x, p_{\perp},\Delta_{\perp},\theta) &=&  ~W^{d[\Gamma](VV)}_{[\Lambda^{N_i}\Lambda^{N_f}]}(x, p_{\perp},\Delta_{\perp},\theta)\,.
\end{eqnarray} 
For various values of Dirac matrix structure $(\Gamma)$ at twist-4, there are 16 quark GTMDs in total $F_{3,1},~F_{3,2},~F_{3,3},~F_{3,4},~G_{3,1},~G_{3,2},~G_{3,3},~G_{3,4},~H_{3,1},~H_{3,2},~H_{3,3},~H_{3,4},~H_{3,5},~H_{3,6},~H_{3,7}$ and $H_{3,8}$, and they can be projected as \cite{Meissner:2009ww}
\begin{equation}
	W_{[\Lambda^{N_i}\Lambda^{N_f}]}^{[\gamma^-]}
	= \frac{M}{2(P^+)^2} \, \bar{u}(P^{f}, \Lambda^{N_F}) \, \bigg[
	F_{3,1}
	+ \frac{i\sigma^{i+} p_T^i}{P^+} \, F_{3,2}
	+ \frac{i\sigma^{i+} \Delta_T^i}{P^+} \, F_{3,3}   + \frac{i\sigma^{ij} p_T^i \Delta_T^j}{M^2} \, F_{3,4}
	\bigg] \, u(P^{i}, \Lambda^{N_i})
	\,, \label{par1} \end{equation}
\begin{equation}
	W_{[\Lambda^{N_i}\Lambda^{N_f}]}^{[\gamma^-\gamma_5]}
	= \frac{M}{2(P^+)^2} \, \bar{u}(P^{f}, \Lambda^{N_F}) \, \bigg[
	- \frac{i\varepsilon_T^{ij} p_T^i \Delta_T^j}{M^2} \, G_{3,1}
	+ \frac{i\sigma^{i+}\gamma_5 p_T^i}{P^+} \, G_{3,2}
	+ \frac{i\sigma^{i+}\gamma_5 \Delta_T^i}{P^+} \, G_{3,3} + i\sigma^{+-}\gamma_5 \, G_{3,4}
	\bigg] \, u(P^{i}, \Lambda^{N_i})
	\,, \label{par2}\end{equation}

\[	W_{[\Lambda^{N_i}\Lambda^{N_f}]}^{[i\sigma^{j-}\gamma_5]}
	= \frac{M}{2(P^+)^2} \, \bar{u}(P^{f}, \Lambda^{N_F}) \, \bigg[
	- \frac{i\varepsilon_T^{ij} p_T^i}{M} \, H_{3,1}
	- \frac{i\varepsilon_T^{ij} \Delta_T^i}{M} \, H_{3,2}
	+ \frac{M \, i\sigma^{j+}\gamma_5}{P^+} \, H_{3,3} + \frac{p_T^j \, i\sigma^{p+}\gamma_5 p_T^p}{M \, P^+} \, H_{3,4} \]
\begin{equation}	+ \frac{\Delta_T^j \, i\sigma^{p+}\gamma_5 p_T^p}{M \, P^+} \, H_{3,5}
	+ \frac{\Delta_T^j \, i\sigma^{p+}\gamma_5 \Delta_T^p}{M \, P^+} \, H_{3,6}  + \frac{p_T^j \, i\sigma^{+-}\gamma_5}{M} \, H_{3,7}
	+ \frac{\Delta_T^j \, i\sigma^{+-}\gamma_5}{M} \, H_{3,8}
	\bigg] \, u(P^{i}, \Lambda^{N_i})
	\,. \label{par3}
\end{equation}
In computing the above expressions, we have used the relations $\varepsilon^{0123} = 1$, $\varepsilon_T^{ij} = \varepsilon^{-+ij}$ and $\sigma^{k l}=i\left[\gamma^{k}, \gamma^{l}\right] / 2$, where $p_T^2=|\vec{p}_T|^2$. The indices $i$ and $j$ are used to denote transverse directions.
%_____________________________________________________________________________
% tr4 end
%====================================================
\section{Results}\label{secresults}
%\subsection{Overlap Form}
To get the expressions of the GTMDs for both the types of diquark, we have to substitute Eq. \eqref{fockSD} and Eq. \eqref{fockVD} with appropriate polarizations in Eq. \eqref{corr} via Eq. \eqref{PS_state}. The determination of a specific GTMD is done by choosing one correlation (for $\Gamma= \gamma^-,  \gamma^-\gamma_5$ and $\sigma^{i-}\gamma_5$) from Eqs. {\eqref{par1}} to {\eqref{par3}}, and proper mash-up of proton's polarization. For the sake of clarity and completeness, as an example, the GTMDs $F_{3,1}^{\nu}, G_{3,4}^{\nu}$ and $H_{3,7}^{\nu}$, corresponding to Dirac matrix structure $\gamma^-,  \gamma^-\gamma_5$ and $\sigma^{i-}\gamma_5$, can be obtained  respectively from Eq. {\eqref{par1}}, {\eqref{par2}} and {\eqref{par3}} and are expressed as 
%
%For example, GTMDs
%$F_{3,1}^{\nu}, G_{3,4}^{\nu}$ and $H_{3,7}^{\nu}$ can be obtained from  as
\begin{eqnarray}
F_{3,1}^{\nu}(x, p_{\perp},\Delta_{\perp},\theta) &=& \frac{(P^+)^2}{2 M^2}\Bigg( ~W^{\nu[\gamma^-]}_{[++]}(x, p_{\perp},\Delta_{\perp},\theta)+W^{\nu[\gamma^-]}_{[--]}(x, p_{\perp},\Delta_{\perp},\theta)\Bigg)\, ,\\
G_{3,4}^{\nu}(x, p_{\perp},\Delta_{\perp},\theta) &=& \frac{(P^+)^2}{4 M^2}\Bigg( ~W^{\nu[\gamma^-\gamma_5]}_{[++]}(x, p_{\perp},\Delta_{\perp},\theta)-W^{\nu[\gamma^-\gamma_5]}_{[--]}(x, p_{\perp},\Delta_{\perp},\theta)\Bigg)\, ,\\
H_{3,7}^{\nu}(x, p_{\perp},\Delta_{\perp},\theta) &=& \frac{(P^+)^2}{4 M} 
\Bigg(
 \Delta_y
\Big( ~W^{\nu[\sigma^{1-}\gamma_5]}_{[++]}(x, p_{\perp},\Delta_{\perp},\theta)-W^{\nu[\sigma^{1-}\gamma_5]}_{[--]}(x, p_{\perp},\Delta_{\perp},\theta)\Big) \nonumber \\ 
&&- \Delta_x
\Big( ~W^{\nu[\sigma^{2-}\gamma_5]}_{[++]}(x, p_{\perp},\Delta_{\perp},\theta)-W^{\nu[\sigma^{2-}\gamma_5]}_{[--]}(x, p_{\perp},\Delta_{\perp},\theta)\Big)
\Bigg)
\,.
\end{eqnarray}
In a similar way, all other GTMDs can be described in terms of GTMD correlator (Eq. \eqref{corr}) and hence in the form product of LFWFs via Eq. \eqref{cors} and  Eq. \eqref{corv}. For convenience, we define
\begin{eqnarray}
	T_{ij}^{(\nu)}(x,p_{\perp},\Delta_{\perp})&=&\varphi_i^{(\nu) \dagger}(x,\bfp+(1-x)\frac{\Dp}{2}) \varphi_j^{(\nu)}(x,\bfp-(1-x)\frac{\Dp}{2})
	\label{Tij1},
\end{eqnarray}
where, $i,j=1,2$. As a direct consequence of Eq. \eqref{Tij1} and Eq. \eqref{LFWF_phi}, this can be written
\begin{eqnarray}
	T_{ij}^{(\nu)}(x,p_{\perp},\Delta_{\perp})&=&T_{ji}^{(\nu)}(x,p_{\perp},\Delta_{\perp})\label{Tij2},\\
	\varphi_i^{(\nu)\dagger}(x,\bfp+(1-x)\frac{\Dp}{2})&=&\varphi_i^{(\nu)}(x,\bfp+(1-x)\frac{\Dp}{2})\label{Tij3}.
\end{eqnarray}
The expressions of GTMDs for the Dirac matrix structure $\Gamma=\gamma^-$ for both scalar and vector diquark can be written as 

%We can choose the appropriate Dirac matrix $(\Gamma)$ by using Eqs. {\eqref{eqtmdlist1}-\eqref{eqtmdlist4}} and the polarization of proton is dependent on the TMD whose expression we want to find. Like for 
%In this way, TMDs in terms of LFWFs for the scalar diquark can be expressed  as
%GTMD k expression $F_{31}^{\nu(S)}(x, p_{\perp},\Delta_{\perp},\theta)$
%$(p_\perp.\Delta_\perp)$
%$(\bfp \cdot \Dp)$
%$p_\perp$
%$\bfp^2$
%$\bfp$
%Further, to get overlap form of TMDs for the vector diquark, we have to substitute Eq. \eqref{fockVD} with suitable polarization in Eq. \eqref{TMDcor} via Eq. \eqref{PS_state}. TMDs in terms of LFWFs for the vector diquark can be written  as
%where $\lambda_A$ runs over $0,\pm$, which is nothing but the summation over vector diquark's helicity.
\begin{eqnarray} 
%	\bigg(\frac{p_\perp^2+m^2}{M^2}\bigg)
%
% F31 SCALAR
%
F_{3,1}^{\nu(S)} &=&  \frac{C_{S}^{2} N_s^2}{16 \pi^3} \frac{1}{x^2 M^2}\Bigg[ \bigg(m^2+\bfp^2-\frac{\Dp^2}{4} \bigg) T_{11}^{\nu}+ \bigg(\Big(m^2+\bfp^2-\frac{\Dp^2}{4} \bigg)\bigg(\bfp^2-(1-x)^2\frac{\Dp^2}{4} \bigg)+(1-x)\bigg(\bfp^2 \Dp^2 \nonumber\\
&&- (\bfp \cdot \Dp)^2 \Big)\bigg) \frac{T_{22}^{\nu}}{x^2 M^2}+ m (1-x)\Dp^2 \frac{T_{12}^{\nu}}{x M}\Bigg], \label{f31s}\\
%
% F31 VECTOR
%
F_{3,1}^{\nu(A)} &=&  \frac{C_{A}^{2}}{16 \pi^3}  \bigg(\frac{1}{3} |N_0^\nu|^2+\frac{2}{3}|N_1^\nu|^2 \bigg)\frac{1}{x^2 M^2}\Bigg[ \bigg(m^2+\bfp^2-\frac{\Dp^2}{4} \bigg) T_{11}^{\nu}+ \bigg(\Big(m^2+\bfp^2-\frac{\Dp^2}{4} \bigg)\bigg(\bfp^2-(1-x)^2\frac{\Dp^2}{4} \bigg)\nonumber\\
&&+(1-x)\bigg(\bfp^2 \Dp^2 - (\bfp \cdot \Dp)^2 \Big)\bigg) \frac{T_{22}^{\nu}}{x^2 M^2}+ m (1-x)\Dp^2 \frac{T_{12}^{\nu}}{x M}\Bigg], \label{f31a}\\
%
% F32 SCALAR
%
F_{3,2}^{\nu(S)} &=&  -\frac{C_{S}^{2} N_s^2}{8 \pi^3} \frac{\bfp \cdot \Dp}{x^2 M}\Bigg[ m \frac{T_{22}^{\nu}}{x^2 M^2}-\frac{T_{12}^{\nu}}{x M}\Bigg], \label{f32s}\\
%
% F32 VECTOR
%
F_{3,2}^{\nu(A)} &=&  \frac{C_{A}^{2}}{8 \pi^3}  \bigg(\frac{1}{3} |N_0^\nu|^2 \bigg) \frac{\bfp \cdot \Dp}{x^2 M}\Bigg[ m \frac{T_{22}^{\nu}}{x^2 M^2}-\frac{T_{12}^{\nu}}{x M}\Bigg], \label{f32a}\\
%
%
% F33 SCALAR
%
F_{3,3}^{\nu(S)} &=&  \frac{C_{S}^{2} N_s^2}{32 \pi^3} \frac{1}{x^2 M^2}\Bigg[ \bigg(\big(m^2+\bfp^2-\frac{\Dp^2}{4}\big)-2m M \bigg) T_{11}^{\nu}+ \Bigg(\bigg(m^2+\bfp^2-\frac{\Dp^2}{4} \bigg)\bigg(\bfp^2-(1-x)^2\frac{\Dp^2}{4} \bigg)+2 m M\bigg(\bfp^2\nonumber\\
&&+(1-x)^2\frac{\Dp^2}{4} \bigg)+(1-x)\bigg(\bfp^2 \Dp^2 - (\bfp \cdot \Dp)^2 \bigg)\Bigg) \frac{T_{22}^{\nu}}{x^2 M^2}+ \bigg((1-x)\Big(m \Dp^2 + 2 M \big( m^2+\bfp^2-\frac{\Dp^2}{4}\big)\Big)\nonumber\\
&&-4 \bfp^2 M\bigg) \frac{T_{12}^{\nu}}{x M}\Bigg], \label{f33s}\\  
%
% F33 VECTOR
%
F_{3,3}^{\nu(A)} &=&  \frac{C_{A}^{2}}{32 \pi^3}  \bigg(\frac{1}{3} |N_0^\nu|^2+\frac{2}{3}|N_1^\nu|^2 \bigg)\frac{1}{x^2 M^2}\Bigg[ \bigg(m^2+\bfp^2-\frac{\Dp^2}{4} \bigg) T_{11}^{\nu}+ \Bigg(\bigg(m^2+\bfp^2-\frac{\Dp^2}{4} \bigg)\bigg(\bfp^2-(1-x)^2\frac{\Dp^2}{4} \bigg)\nonumber\\
&&+(1-x)\bigg(\bfp^2 \Dp^2 - (\bfp \cdot \Dp)^2 \bigg)\Bigg) \frac{T_{22}^{\nu}}{x^2 M^2}+ (1-x)m \Dp^2 \frac{T_{12}^{\nu}}{x M}\Bigg]+\frac{C_{A}^{2}}{32 \pi^3}  \bigg(\frac{1}{3} |N_0^\nu|^2 \bigg)\frac{1}{x^2 M^2}\Bigg[ 2m M \Bigg( T_{11}^{\nu}\nonumber\\
&&-\bigg(\bfp^2+(1-x)^2\frac{\Dp^2}{4} \bigg) \frac{T_{22}^{\nu}}{x^2 M^2}\Bigg)+ \Big(4 \bfp^2 M- 2 M(1-x) \big( m^2+\bfp^2-\frac{\Dp^2}{4}\big)\Big) \frac{T_{12}^{\nu}}{x M}\Bigg], \label{f33a}\\
%
%
% F34 SCALAR
%
F_{3,4}^{\nu(S)} &=&  -\frac{C_{S}^{2} N_s^2}{16 \pi^3} \frac{1}{x^2 }\Bigg[ T_{11}^{\nu}+ \bigg((1-x)\Big(m^2+\bfp^2-\frac{\Dp^2}{4} \Big)-\Big(\bfp^2-(1-x)^2\frac{\Dp^2}{4} \Big)
%\nonumber\\
%&&
\bigg) \frac{T_{22}^{\nu}}{x^2 M^2}-2 m \frac{T_{12}^{\nu}}{x M}\Bigg], \label{f34s}\\
%
%
% F34 VECTOR
%
F_{3,4}^{\nu(A)} &=&  -\frac{C_{A}^{2}}{16 \pi^3}  \bigg(\frac{1}{3} |N_0^\nu|^2-\frac{2}{3}|N_1^\nu|^2 \bigg) \frac{1}{x^2 }\Bigg[ T_{11}^{\nu}+ \bigg((1-x)\Big(m^2+\bfp^2-\frac{\Dp^2}{4} \Big)-\Big(\bfp^2-(1-x)^2\frac{\Dp^2}{4} \Big)\bigg) \frac{T_{22}^{\nu}}{x^2 M^2}
-2 m \frac{T_{12}^{\nu}}{x M}\Bigg], \label{f34a}\nonumber\\
\end{eqnarray}
Similarly, for the Dirac matrix structure $\Gamma=\gamma^-\gamma_5$, the expressions of GTMDs  for both scalar and vector diquark can be written as 
\begin{eqnarray}
%
% G31 SCALAR
%
G_{3,1}^{\nu(S)} &=& \frac{C_{S}^{2} N_s^2}{16 \pi^3} \frac{1}{x^2 }\Bigg[ T_{11}^{\nu}+ \bigg((1-x)\Big(m^2-\bfp^2+\frac{\Dp^2}{4} \Big)+\Big(\bfp^2-(1-x)^2\frac{\Dp^2}{4} \Big)
%\nonumber\\
%&&
\bigg) \frac{T_{22}^{\nu}}{x^2 M^2}-2 m(1-x) \frac{T_{12}^{\nu}}{x M}\Bigg], \nonumber\\ \label{g31s}\\
%
% G31 VECTOR
%
G_{3,1}^{\nu(A)} &=&  \frac{C_{A}^{2}}{16 \pi^3}  \bigg(\frac{1}{3} |N_0^\nu|^2+\frac{2}{3}|N_1^\nu|^2 \bigg)\frac{1}{x^2 }\Bigg[ T_{11}^{\nu}+ \bigg((1-x)\Big(m^2-\bfp^2+\frac{\Dp^2}{4} \Big)+\Big(\bfp^2-(1-x)^2\frac{\Dp^2}{4} \Big)\bigg) \frac{T_{22}^{\nu}}{x^2 M^2}
\nonumber\\
&&-2 m(1-x) \frac{T_{12}^{\nu}}{x M}\Bigg], \label{g31a}\\
%
% G32 SCALAR
%
G_{3,2}^{\nu(S)} &=& \frac{C_{S}^{2} N_s^2}{8 \pi^3} \frac{1}{x^2 M^2}\Bigg[ \bigg(m M+\frac{\Dp^2}{4} \bigg) T_{11}^{\nu}+ \Bigg(-m M\bigg(\bfp^2+(1-x)^2\frac{\Dp^2}{4}\bigg)+\frac{\Dp^2}{4}\bigg(\bfp^2-(1-x)^2\frac{\Dp^2}{4}  \bigg)\nonumber\\
&&+(1-x)\frac{\Dp^2}{4}\bigg(m^2-\bfp^2+\frac{\Dp^2}{4} \bigg)\Bigg) \frac{T_{22}^{\nu}}{x^2 M^2}+ \bigg(2(M-m)(1-x)\frac{\Dp^2}{4}-M\Big(m^2-\bfp^2+\frac{\Dp^2}{4}\Big)\bigg)	 \frac{T_{12}^{\nu}}{x M}\Bigg], \label{g32s}\\
%
% G32 VECTOR
%
G_{3,2}^{\nu(A)} &=& \frac{C_{A}^{2}}{32 \pi^3}  \bigg(\frac{1}{3} |N_0^\nu|^2+\frac{2}{3}|N_1^\nu|^2 \bigg) \frac{\Dp^2}{x^2 M^2}\Bigg[T_{11}^{\nu}+ \Bigg(\bigg(\bfp^2-(1-x)^2\frac{\Dp^2}{4}  \bigg)+(1-x)\bigg(m^2-\bfp^2+\frac{\Dp^2}{4} \bigg)\Bigg) \frac{T_{22}^{\nu}}{x^2 M^2}\nonumber\\
&&-2m(1-x)	 \frac{T_{12}^{\nu}}{x M}\Bigg]+ \frac{C_{A}^{2}}{8 \pi^3}  \bigg(\frac{1}{3} |N_0^\nu|^2 \bigg) \frac{1}{x^2 M^2}\Bigg[ -m M \bigg( T_{11}^{\nu}-\bigg(\bfp^2+(1-x)^2\frac{\Dp^2}{4}\bigg)\Bigg) \frac{T_{22}^{\nu}}{x^2 M^2}\bigg) \nonumber\\
&&+ \bigg(-2M(1-x)\frac{\Dp^2}{4}+M\Big(m^2-\bfp^2+\frac{\Dp^2}{4}\Big)\bigg)	 \frac{T_{12}^{\nu}}{x M}\Bigg], \label{g32a}\\
%
%
% G33 SCALAR
%
G_{3,3}^{\nu(S)} &=&  -\frac{C_{S}^{2} N_s^2}{32 \pi^3} \frac{\bfp \cdot \Dp}{x^2 M^2}\Bigg[T_{11}^{\nu}+ \Bigg(\bigg(\bfp^2-(1-x)^2\frac{\Dp^2}{4} \bigg)+(1-x)\bigg(m^2-\bfp^2+\frac{\Dp^2}{4} \bigg)-2 m M (1-x)^2\Bigg) \frac{T_{22}^{\nu}}{x^2 M^2}\nonumber\\
&&+ 2(1-x)(M-m) \frac{T_{12}^{\nu}}{x M}\Bigg], \label{g33s}\\
%
% G33 VECTOR
%
G_{3,3}^{\nu(A)} &=&  - \frac{C_{A}^{2}}{32 \pi^3}  \bigg(\frac{1}{3} |N_0^\nu|^2+\frac{2}{3}|N_1^\nu|^2 \bigg) \frac{\bfp \cdot \Dp}{x^2 M^2}\Bigg[T_{11}^{\nu}+ \Bigg(\bigg(\bfp^2-(1-x)^2\frac{\Dp^2}{4} \bigg)+(1-x)\bigg(m^2-\bfp^2+\frac{\Dp^2}{4} \bigg)\Bigg) \frac{T_{22}^{\nu}}{x^2 M^2}\nonumber\\
&&-2m(1-x) \frac{T_{12}^{\nu}}{x M}\Bigg]- \frac{C_{A}^{2}}{32 \pi^3}  \bigg(\frac{1}{3} |N_0^\nu|^2 \bigg) \frac{\bfp \cdot \Dp}{x^2 M^2}\Bigg[ \Bigg(2 m M (1-x)^2\Bigg) \frac{T_{22}^{\nu}}{x^2 M^2}-2M (1-x) \frac{T_{12}^{\nu}}{x M}\Bigg], \label{g33a}\\
%
%
% G34 SCALAR
%
G_{3,4}^{\nu(S)} &=&-  \frac{C_{S}^{2} N_s^2}{32 \pi^3} \frac{1}{x^2 M^2}\Bigg[ \bigg(m^2-\bfp^2+\frac{\Dp^2}{4} \bigg)T_{11}^{\nu}+ \bigg((1-x)\bigg(\bfp^2 \Dp^2 - (\bfp \cdot \Dp)^2  \bigg)-\Big(m^2-\bfp^2+\frac{\Dp^2}{4}\Big)\nonumber\\
&&\bigg(\bfp^2-(1-x)^2\frac{\Dp^2}{4} \bigg)\bigg) \frac{T_{22}^{\nu}}{x^2 M^2}+ 4m\bfp^2 \frac{T_{12}^{\nu}}{x M}\Bigg], \label{g34s}\\
%
%
% G34 VECTOR
%
G_{3,4}^{\nu(A)} &=&  -\frac{C_{A}^{2}}{32\pi^3}  \bigg(\frac{1}{3} |N_0^\nu|^2-\frac{2}{3}|N_1^\nu|^2 \bigg) \frac{1}{x^2 M^2}\Bigg[ \bigg(m^2-\bfp^2+\frac{\Dp^2}{4} \bigg)T_{11}^{\nu}+ \bigg((1-x)\bigg(\bfp^2 \Dp^2 - (\bfp \cdot \Dp)^2  \bigg)\nonumber\\
&&-\Big(m^2-\bfp^2+\frac{\Dp^2}{4}\Big)\bigg(\bfp^2-(1-x)^2\frac{\Dp^2}{4} \bigg)\bigg) \frac{T_{22}^{\nu}}{x^2 M^2}+ 4m\bfp^2 \frac{T_{12}^{\nu}}{x M}\Bigg],  \label{g34a}
\end{eqnarray}
Finally, the expressions of GTMDs for both scalar and vector diquark for the Dirac matrix structure $\Gamma=i\sigma^{j-}\gamma_5$, can be written as 
\begin{eqnarray}
	%
	% H31 SCALAR
	%
	H_{3,1}^{\nu(S)} &=&  -\frac{C_{S}^{2} N_s^2}{8 \pi^3} \frac{\bfp \cdot \Dp}{x^2 M}(1-x)\Bigg[ m \frac{T_{22}^{\nu}}{x^2 M^2}-\frac{T_{12}^{\nu}}{x M}\Bigg], \label{h31s}\\
	%
	% H31 VECTOR
	%
	H_{3,1}^{\nu(A)}&=& -\frac{C_{A}^{2}}{8 \pi^3}  \bigg(\frac{1}{3} |N_0^\nu|^2+\frac{2}{3}|N_1^\nu|^2 \bigg) \frac{\bfp \cdot \Dp}{x^2 M}(1-x)\Bigg[ m \frac{T_{22}^{\nu}}{x^2 M^2}-\frac{T_{12}^{\nu}}{x M}\Bigg], \label{h31a}\\
	%
	% H32 SCALAR
	%
	H_{3,2}^{\nu(S)} &=&    -\frac{C_{S}^{2} N_s^2}{16 \pi^3} \frac{1}{x^2 M}\Bigg[m \bigg( T_{11}^{\nu}+ \bigg(\bfp^2-(1-x)^2\frac{\Dp^2}{4}-2(1-x){\bfp^2} \bigg)\frac{T_{22}^{\nu}}{x^2 M^2}\bigg) -(1-x) \bigg(m^2-\bfp^2\nonumber\\
	&&-\frac{\Dp^2}{4} \bigg) \frac{T_{12}^{\nu}}{x M}\Bigg],  \label{h32s}\\
	%
	% H32 VECTOR
	%
	H_{3,2}^{\nu(A)} &=&  -\frac{C_{A}^{2}}{16 \pi^3}  \bigg(\frac{1}{3} |N_0^\nu|^2+\frac{2}{3}|N_1^\nu|^2 \bigg)\frac{1}{x^2 M}\Bigg[m \bigg( T_{11}^{\nu}+ \bigg(\bfp^2-(1-x)^2\frac{\Dp^2}{4}-2(1-x){\bfp^2} \bigg)\frac{T_{22}^{\nu}}{x^2 M^2}\bigg) - (1-x)\nonumber\\
	&&\bigg(m^2-\bfp^2-\frac{\Dp^2}{4} \bigg) \frac{T_{12}^{\nu}}{x M}\Bigg],  \label{h32a}\\
	%
	%
	% H33 SCALAR
	%
	H_{3,3}^{\nu(S)} &=&  \frac{C_{S}^{2} N_s^2}{16 \pi^3} \frac{1}{x^2 M^2}\Bigg[ \bigg(\big(m^2+\bfp^2-\frac{\Dp^2}{4}\big)+\frac{m }{2 M}\Dp^2 \bigg) T_{11}^{\nu}+ \Big[(1-x)\frac{m}{M}(\bfp \cdot \Dp)^2+\frac{m}{2M}\Dp^2\bigg(\bfp^2\nonumber\\
	&&-(1-x)^2\frac{\Dp^2}{4} -2(1-x)\bfp^2\bigg)+\big(m^2+\bfp^2-\frac{\Dp^2}{4}\big){\bfp^2}-\frac{(1-x)^2}{4}\bigg(m^2 \Dp^2-\frac{\Dp^4}{4}-\big(3\bfp^2 \Dp^2 \nonumber\\
	&&- 4(\bfp \cdot \Dp)^2 \big)\bigg)\Big] \frac{T_{22}^{\nu}}{x^2 M^2}+  (1-x)\bigg(m\Dp^2 -\big(m^2-\bfp^2-\frac{\Dp^2}{4}\big)\frac{\Dp^2}{2 M}
-	\frac{1}{M}(\bfp \cdot \Dp)^2
	\bigg)	 \frac{T_{12}^{\nu}}{x M}\Bigg], \nonumber\\ \label{h33s}\\
	%
	% H33 VECTOR
	%
	H_{3,3}^{\nu(A)} &=&  -\frac{C_{A}^{2}}{16 \pi^3}  \bigg(\frac{1}{3} |N_0^\nu|^2\bigg)\frac{1}{x^2 M^2}\Bigg[ \bigg(m^2+\bfp^2-\frac{\Dp^2}{4}\bigg) T_{11}^{\nu}+\Bigg(\big(m^2+\bfp^2-\frac{\Dp^2}{4}\big){\bfp^2}-\frac{(1-x)^2}{4}\bigg(m^2 \Dp^2-\frac{\Dp^4}{4}\nonumber\\
	&&-\big(3\bfp^2 \Dp^2 - 4(\bfp \cdot \Dp)^2 \big)\bigg)\Bigg) \frac{T_{22}^{\nu}}{x^2 M^2}+ \bigg( (1-x)m\Dp^2 \bigg)	 \frac{T_{12}^{\nu}}{x M}\Bigg]+	\frac{C_{A}^{2}}{16 \pi^3}  \bigg(\frac{1}{3} |N_0^\nu|^2+\frac{2}{3}|N_1^\nu|^2 \bigg)\nonumber\\
	&&\frac{(\bfp \cdot \Dp)^2}{x^2 M^3}(1-x)\Bigg[m \frac{T_{22}^{\nu}}{x^2 M^2}- \frac{T_{12}^{\nu}}{x M}\Bigg]+	\frac{C_{A}^{2}}{32 \pi^3}  \bigg(\frac{1}{3} |N_0^\nu|^2+\frac{2}{3}|N_1^\nu|^2 \bigg)\frac{\Dp^2}{x^2 M^3}\Bigg[m\Bigg( T_{11}^{\nu}\nonumber\\
	&&+ \bigg(\bfp^2-(1-x)^2\frac{\Dp^2}{4} -2(1-x)\bfp^2\bigg) \frac{T_{22}^{\nu}}{x^2 M^2}\Bigg)-(1-x)\bigg(m^2-\bfp^2-\frac{\Dp^2}{4}
	\bigg)	 \frac{T_{12}^{\nu}}{x M}\Bigg],
	 \label{h33a}\\
	%
	%
	% H34 SCALAR
	%
	%
	H_{3,4}^{\nu(S)} &=& -\frac{C_{S}^{2} N_s^2}{8 \pi^3} \frac{1}{x^2}\Bigg[T_{11}^{\nu}+\bigg(m^2+ \Big( \big(1-x \big)^2-1 \Big)\frac{\Dp^2}{4}\bigg) \frac{T_{22}^{\nu}}{x^2 M^2}-2m \frac{T_{12}^{\nu}}{x M}\Bigg], \label{h34s}\\
	%
	% H34 VECTOR
	%
	H_{3,4}^{\nu(A)} &=&  \frac{C_{A}^{2}}{8 \pi^3}  \bigg(\frac{1}{3} |N_0^\nu|^2 \bigg)\frac{1}{x^2}\Bigg[T_{11}^{\nu}+\bigg(m^2+ \Big( \big(1-x \big)^2-1 \Big)\frac{\Dp^2}{4}\bigg) \frac{T_{22}^{\nu}}{x^2 M^2}-2m \frac{T_{12}^{\nu}}{x M}\Bigg], \label{h34a}
\end{eqnarray}
\begin{eqnarray} 
	%	\bigg(\frac{\bfp^2+m^2}{M^2}\bigg)
	%
	% H35 SCALAR
	%
	H_{3,5}^{\nu(S)} &=&  \frac{C_{S}^{2} N_s^2}{16 \pi^3} \frac{\bfp \cdot \Dp}{x^2 M}\Bigg[\bigg( \Big( \big(1-x \big)^2-1 \Big)M- (1-x)m\bigg) \frac{T_{22}^{\nu}}{x^2 M^2}+ (1-x) \frac{T_{12}^{\nu}}{x M}\Bigg], \label{h35s}\\
	%
	% H35 VECTOR
	%
	H_{3,5}^{\nu(A)} &=&  -\frac{C_{A}^{2}}{16 \pi^3} \frac{\bfp \cdot \Dp}{x^2 M}\Bigg[\frac{1}{3} |N_0^\nu|^2\bigg( \Big( \big(1-x \big)^2-1 \Big)M\bigg) \frac{T_{22}^{\nu}}{x^2 M^2}+\bigg(\frac{1}{3} |N_0^\nu|^2+\frac{2}{3}|N_1^\nu|^2 \bigg) (1-x) \Bigg(m \frac{T_{22}^{\nu}}{x^2 M^2}\nonumber\\
	&&- \frac{T_{12}^{\nu}}{x M}\Bigg)\Bigg], \label{h35a}\\
	%
	% H36 SCALAR
	%
	H_{3,6}^{\nu(S)} &=&  \frac{C_{S}^{2} N_s^2}{32 \pi^3} \frac{1}{x^2 M}\Bigg[ (M-m)T_{11}^{\nu}+ \bigg(\bfp^2(M-m)+(1-x)^2 \Big( M(m^2-\bfp^2)+m \frac{\Dp^2}{4}\Big)+2(1-x)m \bfp^2\bigg)\nonumber\\
	&& \frac{T_{22}^{\nu}}{x^2 M^2}+(1-x) \bigg(m^2-\bfp^2-\frac{\Dp^2}{4}-2 m M \bigg) \frac{T_{12}^{\nu}}{x M}\Bigg], \label{h36s}\\
	%
	% H36 VECTOR
	%
	H_{3,6}^{\nu(A)} &=&  -\frac{C_{A}^{2}}{32 \pi^3}  \bigg(\frac{1}{3} |N_0^\nu|^2+\frac{2}{3}|N_1^\nu|^2 \bigg) \frac{1}{x^2 M}\Bigg[ m\bigg(T_{11}^{\nu}+ \bigg(\bfp^2-(1-x)^2\frac{\Dp^2}{4}-2(1-x) \bfp^2\bigg) \frac{T_{22}^{\nu}}{x^2 M^2}\bigg)\nonumber\\
	&&-(1-x) \bigg(m^2-\bfp^2-\frac{\Dp^2}{4}\bigg) \frac{T_{12}^{\nu}}{x M}\Bigg]-\frac{C_{A}^{2}}{32 \pi^3}  \bigg(\frac{1}{3} |N_0^\nu|^2 \bigg)\frac{1}{x^2}\Bigg[T_{11}^{\nu}+ \bigg(\bfp^2+(1-x)^2  (m^2-\bfp^2)\bigg) \frac{T_{22}^{\nu}}{x^2 M^2}\nonumber\\
	&&-2m(1-x) \frac{T_{12}^{\nu}}{x M}\Bigg], \label{h36a}\\
	%
	%
	% H37 SCALAR
	%
	H_{3,7}^{\nu(S)} &=&  \frac{C_{S}^{2} N_s^2}{16 \pi^3} \frac{1}{x^2 M}\Bigg[m \bigg( T_{11}^{\nu}- \bigg(\bfp^2-(1-x)^2\frac{\Dp^2}{4}+(1-x)\frac{\Dp^2}{2} \bigg)\frac{T_{22}^{\nu}}{x^2 M^2}\bigg) - \bigg(m^2-\bfp^2-\frac{\Dp^2}{4} \bigg) \frac{T_{12}^{\nu}}{x M}\Bigg], \nonumber\\ \label{h37s}\\
	%
	% H37 VECTOR
	%
	H_{3,7}^{\nu(A)} &=&  \frac{C_{A}^{2}}{16 \pi^3}  \bigg(\frac{1}{3} |N_0^\nu|^2-\frac{2}{3}|N_1^\nu|^2 \bigg)\frac{1}{x^2 M}\Bigg[m \bigg( T_{11}^{\nu}- \bigg(\bfp^2-(1-x)^2\frac{\Dp^2}{4}+(1-x)\frac{\Dp^2}{2} \bigg)\frac{T_{22}^{\nu}}{x^2 M^2}\bigg) - \bigg(m^2\nonumber\\
	&&-\bfp^2-\frac{\Dp^2}{4} \bigg) \frac{T_{12}^{\nu}}{x M}\Bigg], \label{h37a}\\
	%
	%
	% H38 SCALAR
	%
	H_{3,8}^{\nu(S)} &=&  \frac{C_{S}^{2} N_s^2}{32 \pi^3} \frac{\bfp \cdot \Dp}{x^2 M}\Bigg[m(1-x) \frac{T_{22}^{\nu}}{x^2 M^2}- \frac{T_{12}^{\nu}}{x M}\Bigg], \label{h38s}\\
	%
	%
	% H38 SCALAR
	%
	%
	H_{3,8}^{\nu(A)} &=&  \frac{C_{A}^{2}}{32 \pi^3} \bigg(\frac{1}{3} |N_0^\nu|^2-\frac{2}{3}|N_1^\nu|^2 \bigg) \frac{\bfp \cdot \Dp}{x^2 M}\Bigg[m(1-x) \frac{T_{22}^{\nu}}{x^2 M^2}- \frac{T_{12}^{\nu}}{x M}\Bigg]. \label{h38v}
\end{eqnarray}
For any GTMD $X(x, p_{\perp},\Delta_{\perp},\theta)$ (say), if $X^{u(S)}$, $X^{u(V)}$ and $X^{d(VV)}$ denotes the contribution from isoscalar-scalar, isoscalar-vector and isovector-vector diquark parts respectively, then twist-4 GTMDs in the LFQDM model for $u$ and $d$ quarks can be written as
%For any GTMD $X(x, p_{\perp},\Delta_{\perp},\theta)$ (say) twist-4 GTMDs in the LFQDM model for $u$ and $d$ quarks can be written by using the scalar and the vector diquark parts as
\begin{eqnarray} 
 	X^{u}(x, p_{\perp},\Delta_{\perp},\theta) &=&  ~X^{u(S)}(x, p_{\perp},\Delta_{\perp},\theta) + ~X^{u(V)}(x, p_{\perp},\Delta_{\perp},\theta)\,,\label{gtmdu} \\
 	X^{d}(x, p_{\perp},\Delta_{\perp},\theta) &=&  ~X^{d(VV)}(x, p_{\perp},\Delta_{\perp},\theta)\,\label{gtmdd}.
\end{eqnarray} 
%
% $	H_{3,8}^{\nu(A)}$
%_____________________________________________________________________________
% ts1 end
%====================================================

%\subsection{Explicit Expressions of TMDs}

\section{Discussion}\label{secdiscussion}
\begin{figure*}
	\centering
	\begin{minipage}[c]{0.98\textwidth}
		(a)\includegraphics[width=7.3cm]{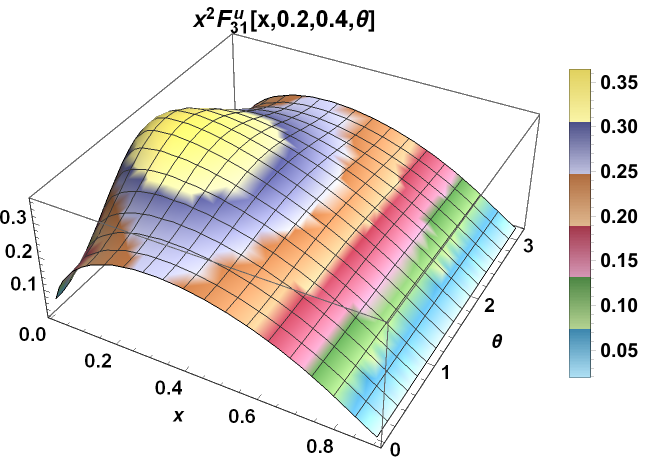}
		\hspace{0.05cm}
		(b)\includegraphics[width=7.3cm]{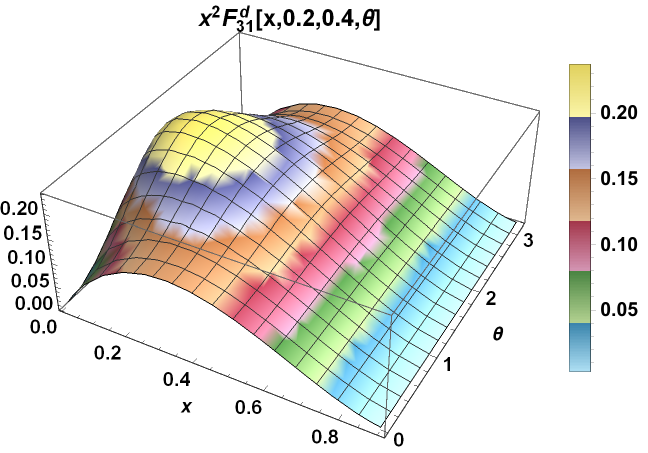}
		\hspace{0.05cm}\\
	\end{minipage}
	\caption{\label{fig3dth} (Color online) The twist-4 GTMD 		
		$x^2 F_{3,1}^{\nu}(x, p_{\perp},\Delta_{\perp},\theta)$
		plotted with respect to $x$ and ${\theta}$ for ${{ p_\perp}=0.2~\mathrm{GeV}}$ and ${ \Delta_\perp}= 0.4~\mathrm{GeV}$. The left and right column correspond to $u$ and $d$ quarks sequentially.
	}
\end{figure*}

 Being mother distributions, the GTMDs show multitude of distribution functions at their respective limits. At skewness $\xi=0$, the GTMDs have dependence on four variables $X^{\nu}(x, p_{\perp},\Delta_{\perp},\theta)$. Since, it is not possible to show the variation on four variables simultaneously, we will study their variation with two variables at a time while keeping the others fixed or integrated. Even though the angle $\theta$ between $\bfp$ and $\Dp$ plane ranges from $0$ to $\pi$ whereas $\bfp \cdot \Dp$ is ranging from $p_{\perp} \Delta_{\perp}$ to $-p_{\perp} \Delta_{\perp}$, we have fixed $\theta$ at $0$ throughout the discussion for our study since it is observed that when $\bfp \parallel \Dp$, all the GTMDs from Eqn. (\ref{f31s}) to Eqn. (\ref{h38v}) are non zero. For other values of $\theta$, some of the GTMDs vanish.Because $\theta$ is a crucial kinematic variable that contributes to the angular dependence, factorization, and the overall behavior of cross sections in scattering processes, for sake of completion, we have provided the variation of GTMD $F_{3,1}$ with the angle $\theta$ and longitudinal momentum fraction $x$  for ${p_\perp=0.2~\mathrm{GeV}}$ and $\Delta_{\perp}= 0.4~\mathrm{GeV}$ in Fig. (\ref{fig3dth}) for both the $u$ and $d$ quarks. The maximum rise in the contribution of GTMD $F_{3,1}$ at $\theta=\pi/2$ signifies the alignment of non flipping quark helicities with the proton spin in the same and opposite direction for $\pm1$ and $0$ diquark helicities respectively.
\subsection{GTMDs}\label{ssgtmds}
\subsubsection{Hermiticity}\label{sssher}
From the direct consequence of the relation between GTMDs and GPCFs \cite{Meissner:2009ww}, it has been shown that GTMDs follows the hermiticity constraint
\begin{equation}
	X^{\nu*}(x, p_{\perp},\Delta_{\perp},\theta)
	=\pm X^{\nu}(x, p_{\perp},\Delta_{\perp},\pi-\theta) \,,
	\label{e:gtmd_hermiticity}
\end{equation}
with a plus sign for $X=
F_{3,1}, F_{3,3}, F_{3,4}, G_{3,1}, G_{3,2}, G_{3,4}, H_{3,2}, H_{3,3},
H_{3,4}, H_{3,6}, H_{3,7}$ and a minus sign for $F_{3,2},
G_{3,3}, H_{3,1}, H_{3,5}, H_{3,8}$. We have checked it for our results and these constraints have been followed by our calculations as well.

%\subsection{Light-Front Parity}\label{sslfpar}

\subsubsection{Variation with $x$ and ${ p_\perp}$}\label{ssxp}
\begin{figure*}
	\centering
	\begin{minipage}[c]{0.98\textwidth}
		(a)\includegraphics[width=7.3cm]{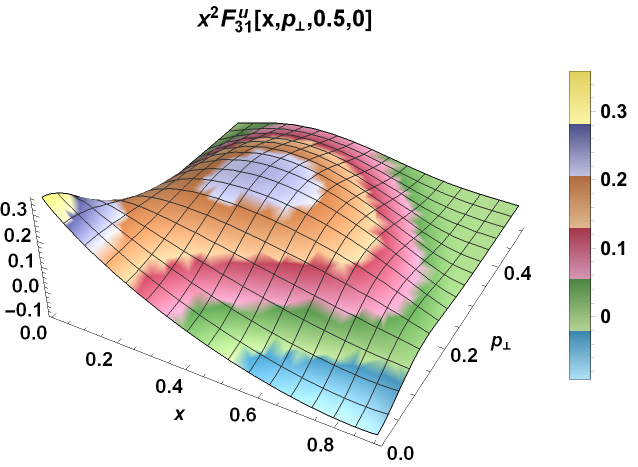}
		\hspace{0.05cm}
		(b)\includegraphics[width=7.3cm]{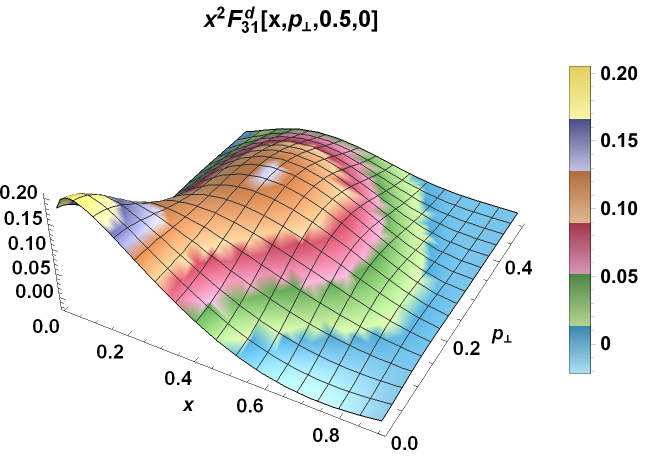}
		\hspace{0.05cm}
		(c)\includegraphics[width=7.3cm]{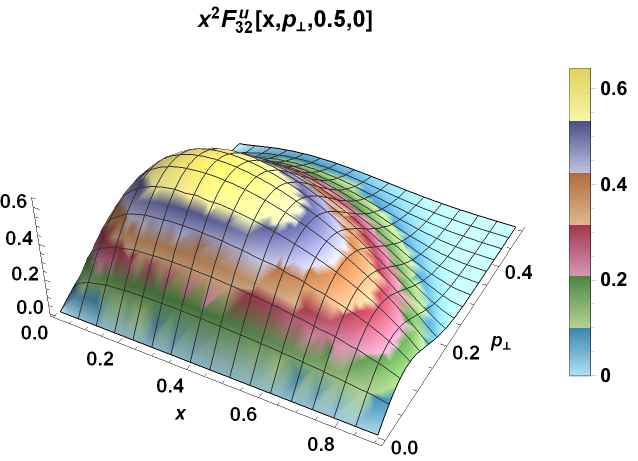}
		\hspace{0.05cm}
		(d)\includegraphics[width=7.3cm]{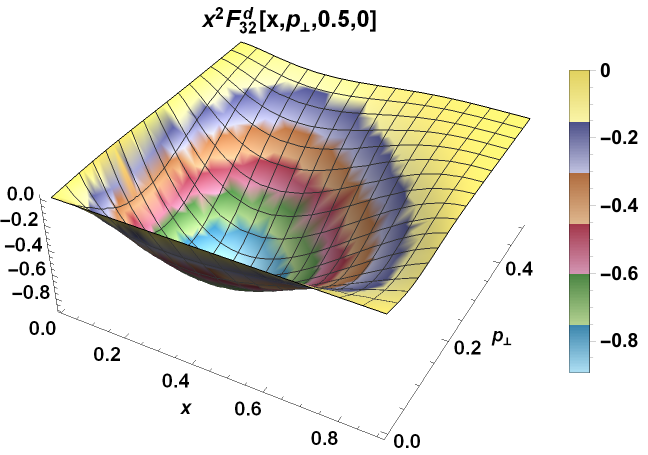}
		\hspace{0.05cm}
		(e)\includegraphics[width=7.3cm]{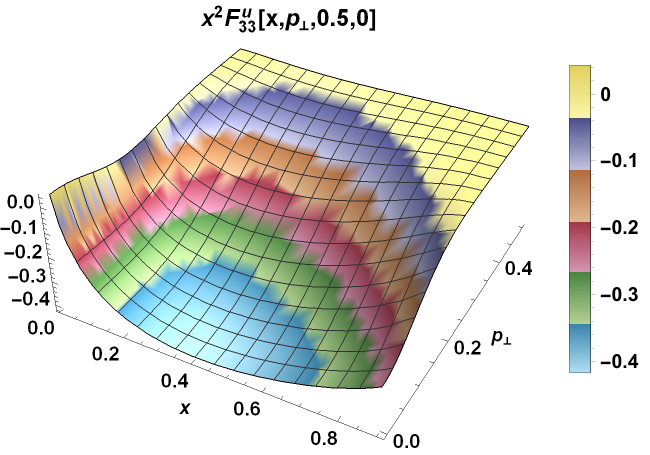}
		\hspace{0.05cm}
		(f)\includegraphics[width=7.3cm]{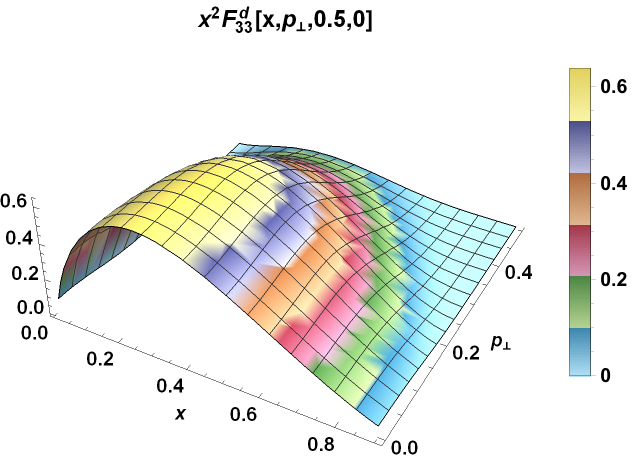}
		\hspace{0.05cm}
		(g)\includegraphics[width=7.3cm]{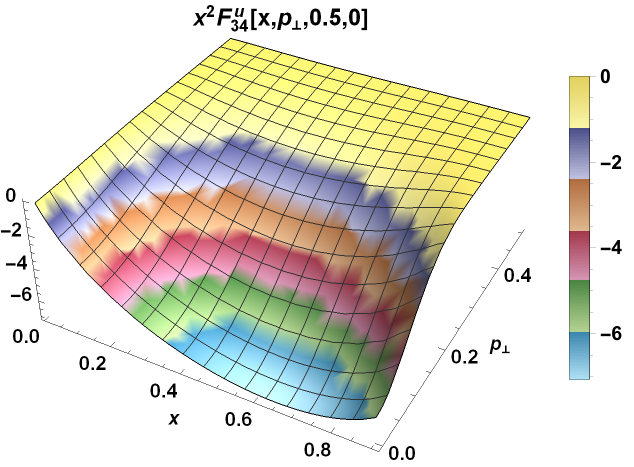}
		\hspace{0.05cm}
		(h)\includegraphics[width=7.3cm]{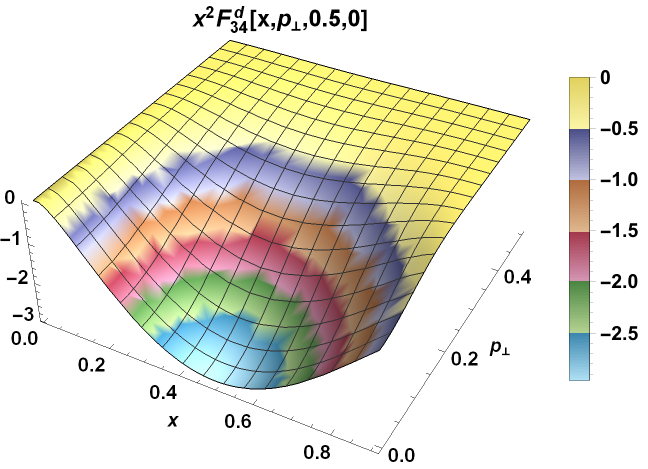}
		\hspace{0.05cm}\\
	\end{minipage}
	\caption{\label{fig3dXPF} (Color online) The twist-4 GTMDs 		
		$x^2 F_{3,1}^{\nu}(x, p_{\perp},\Delta_{\perp},\theta)$,
		$x^2 F_{3,2}^{\nu}(x, p_{\perp},\Delta_{\perp},\theta)$,
		$x^2 F_{3,3}^{\nu}(x, p_{\perp},\Delta_{\perp},\theta)$ and
		$x^2 F_{3,4}^{\nu}(x, p_{\perp},\Delta_{\perp},\theta)$
	 plotted with respect to $x$ and ${{ p_\perp}}$ for ${ \Delta_\perp}= 0.5~\mathrm{GeV}$ for ${\bfp} \parallel {\Dp}$. The left and right column correspond to $u$ and $d$ quarks sequentially.
}
\end{figure*}

\begin{figure*}
	\centering
	\begin{minipage}[c]{0.98\textwidth}
		(a)\includegraphics[width=7.3cm]{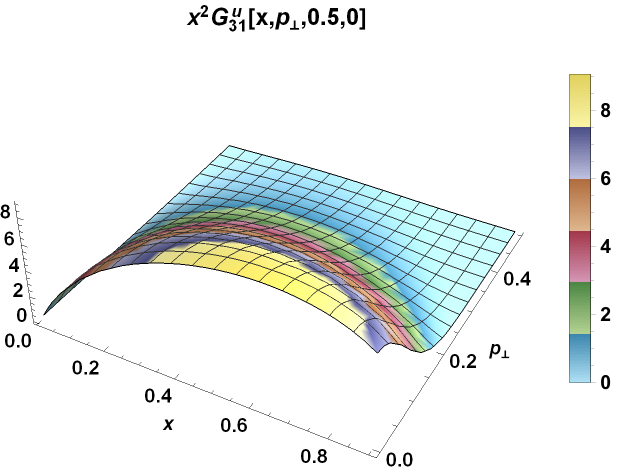}
		\hspace{0.05cm}
		(b)\includegraphics[width=7.3cm]{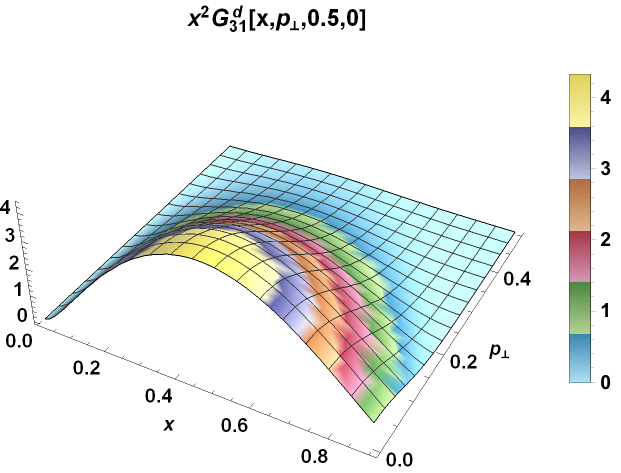}
		\hspace{0.05cm}
		(c)\includegraphics[width=7.3cm]{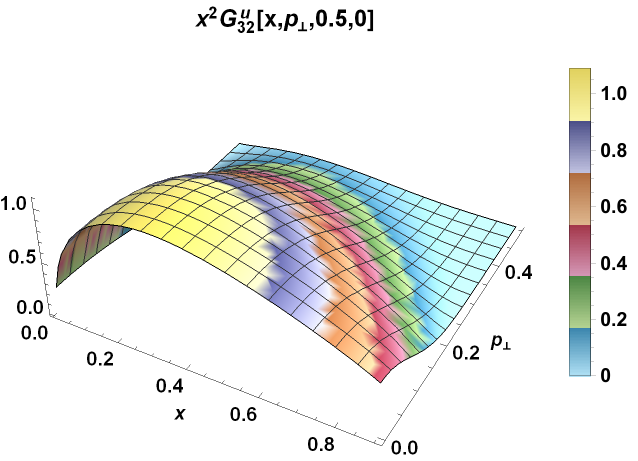}
		\hspace{0.05cm}
		(d)\includegraphics[width=7.3cm]{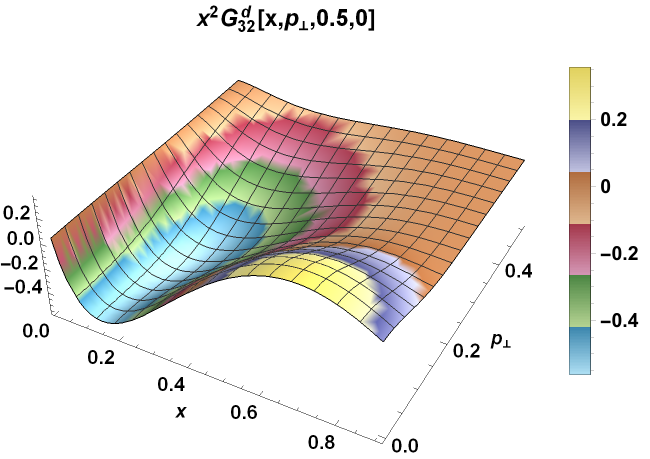}
		\hspace{0.05cm}
		(e)\includegraphics[width=7.3cm]{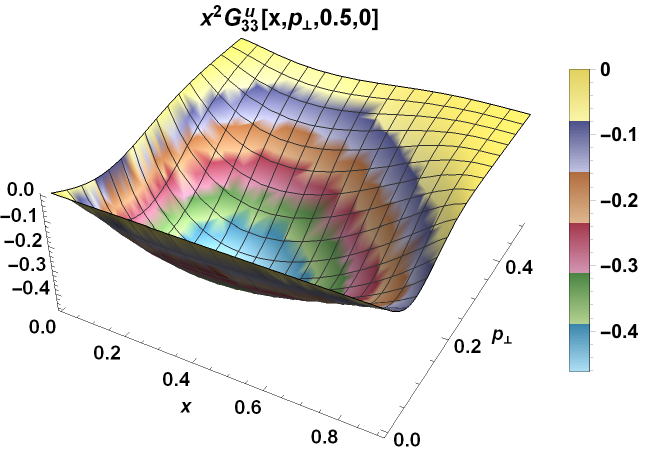}
		\hspace{0.05cm}
		(f)\includegraphics[width=7.3cm]{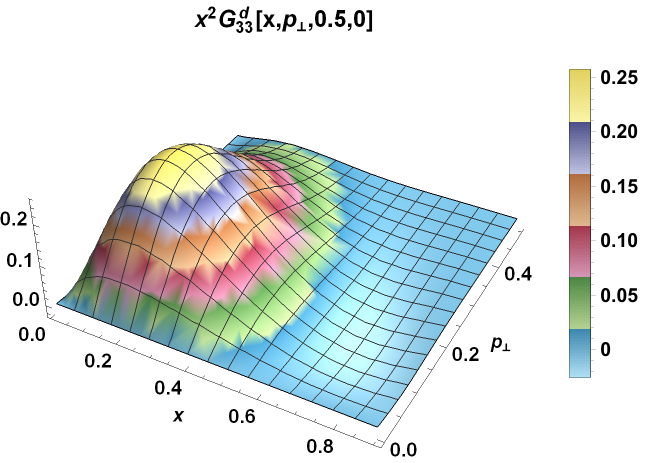}
		\hspace{0.05cm}
		(g)\includegraphics[width=7.3cm]{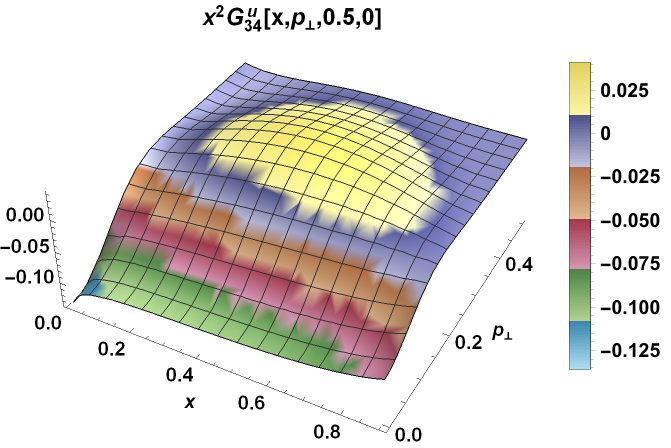}
		\hspace{0.05cm}
		(h)\includegraphics[width=7.3cm]{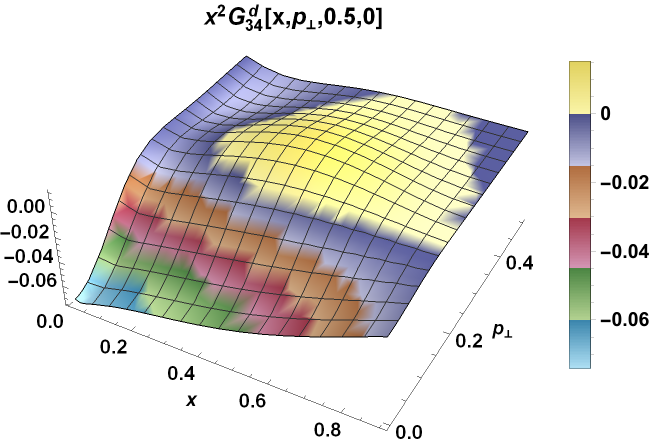}
		\hspace{0.05cm}\\
	\end{minipage}
	\caption{\label{fig3dXPG} (Color online) The twist-4 GTMDs 		
		$x^2 G_{3,1}^{\nu}(x, p_{\perp},\Delta_{\perp},\theta)$,
		$x^2 G_{3,2}^{\nu}(x, p_{\perp},\Delta_{\perp},\theta)$,
		$x^2 G_{3,3}^{\nu}(x, p_{\perp},\Delta_{\perp},\theta)$ and
		$x^2 G_{3,4}^{\nu}(x, p_{\perp},\Delta_{\perp},\theta)$
		plotted with respect to $x$ and ${{ p_\perp}}$ for ${ \Delta_\perp}= 0.5~\mathrm{GeV}$ for ${\bfp} \parallel {\Dp}$. The left and right column correspond to $u$ and $d$ quarks sequentially.
	}
\end{figure*}

\begin{figure*}
	\centering
	\begin{minipage}[c]{0.98\textwidth}
		(a)\includegraphics[width=7.3cm]{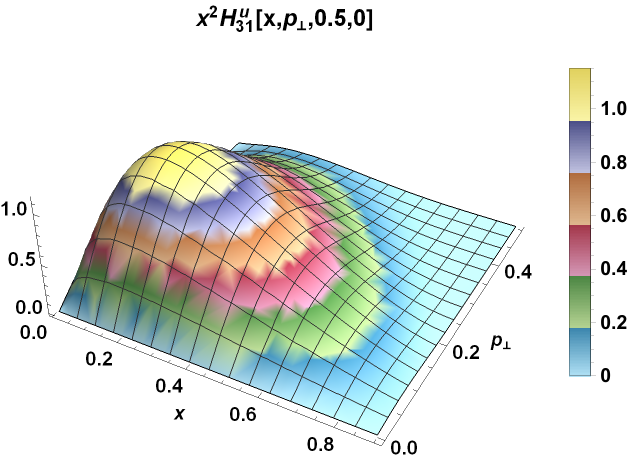}
		\hspace{0.05cm}
		(b)\includegraphics[width=7.3cm]{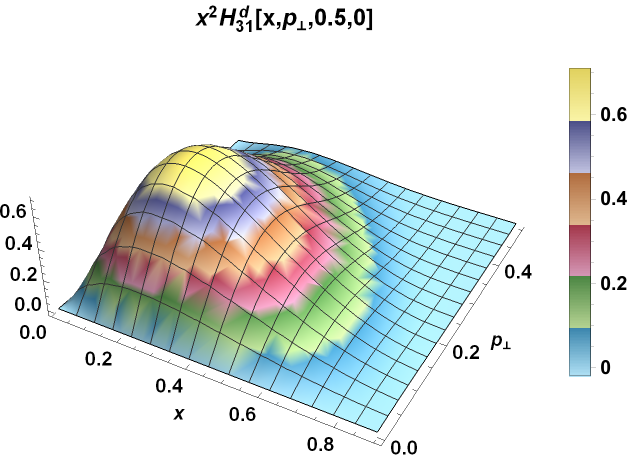}
		\hspace{0.05cm}
		(c)\includegraphics[width=7.3cm]{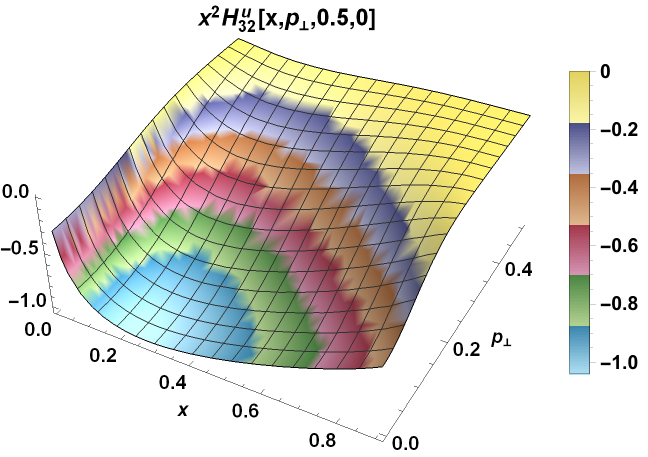}
		\hspace{0.05cm}
		(d)\includegraphics[width=7.3cm]{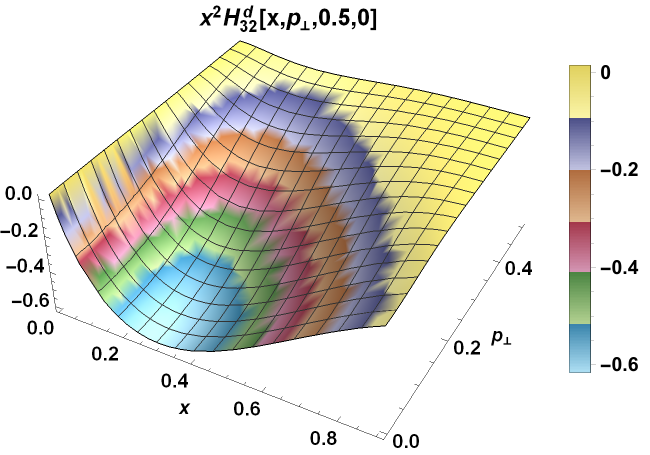}
		\hspace{0.05cm}
		(e)\includegraphics[width=7.3cm]{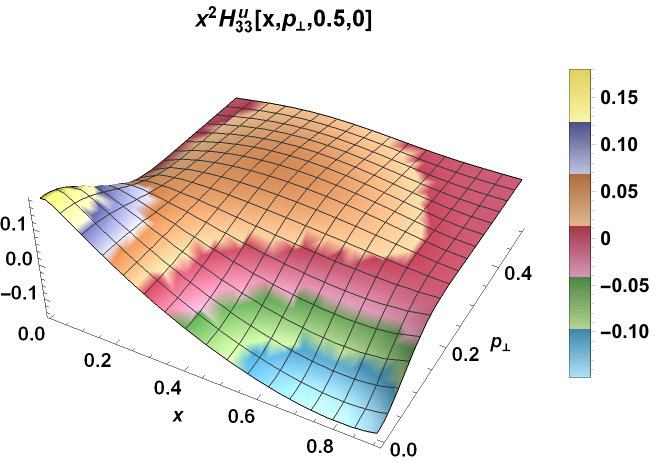}
		\hspace{0.05cm}
		(f)\includegraphics[width=7.3cm]{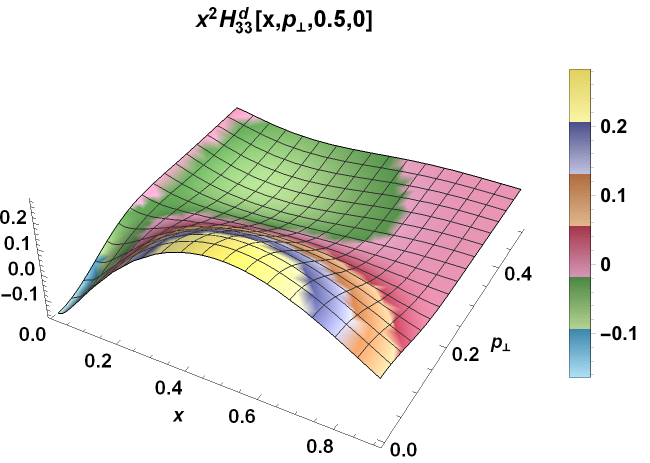}
		\hspace{0.05cm}
		(g)\includegraphics[width=7.3cm]{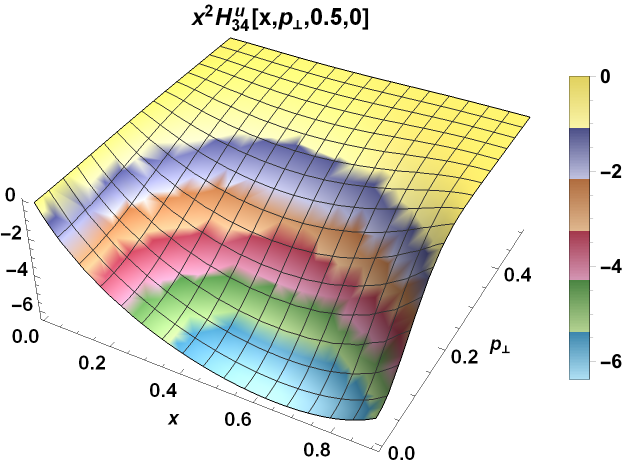}
		\hspace{0.05cm}
		(h)\includegraphics[width=7.3cm]{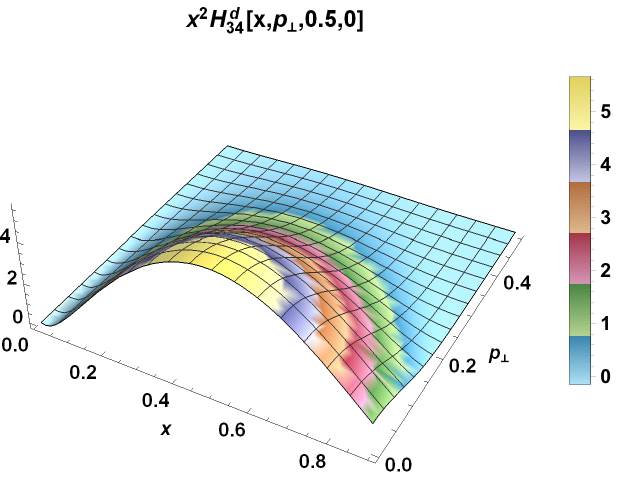}
		\hspace{0.05cm}\\
	\end{minipage}
	\caption{\label{fig3dXPH1} (Color online) The twist-4 GTMDs 		
		$x^2 H_{3,1}^{\nu}(x, p_{\perp},\Delta_{\perp},\theta)$,
		$x^2 H_{3,2}^{\nu}(x, p_{\perp},\Delta_{\perp},\theta)$,
		$x^2 H_{3,3}^{\nu}(x, p_{\perp},\Delta_{\perp},\theta)$ and
		$x^2 H_{3,4}^{\nu}(x, p_{\perp},\Delta_{\perp},\theta)$
		plotted with respect to $x$ and ${{ p_\perp}}$ for ${ \Delta_\perp}= 0.5~\mathrm{GeV}$ for ${\bfp} \parallel {\Dp}$. The left and right column correspond to $u$ and $d$ quarks sequentially.
	}
\end{figure*}

\begin{figure*}
	\centering
	\begin{minipage}[c]{0.98\textwidth}
		(a)\includegraphics[width=7.3cm]{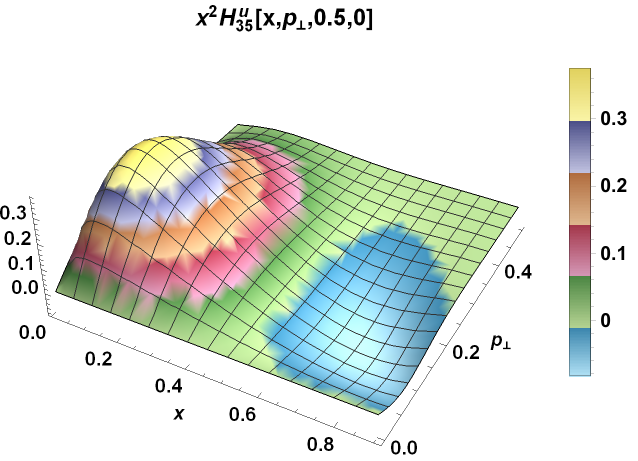}
		\hspace{0.05cm}
		(b)\includegraphics[width=7.3cm]{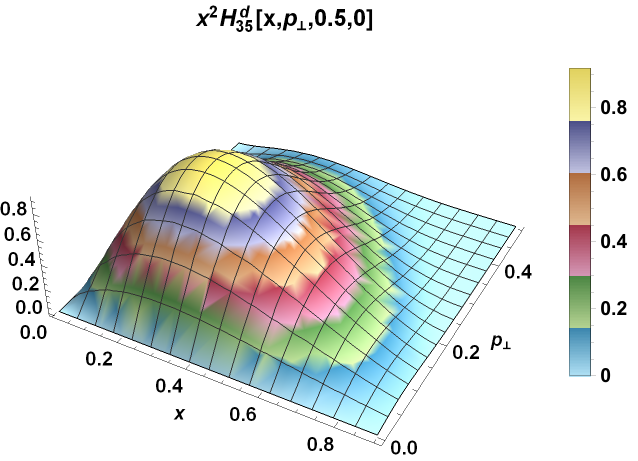}
		\hspace{0.05cm}
		(c)\includegraphics[width=7.3cm]{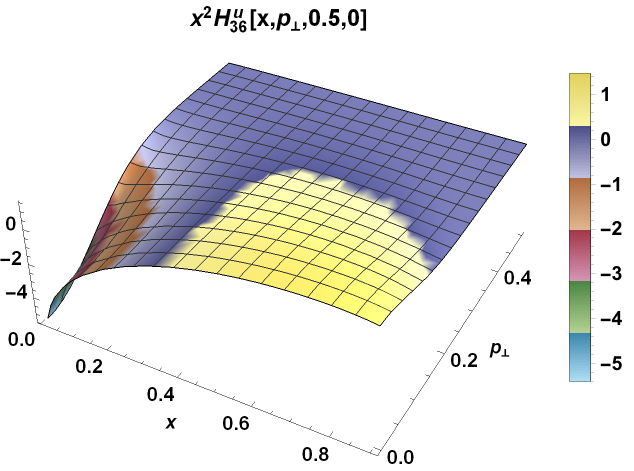}
		\hspace{0.05cm}
		(d)\includegraphics[width=7.3cm]{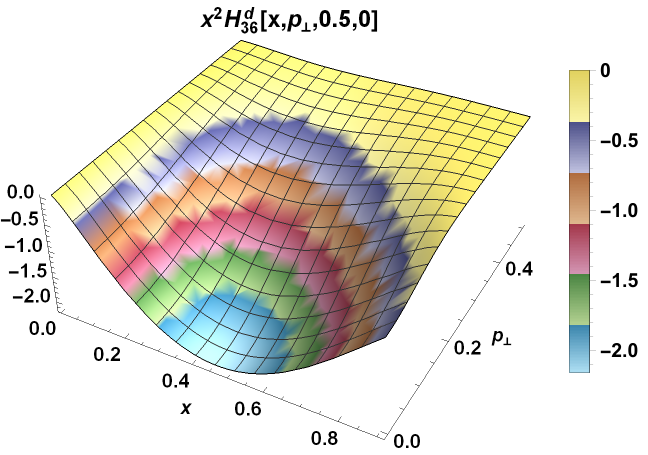}
		\hspace{0.05cm}
		(e)\includegraphics[width=7.3cm]{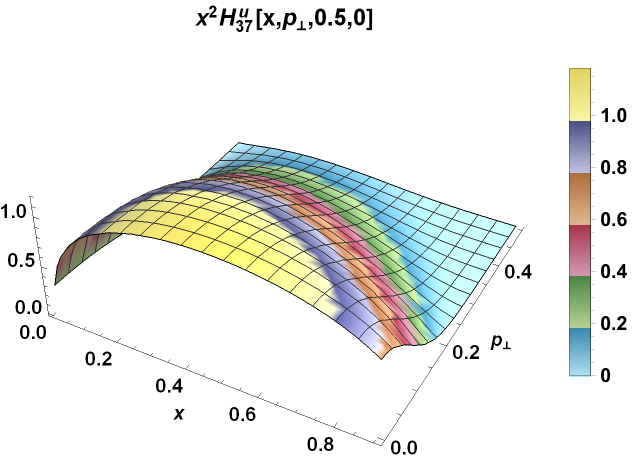}
		\hspace{0.05cm}
		(f)\includegraphics[width=7.3cm]{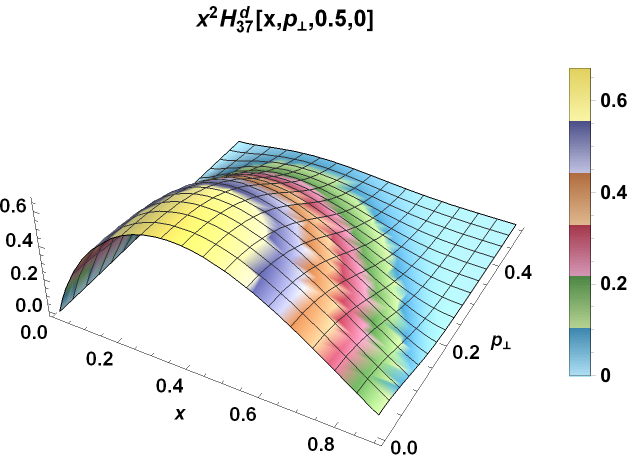}
		\hspace{0.05cm}
		(g)\includegraphics[width=7.3cm]{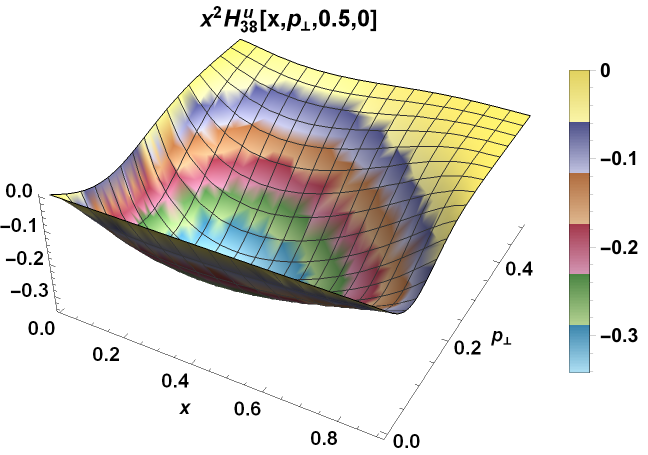}
		\hspace{0.05cm}
		(h)\includegraphics[width=7.3cm]{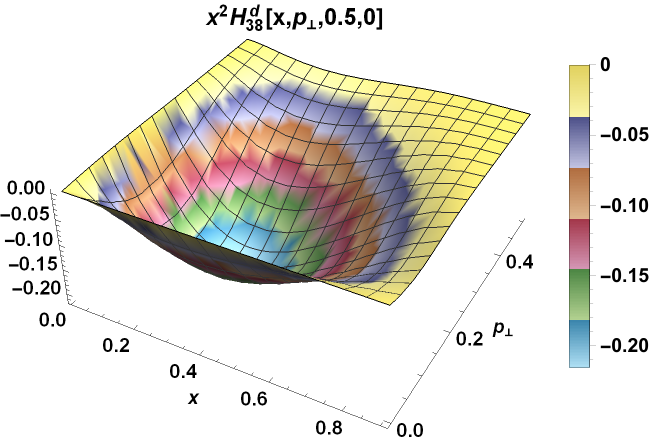}
		\hspace{0.05cm}\\
	\end{minipage}
	\caption{\label{fig3dXPH2}  (Color online) The twist-4 GTMDs 		
		$x^2 H_{3,5}^{\nu}(x, p_{\perp},\Delta_{\perp},\theta)$,
		$x^2 H_{3,6}^{\nu}(x, p_{\perp},\Delta_{\perp},\theta)$,
		$x^2 H_{3,7}^{\nu}(x, p_{\perp},\Delta_{\perp},\theta)$ and
		$x^2 H_{3,8}^{\nu}(x, p_{\perp},\Delta_{\perp},\theta)$
		plotted with respect to $x$ and ${{ p_\perp}}$ for ${ \Delta_\perp}= 0.5~\mathrm{GeV}$ for ${\bfp} \parallel {\Dp}$. The left and right column correspond to $u$ and $d$ quarks sequentially.
	}
\end{figure*}

When there is no momentum transfer between the initial and the final state proton i.e. at the TMD limit $\Delta=0$, twist-4 T-even TMDs can be obtained from GTMDs as \cite{Meissner:2009ww}
\begin{eqnarray}
	f_3^{\nu}(x,{ p_\perp}) & = & F_{3,1}^{\nu}(x, { p_\perp},0,\theta) \,, \label{tmd1} \\
		g_{3T}^{\nu}(x,{ p_\perp}) & = & G_{3,2}^{\nu}(x, { p_\perp},0,\theta) \,, \label{tmd2} \\
	g_{3L}^{\nu}(x,{ p_\perp}) & = & G_{3,4}^{\nu}(x, { p_\perp},0,\theta) \,, \label{tmd3} \\
	h_{3T}^{\nu}(x,{ p_\perp}) & = & H_{3,3}^{\nu}(x, { p_\perp},0,\theta) \,, \label{tmd4}\\
	h_{3T}^{\bot \nu}(x,{ p_\perp}) & = & H_{3,4}^{\nu}(x, { p_\perp},0,\theta) \,, \label{tmd5}\\
		h_{3L}^{\bot \nu}(x,{ p_\perp}) & = & H_{3,7}^{\nu}(x, { p_\perp},0,\theta) \,. \label{tmd6} 
\end{eqnarray}
Twist-4 T-even TMDs and corresponding PDFs in LFQDM have already been discussed in detail in Ref. \cite{sstwist4}. Here, we will study the behavior of twist-4 GTMDs at non zero $\Delta$ with variation in the longitudinal momentum fraction $x$ and the transverse momentum of quark ${ p_\perp}$. In Figs. (\ref{fig3dXPF}), (\ref{fig3dXPG}), (\ref{fig3dXPH1}) and (\ref{fig3dXPH2}), the GTMDs
corresponding to different Dirac matrix structures ($\Gamma= \gamma^-,  \gamma^-\gamma_5$ and $\sigma^{i-}\gamma_5$) given in 
 Eqs. {\eqref{f31s}} to {\eqref{h38v}} $F_{3,1}, F_{3,2}, F_{3,3}, F_{3,4}, G_{3,1}, G_{3,2}, G_{3,3}, G_{3,4}, H_{3,1}, H_{3,2}, H_{3,3}, H_{3,4},  H_{3,5}, H_{3,6}, H_{3,7}$ and  $H_{3,8}$ have been plotted with respect to $x$ and ${ p_\perp}$ at ${ \Delta_\perp}= 0.5~\mathrm{GeV}$ keeping $\theta=0$. \par
Firstly, corresponding to the Dirac matrix structure $\Gamma=\gamma^-$, GTMDs $F_{3,1}, F_{3,2}, F_{3,3}$ and $F_{3,4}$ have been plotted in Fig. (\ref{fig3dXPF}). In Fig. \ref{fig3dXPF} (a) and \ref{fig3dXPF} (b), the GTMD $F_{3,1}^{\nu}$ $(\nu=u, d)$ has been plotted which is related to the twist-4 TMD $f_3^{\nu}(x,{ p_\perp})$ via Eq. (\ref{tmd1}). The GTMD $F_{3,1}^{\nu}$ possesses a similar behavior for $u$ as well as $d$ quarks with variation only in their amplitudes. Similar to its TMD counterpart \cite{sstwist4}, the maximum possibility of having quark and proton polarization corresponding to this GTMD occurs when the quark carries about $20\%-30\%$ of the longitudinal momentum from proton. The presence of non zero momentum transfer $\Delta$ is responsible for low magnitude of GTMD in comparison the TMD $f_3^{\nu}(x,{ p_\perp})$. The GTMD $F_{3,2}^{\nu}$ has been plotted in Fig. \ref{fig3dXPF} (c) and \ref{fig3dXPF} (d) for $\nu$ being $u$ and $d$ quarks respectively. It has been observed that, on changing the quark flavor from $u$ to $d$, the sign of GTMD reverses . The magnitudes are also different. This is in sync with the GTMD expression for $u$ and $d$ quarks obtained from scalar and vector diquark parts Eq. (\ref{f32s}) and Eq. (\ref{f32a}) respectively. In Fig. \ref{fig3dXPF} (e) and \ref{fig3dXPF} (f), the GTMD $F_{3,3}^{\nu}$ has been plotted for $\nu=u$ and $d$ quark respectively. Behavior of this GTMD looks similar over the change in quark flavor, however,  in this case it is negative for $u$ quarks and positive for $d$ quarks. Unlike $F_{3,2}^{\nu}$, the expression for both the flavors is not identical for $F_{3,3}^{\nu}$. Contrary to previously studied GTMDs, the maximum amplitude of GTMD in this case is not at a value corresponding to $\frac{1}{3}^{rd}$ longitudinal momentum of proton, instead, it shifts to high and low $x$ for $u$ and $d$ quark flavors respectively. GTMD $F_{3,4}^{\nu}$ has been plotted in Fig. \ref{fig3dXPF} (g) and \ref{fig3dXPF} (h) for $u$ and $d$ quarks respectively. The trend of this GTMD is same for both flavors with variation only in the magnitude. This is due to the varying coefficients in scalar and vector diquark expressions, Eq. (\ref{f34s}) and Eq. (\ref{f34a}) respectively. The possibility of having this GTMD is maximum when average momentum of quark is negligible and share of longitudinal momentum of proton is $40\%$ to $80\%$. Even though there is a comparatively less chance  of having quark with other values of $x$ having non zero transverse momentum but chances are still finite.  
%
%
%---------------G k PLOT---------------------%
%
%
\par Now, for the study of GTMDs having the Dirac matrix structure $\Gamma=\gamma^-\gamma^5$ (i.e., $G_{3,1}, G_{3,2}, G_{3,3}$ and $G_{3,4}$), they have been plotted with respect to $x$ and ${ p_\perp}$ in Fig. (\ref{fig3dXPG}), keeping ${ \Delta_\perp}$ and $\theta$ fixed at $0.5~\mathrm{GeV}$ and $0^{\circ}$ respectively. In Fig. \ref{fig3dXPG} (a) and \ref{fig3dXPG} (b), the GTMD $G_{3,1}^{\nu}$ has been plotted which is positive over the entire range of $x$ and ${ p_\perp}$ with its general behavior being the same for $u$ as well as $d$ quarks. The variation is only in the amplitude which is again due to same dependence on the plotted variables in scalar and vector diquark parts in Eq. (\ref{g31s}) and Eq. (\ref{g31a}) respectively. The GTMD $G_{3,2}^{\nu}$ has been plotted in Fig. \ref{fig3dXPG} (c) and \ref{fig3dXPG} (d) for $u$ and $d$ quarks respectively.
For $u$ quark, the GTMD is positive over the entire range of the plotted variables. It has been observed that, at a particular transverse momentum of quark, with the increase in the longitudinal momentum fraction $x$, the possibility of having GTMD $G_{3,2}^{u}$ first increases and then decreases to give a maxima in between. For the $d$ flavor, the GTMD $G_{3,2}^{d}$ first decreases and then increases to give a minima and after that it again decreases to give maxima with the rise in $x$ at low ${ p_\perp}$ as shown in Fig. \ref{fig3dXPG} (d). The presence of node in this GTMD is due to non-zero momentum transfer ${ \Delta_\perp}$. This can also be ascribed due to presence of non zero orbital angular momentum  and is a mixture of $L_z=\pm1,\pm2,\pm3$ states. This effect does not exist in its TMD partner $g_{3T}^{\nu}(x,{ p_\perp})$ (see Eq. (\ref{tmd2})) where the  amplitude varies. Anti-symmetry is observed over the quark flavor \cite{sstwist4}. The GTMD $G_{3,2}^{\nu}$ is related to the twist-4 TMD $g_{3T}^{\nu}(x,{ p_\perp})$ via Eq. (\ref{tmd2}). $G_{3,3}^{\nu}$ GTMD has been plotted in Fig. \ref{fig3dXPG} (e) and \ref{fig3dXPG} (f). It is negative for $u$ quarks and positive for $d$ quarks, however, due to different contributing expressions of scalar and the vector diquark (Eq. (\ref{g33s}) and Eq. (\ref{g33a})), their behavior is not exactly similar. In Fig. \ref{fig3dXPG} (g) and \ref{fig3dXPG} (h), the GTMD $G_{3,4}^{\nu}$ has been plotted and its behavior is similar to the twist-4 TMD $g_{3L}^{\nu}(x,{ p_\perp})$ \cite{sstwist4} related via Eq. (\ref{tmd3}).
%
%---------------H k PLOT---------------------%
%

\par Finally, considering the Dirac matrix structure $\Gamma=i\sigma^{+-}\gamma_5$, the GTMDs $H_{3,1}$, $H_{3,2}$, $H_{3,3}$, $H_{3,4}$, $H_{3,5}$, $H_{3,6}$, $H_{3,7}$ and $H_{3,8}$ have been plotted in Figs. (\ref{fig3dXPH1}) and (\ref{fig3dXPH2}). 
In the first and second row of Fig. (\ref{fig3dXPH1}), the GTMDs $H_{3,1}^{\nu}$ and  $H_{3,2}^{\nu}$ have been plotted. The behavior of these GTMDs are found to be symmetric under the change in quark flavor. The GTMD $H_{3,2}^{\nu}$ has been plotted in Fig. \ref{fig3dXPH1} (c) and \ref{fig3dXPH1} (d) for $u$ and $d$ quarks respectively. The plot is similar for both quark flavors with the variation  only in their magnitude. In Fig. \ref{fig3dXPH1} (e) and \ref{fig3dXPH1} (f), the GTMD $H_{3,3}^{\nu}$ is plotted which is related to the twist-4 TMD 
$h_{3T}^{\nu}(x,{ p_\perp})$ via Eq. (\ref{tmd4}). No quark flavor symmetry is observed in this GTMD. The GTMD $H_{3,4}^{\nu}$ has been plotted in Fig. \ref{fig3dXPH1} (g) and \ref{fig3dXPH1} (h) where on changing the quark flavor from $u$ to $d$ quarks, the GTMD changes its sign. It is related to the twist-4 TMD $h_{3T}^{\bot \nu}(x,{ p_\perp})$ via Eq. (\ref{tmd5}). In Fig. \ref{fig3dXPH2} (a) to \ref{fig3dXPH2} (d) the GTMDs $H_{3,5}^{\nu}$ and $H_{3,6}^{\nu}$ are plotted for $u$ and $d$ quarks. In both of these GTMDs, no quark flavor symmetry is observed. The GTMD $H_{3,7}^{\nu}$ is plotted in Fig. \ref{fig3dXPH2} (e) and \ref{fig3dXPH2} (f) and is related to twist-4 TMD $	h_{3L}^{\bot \nu}(x,{ p_\perp})$ through Eq. (\ref{tmd6}). The GTMD varies in a similar fashion just like it's TMD partner and it possess quark flavor symmetry. For GTMD  $H_{3,8}^{\nu}$, we observe similar negative plots varying only by the amplitude for both the quark flavors in Fig. \ref{fig3dXPH2} (g) and \ref{fig3dXPH2} (h). This is because the expressions for $u$ and $d$ quark contains the same dependence on variables $x$ and ${ p_\perp}$. 
\subsubsection{Variation with $x$ and ${ \Delta_\perp}$}\label{ssxd}
\begin{figure*}
	\centering
	\begin{minipage}[c]{0.98\textwidth}
		(a)\includegraphics[width=7.3cm]{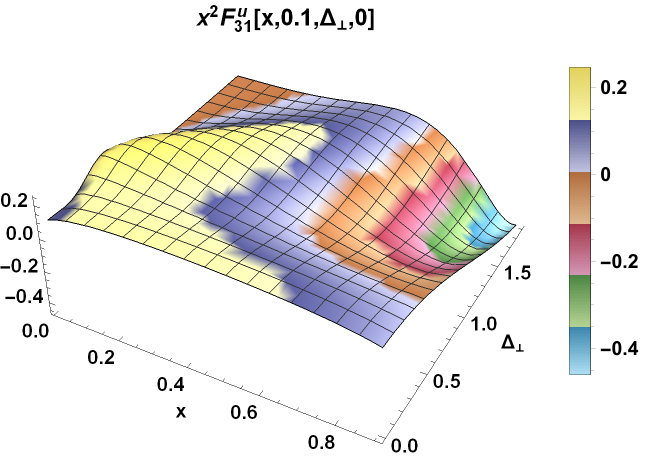}
		\hspace{0.05cm}
		(b)\includegraphics[width=7.3cm]{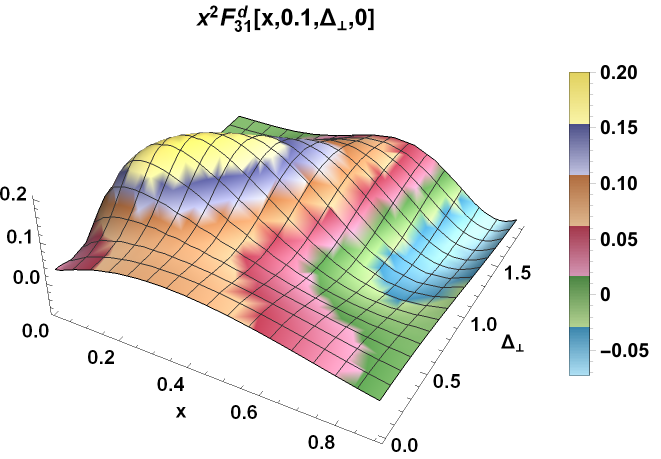}
		\hspace{0.05cm}
		(c)\includegraphics[width=7.3cm]{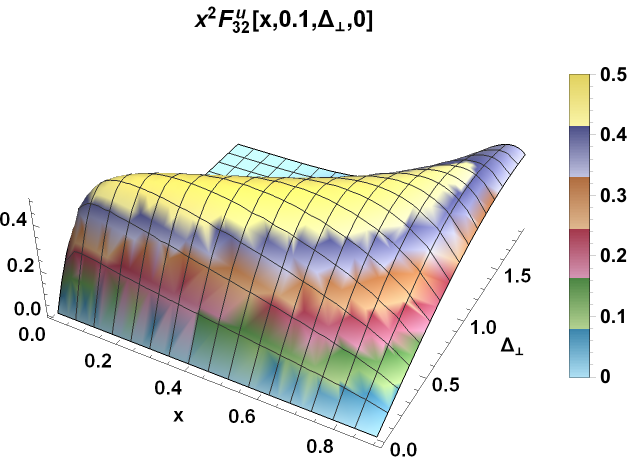}
		\hspace{0.05cm}
		(d)\includegraphics[width=7.3cm]{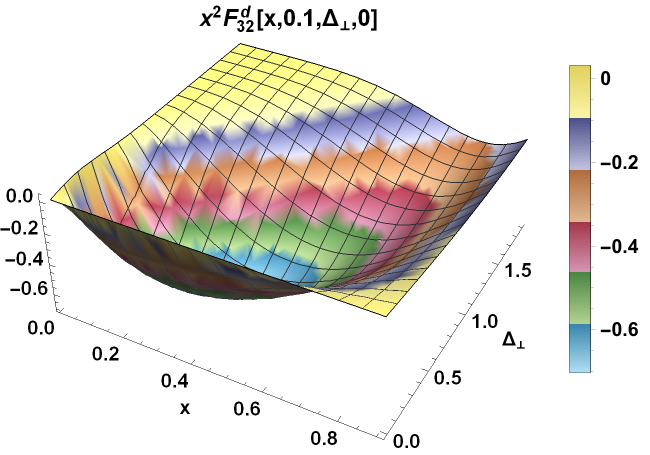}
		\hspace{0.05cm}
		(e)\includegraphics[width=7.3cm]{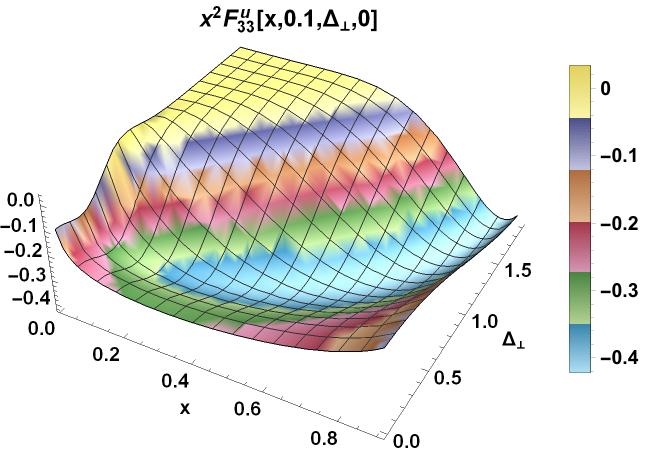}
		\hspace{0.05cm}
		(f)\includegraphics[width=7.3cm]{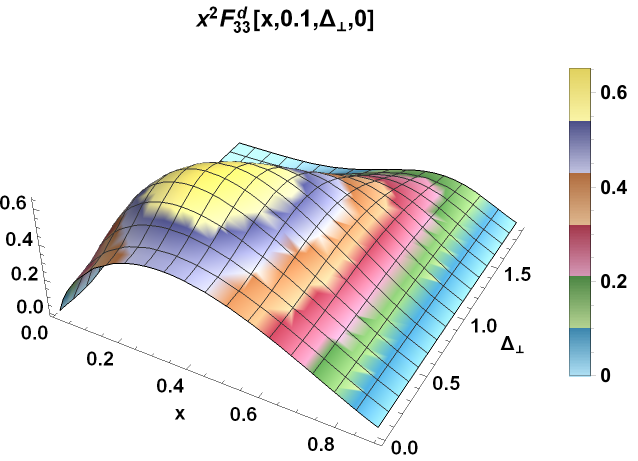}
		\hspace{0.05cm}
		(g)\includegraphics[width=7.3cm]{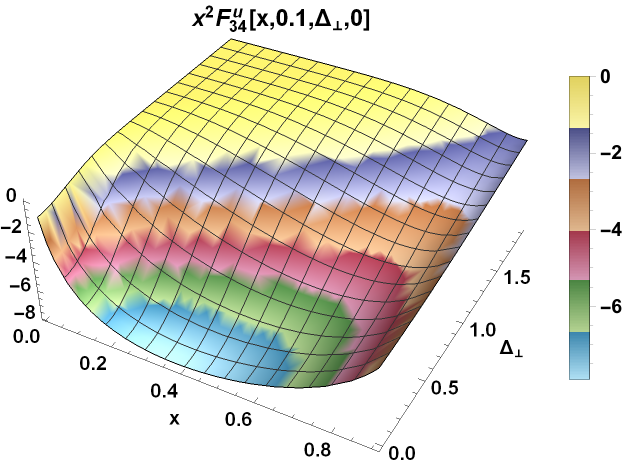}
		\hspace{0.05cm}
		(h)\includegraphics[width=7.3cm]{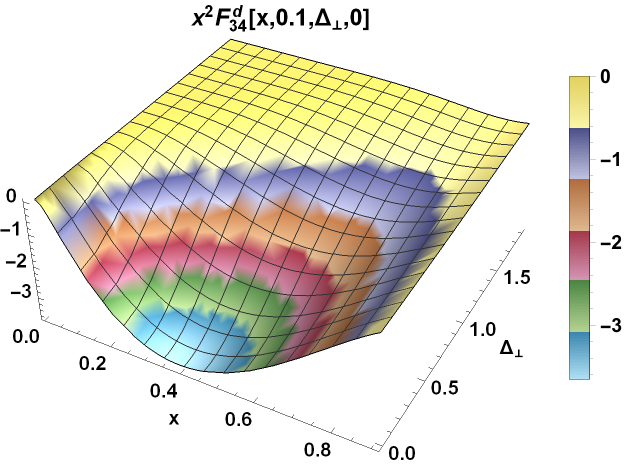}
		\hspace{0.05cm}\\
	\end{minipage}
	\caption{\label{fig3dXDF}(Color online) The twist-4 GTMDs 		
		$x^2 F_{3,1}^{\nu}(x, p_{\perp},\Delta_{\perp},\theta)$,
		$x^2 F_{3,2}^{\nu}(x, p_{\perp},\Delta_{\perp},\theta)$,
		$x^2 F_{3,3}^{\nu}(x, p_{\perp},\Delta_{\perp},\theta)$ and
		$x^2 F_{3,4}^{\nu}(x, p_{\perp},\Delta_{\perp},\theta)$
		plotted with respect to $x$ and ${{ \Delta_\perp}}$ for ${ p_\perp}= 0.1~\mathrm{GeV}$ for ${\bfp} \parallel {\Dp}$. The left and right column correspond to $u$ and $d$ quarks sequentially.
	}
\end{figure*}

\begin{figure*}
	\centering
	\begin{minipage}[c]{0.98\textwidth}
		(a)\includegraphics[width=7.3cm]{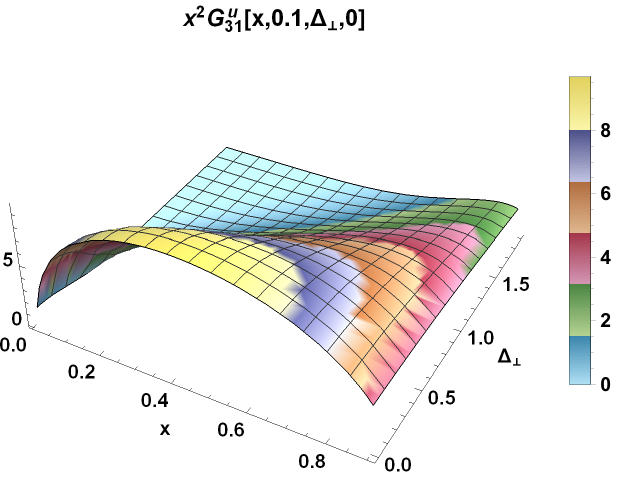}
		\hspace{0.05cm}
		(b)\includegraphics[width=7.3cm]{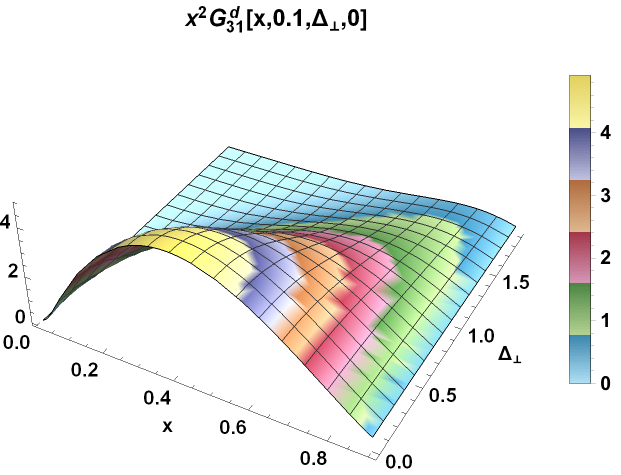}
		\hspace{0.05cm}
		(c)\includegraphics[width=7.3cm]{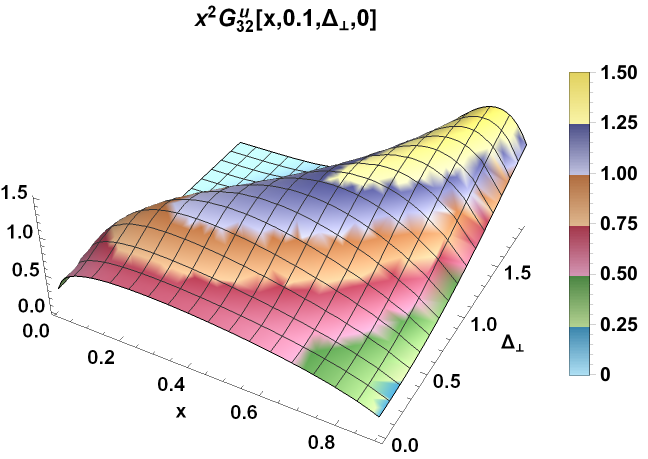}
		\hspace{0.05cm}
		(d)\includegraphics[width=7.3cm]{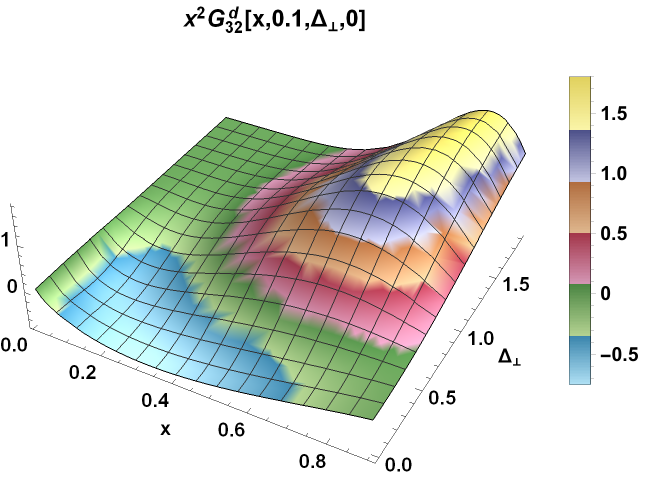}
		\hspace{0.05cm}
		(e)\includegraphics[width=7.3cm]{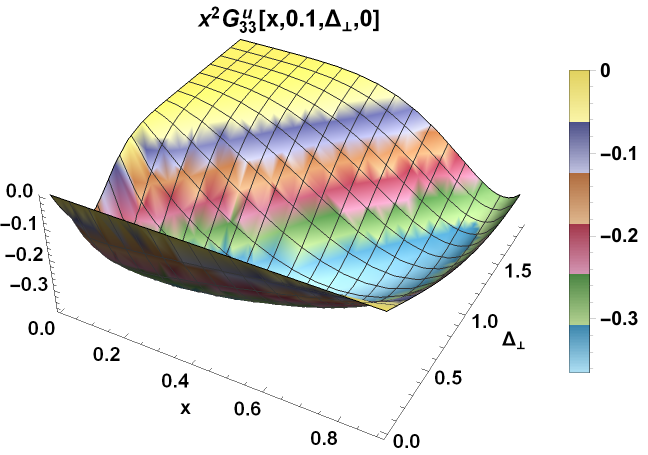}
		\hspace{0.05cm}
		(f)\includegraphics[width=7.3cm]{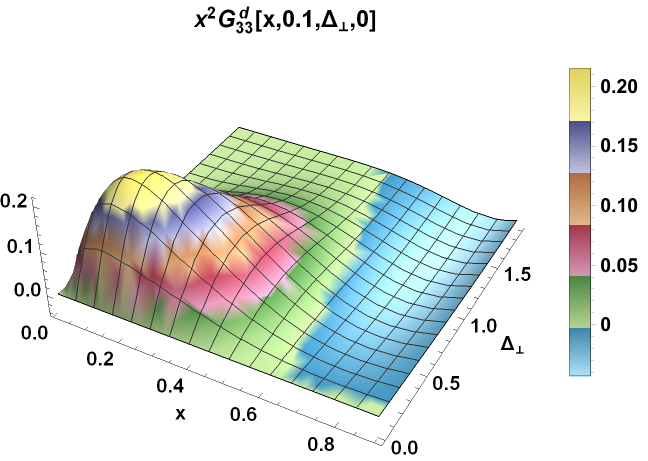}
		\hspace{0.05cm}
		(g)\includegraphics[width=7.3cm]{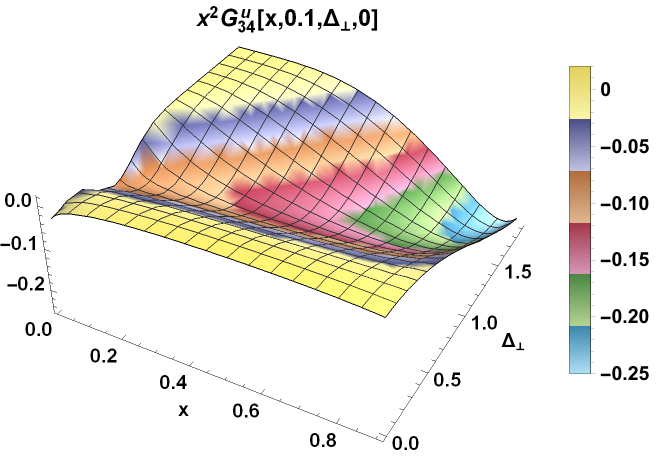}
		\hspace{0.05cm}
		(h)\includegraphics[width=7.3cm]{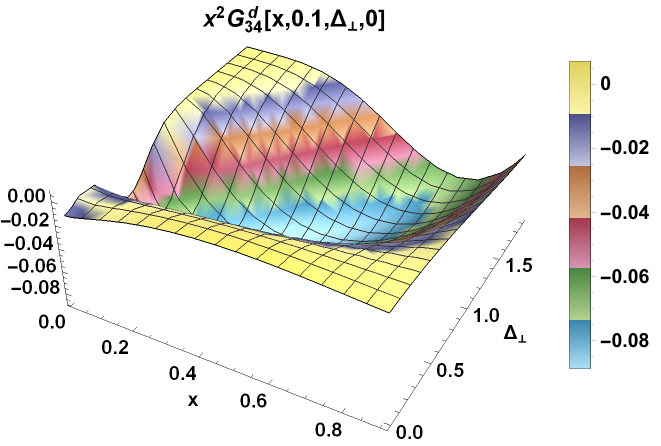}
		\hspace{0.05cm}\\
	\end{minipage}
	\caption{\label{fig3dXDG}  (Color online) The twist-4 GTMDs 		
		$x^2 G_{3,1}^{\nu}(x, p_{\perp},\Delta_{\perp},\theta)$,
		$x^2 G_{3,2}^{\nu}(x, p_{\perp},\Delta_{\perp},\theta)$,
		$x^2 G_{3,3}^{\nu}(x, p_{\perp},\Delta_{\perp},\theta)$ and
		$x^2 G_{3,4}^{\nu}(x, p_{\perp},\Delta_{\perp},\theta)$
		plotted with respect to $x$ and ${{ \Delta_\perp}}$ for ${ p_\perp}= 0.1~\mathrm{GeV}$ for ${\bfp} \parallel {\Dp}$. The left and right column correspond to $u$ and $d$ quarks sequentially.
	}
\end{figure*}

\begin{figure*}
	\centering
	\begin{minipage}[c]{0.98\textwidth}
		(a)\includegraphics[width=7.3cm]{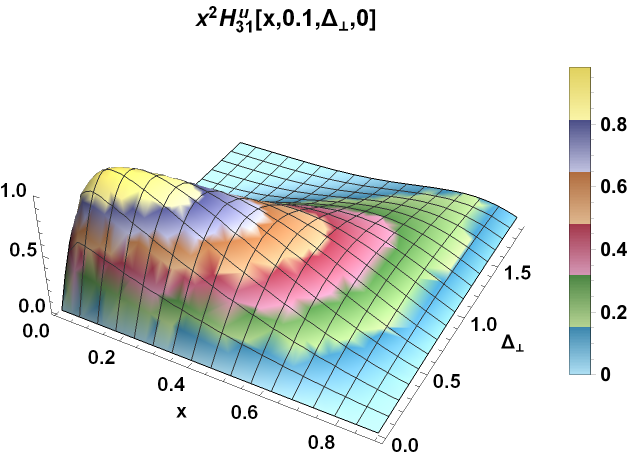}
		\hspace{0.05cm}
		(b)\includegraphics[width=7.3cm]{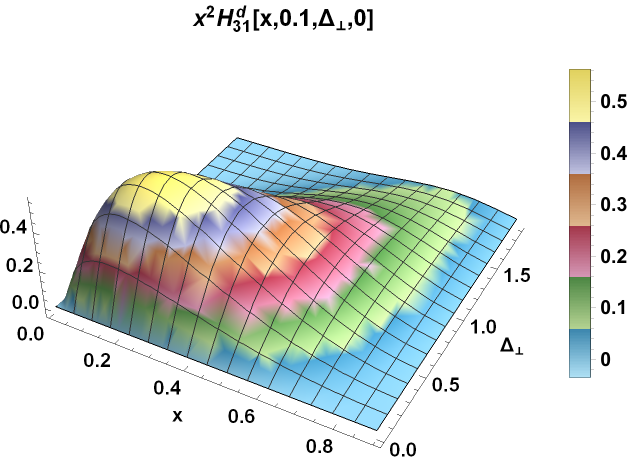}
		\hspace{0.05cm}
		(c)\includegraphics[width=7.3cm]{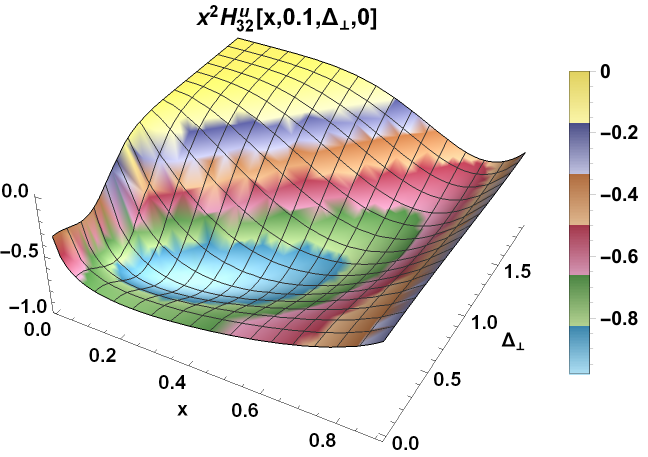}
		\hspace{0.05cm}
		(d)\includegraphics[width=7.3cm]{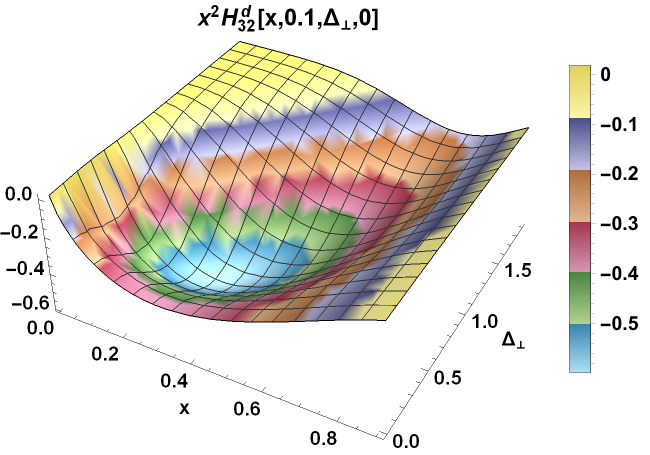}
		\hspace{0.05cm}
		(e)\includegraphics[width=7.3cm]{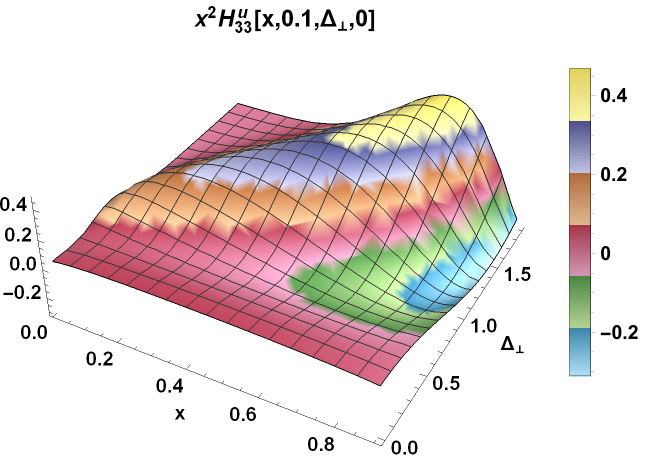}
		\hspace{0.05cm}
		(f)\includegraphics[width=7.3cm]{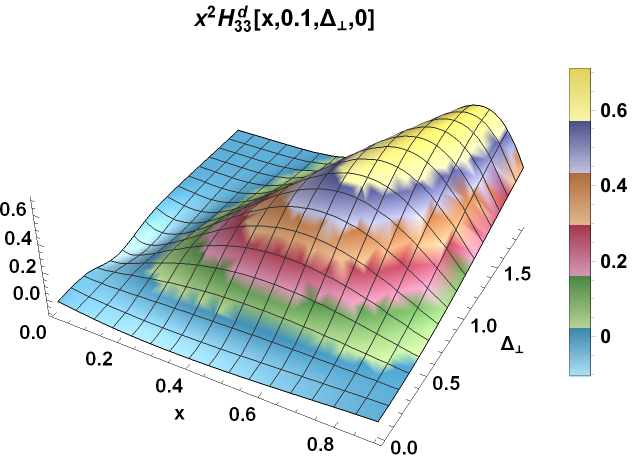}
		\hspace{0.05cm}
		(g)\includegraphics[width=7.3cm]{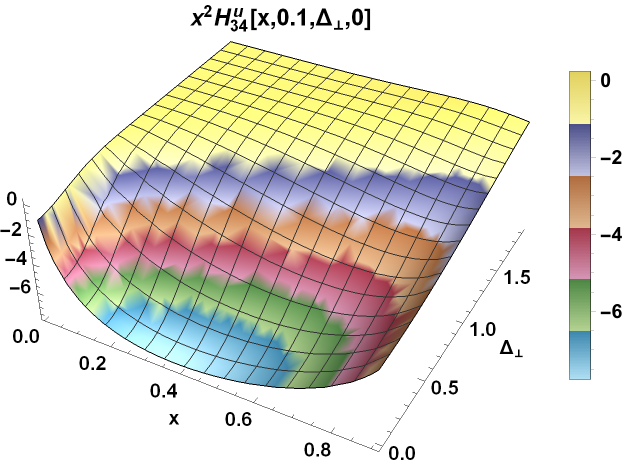}
		\hspace{0.05cm}
		(h)\includegraphics[width=7.3cm]{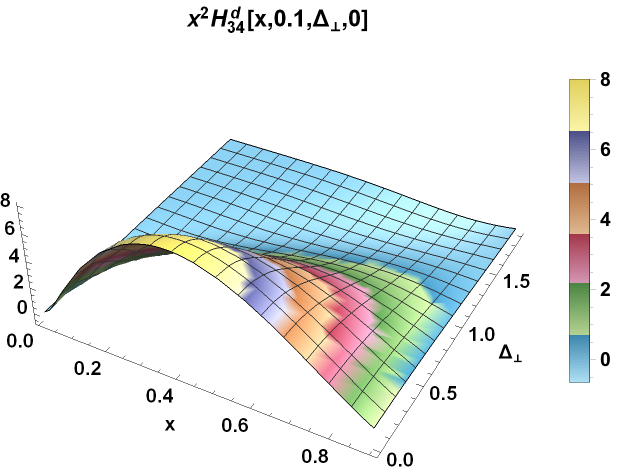}
		\hspace{0.05cm}\\
	\end{minipage}
	\caption{\label{fig3dXDH1} (Color online) The twist-4 GTMDs 		
		$x^2 H_{3,1}^{\nu}(x, p_{\perp},\Delta_{\perp},\theta)$,
		$x^2 H_{3,2}^{\nu}(x, p_{\perp},\Delta_{\perp},\theta)$,
		$x^2 H_{3,3}^{\nu}(x, p_{\perp},\Delta_{\perp},\theta)$ and
		$x^2 H_{3,4}^{\nu}(x, p_{\perp},\Delta_{\perp},\theta)$
		plotted with respect to $x$ and ${{ \Delta_\perp}}$ for ${ p_\perp}= 0.1~\mathrm{GeV}$ for ${\bfp} \parallel {\Dp}$. The left and right column correspond to $u$ and $d$ quarks sequentially.
	}
\end{figure*}

\begin{figure*}
	\centering
	\begin{minipage}[c]{0.98\textwidth}
		(a)\includegraphics[width=7.3cm]{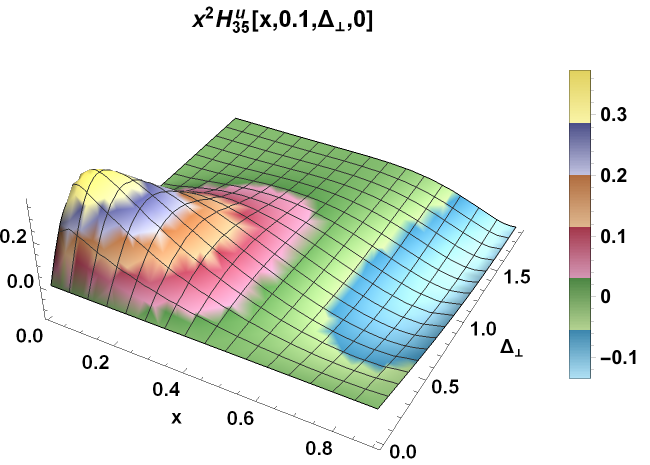}
		\hspace{0.05cm}
		(b)\includegraphics[width=7.3cm]{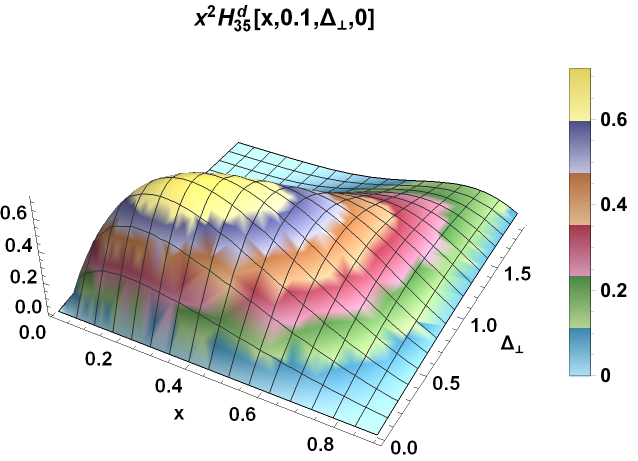}
		\hspace{0.05cm}
		(c)\includegraphics[width=7.3cm]{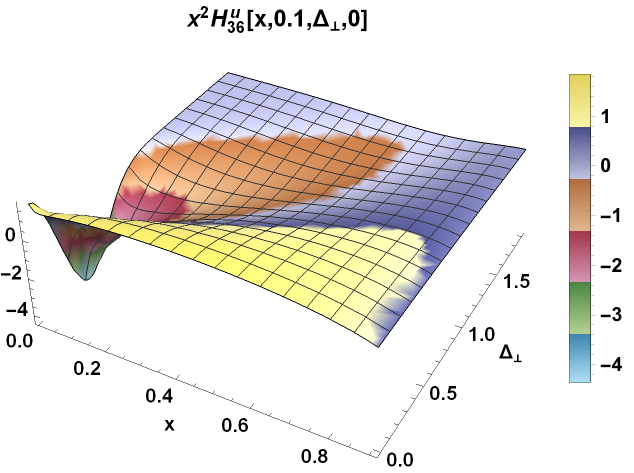}
		\hspace{0.05cm}
		(d)\includegraphics[width=7.3cm]{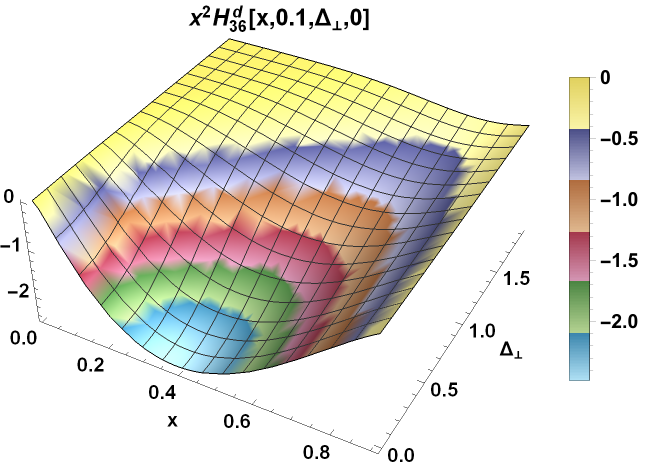}
		\hspace{0.05cm}
		(e)\includegraphics[width=7.3cm]{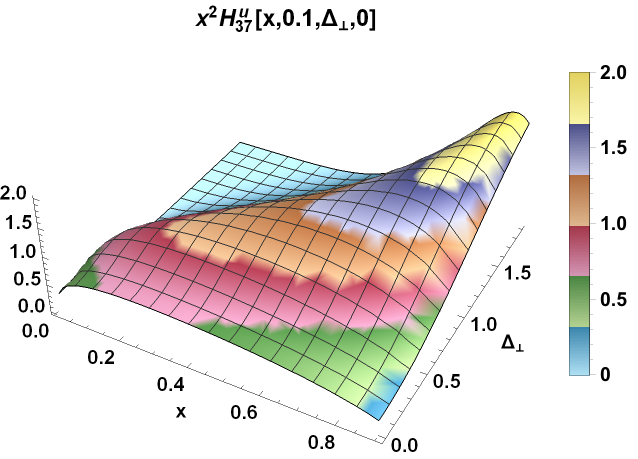}
		\hspace{0.05cm}
		(f)\includegraphics[width=7.3cm]{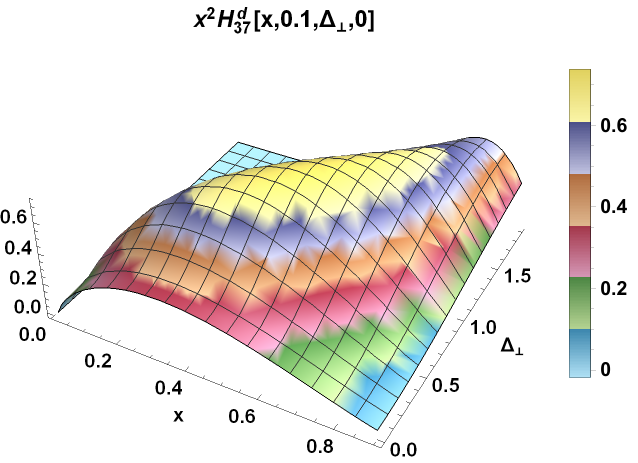}
		\hspace{0.05cm}
		(g)\includegraphics[width=7.3cm]{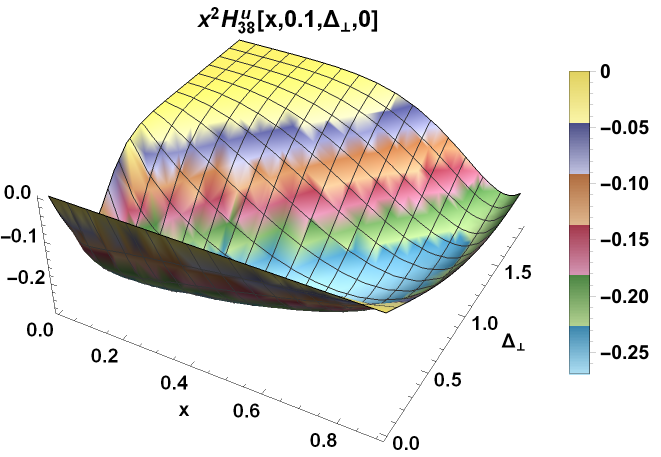}
		\hspace{0.05cm}
		(h)\includegraphics[width=7.3cm]{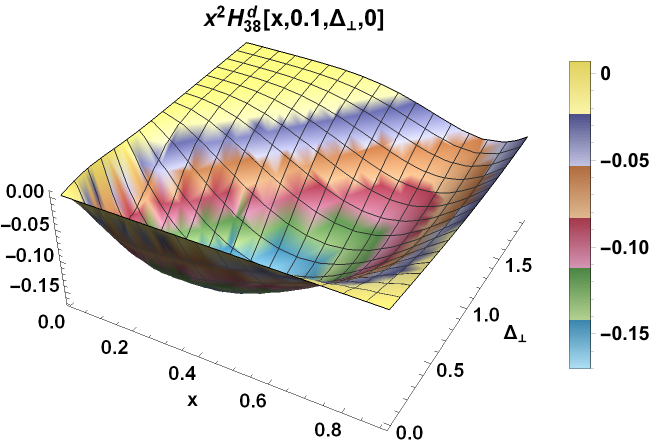}
		\hspace{0.05cm}\\
	\end{minipage}
	\caption{\label{fig3dXDH2} (Color online) The twist-4 GTMDs 		
		$x^2 H_{3,5}^{\nu}(x, p_{\perp},\Delta_{\perp},\theta)$,
		$x^2 H_{3,6}^{\nu}(x, p_{\perp},\Delta_{\perp},\theta)$,
		$x^2 H_{3,7}^{\nu}(x, p_{\perp},\Delta_{\perp},\theta)$ and
		$x^2 H_{3,8}^{\nu}(x, p_{\perp},\Delta_{\perp},\theta)$
		plotted with respect to $x$ and ${{ \Delta_\perp}}$ for ${ p_\perp}= 0.1~\mathrm{GeV}$ for ${\bfp} \parallel {\Dp}$. The left and right column correspond to $u$ and $d$ quarks sequentially.
	}
\end{figure*}

For the study of twist-4 GTMDs with simultaneous change in variables $x$ and ${ \Delta_\perp}$, we have plotted their 3-D variation at ${ p_\perp}=0.1~\mathrm{GeV}$ and $\theta=0^{\circ} $  in Figs. (\ref{fig3dXDF}), (\ref{fig3dXDG}), (\ref{fig3dXDH1}) and (\ref{fig3dXDH2}).
%---------------F x Dp PLOT---------------------%
For the Dirac matrix structure $\Gamma=\gamma^-$, the GTMDs $F_{3,1}, F_{3,2}, F_{3,3}$ and $F_{3,4}$ have been plotted in Fig. (\ref{fig3dXDF}).
In Fig. \ref{fig3dXDF} (a) and \ref{fig3dXDF} (b), the GTMD $F_{3,1}^{\nu}$ has been plotted, the variation of which is not straightforward with longitudinal momentum fraction $x$ and transverse momentum transfer ${ \Delta_\perp}$, however, it is clear that the GTMD possesses the same behavior for $u$ as well as $d$ quarks, varying only by amplitude. The GTMD $F_{3,2}^{\nu}$ has been plotted in Fig. \ref{fig3dXDF} (c) and \ref{fig3dXDF} (d) for $u$ and $d$ quarks respectively. As the transverse momentum of quark has been fixed, the polarization combination corresponding to this GTMD is possible when the longitudinal momentum fraction is low (high) and transverse momentum transfer is high (low). The GTMD $F_{3,3}^{\nu}$ has been plotted in Fig. \ref{fig3dXDF} (e) and \ref{fig3dXDF} (f) for $u$ and $d$ quarks respectively. Due to the outstretched equation of GTMD for vector diquark, no quark flavor symmetry is observed on changing the quark flavor between $u$ and $d$ quarks. The GTMD $F_{3,4}^{\nu}$ has been plotted in Fig. \ref{fig3dXDF} (g) and \ref{fig3dXDF} (h) for $u$ and $d$ quarks respectively.
Due to similar expressions of scalar and vector diquark parts, the trend of this GTMD is same for both flavors and the plots are varying only by magnitude.
%
%---------------G x Dp PLOT---------------------%
%
For the study of GTMDs $G_{3,1}, G_{3,2}, G_{3,3}$ and $G_{3,4}$, the Dirac matrix structure $\Gamma=\gamma^-\gamma^5$ is chosen.
They have been plotted with respect to $x$ and ${ \Delta_\perp}$ in Fig. (\ref{fig3dXDG}), while ${ p_\perp}$ and $\theta$ are fixed at $0.1~\mathrm{GeV}$ and $0^{\circ}$ respectively. In Fig. \ref{fig3dXDG} (a) and \ref{fig3dXDG} (b), the GTMD $G_{3,1}^{\nu}$ has been plotted which is positive over the entire range of $x$ and ${ p_\perp}$ and its behavior is same for $u$ as well as $d$ quarks, varying only by the amplitude. GTMDs $G_{3,2}^{\nu}$ and $G_{3,3}^{\nu}$ has been plotted w.r.t $x$ and ${ \Delta_\perp}$ in Fig. \ref{fig3dXDG} (c) to \ref{fig3dXDG} (f). Due to their lengthy, non symmetric expressions of GTMDs over the quark flavor in scalar and vector diquark parts, part-wise symmetry is observed over the quark flavor. In Fig. \ref{fig3dXDG} (g) and \ref{fig3dXDG} (h), the GTMD $G_{3,4}^{\nu}$ has been plotted and its behavior is observed to be similar for both quark flavors with difference only in their amplitude.
%
%---------------H x Dp PLOT---------------------%
%
By considering the Dirac matrix structure $\Gamma=i\sigma^{+-}\gamma_5$, GTMDs $H_{3,1}$, $H_{3,2}$, $H_{3,3}$, $H_{3,4}$, $H_{3,5}$, $H_{3,6}$, $H_{3,7}$ and $H_{3,8}$ have been plotted in Figs. (\ref{fig3dXDH1}) and (\ref{fig3dXDH2}). 
In Fig. \ref{fig3dXDH1} (a) to \ref{fig3dXDH1} (d), the GTMD $H_{3,1}^{\nu}$ and $H_{3,2}^{\nu}$ have been plotted. These GTMDs do not change their sign on changing the quark flavor between $u$ and $d$ quarks. Keeping amplitude aside, the trend of its variation is same in both cases. In Fig. \ref{fig3dXDH1} (e) and \ref{fig3dXDH1} (f), the GTMD $H_{3,3}^{\nu}$ is plotted. Due to the lengthy equation of vector diquark, no direct correlation is derived between the behavior of $u$ and $d$ quarks. 
The behavior of GTMD $H_{3,4}^{\nu}$ is same for both flavors of quark with only difference lies in the amplitude as observed in Fig. \ref{fig3dXDH1} (g) and \ref{fig3dXDH1} (h). In Fig. \ref{fig3dXDH2} (a) to \ref{fig3dXDH2} (d) the GTMDs $H_{3,5}^{\nu}$ and $H_{3,6}^{\nu}$ have been plotted. In both of these GTMDs, no quark flavor symmetry is observed. The GTMDs $H_{3,7}^{\nu}$ and $H_{3,8}^{\nu}$ have been plotted in Fig. \ref{fig3dXDH2} (e) to \ref{fig3dXDH2} (h) and they beautifully present the quark flavor symmetry. 
\subsubsection{Variation with ${ p_\perp}$ and ${ \Delta_\perp}$}\label{sspd}
\begin{figure*}
	\centering
	\begin{minipage}[c]{0.98\textwidth}
		(a)\includegraphics[width=7.3cm]{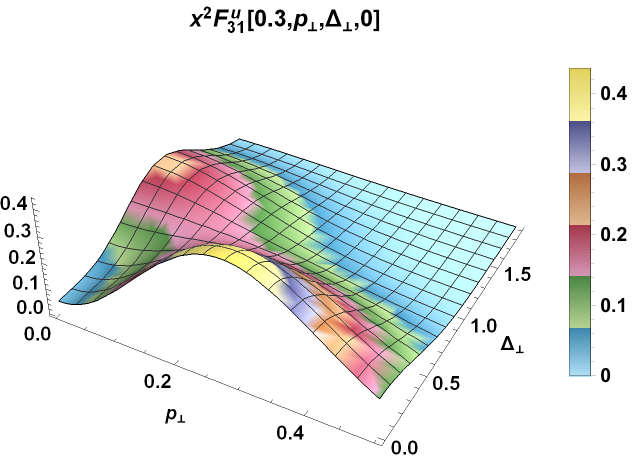}
		\hspace{0.05cm}
		(b)\includegraphics[width=7.3cm]{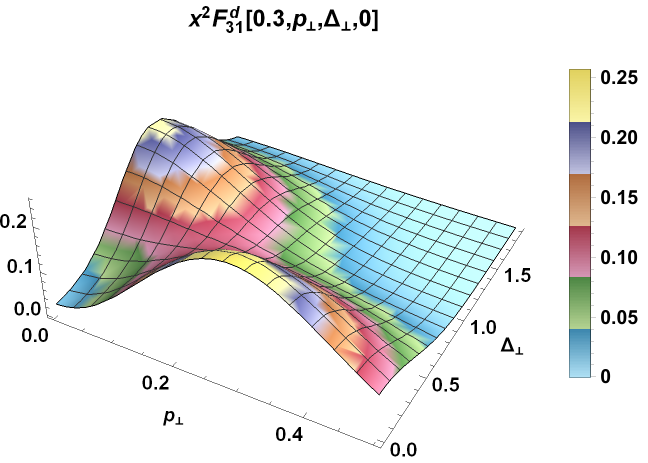}
		\hspace{0.05cm}
		(c)\includegraphics[width=7.3cm]{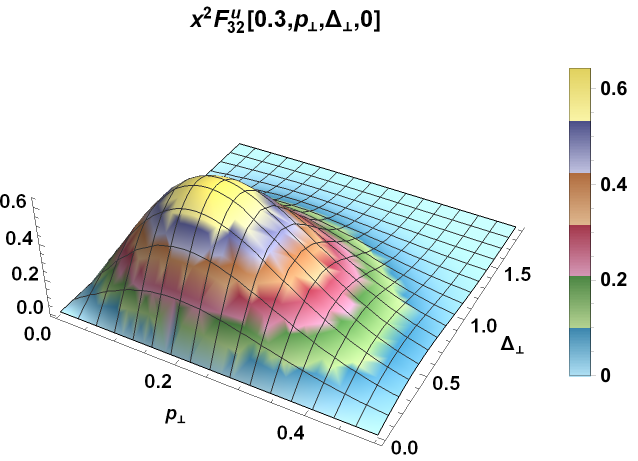}
		\hspace{0.05cm}
		(d)\includegraphics[width=7.3cm]{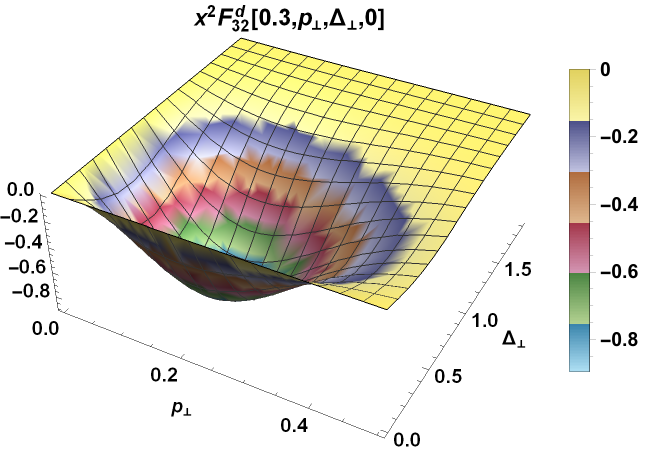}
		\hspace{0.05cm}
		(e)\includegraphics[width=7.3cm]{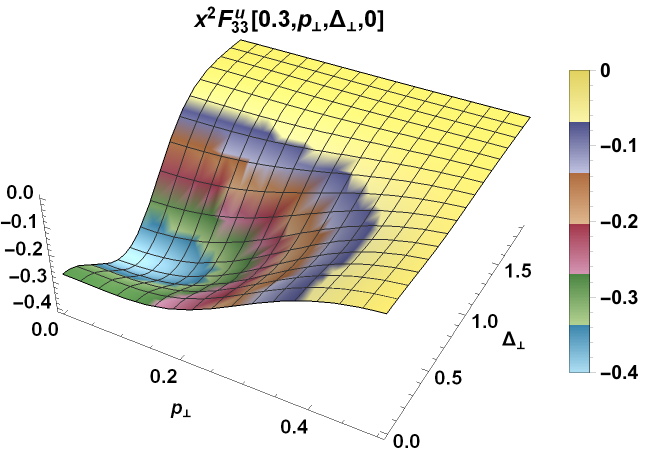}
		\hspace{0.05cm}
		(f)\includegraphics[width=7.3cm]{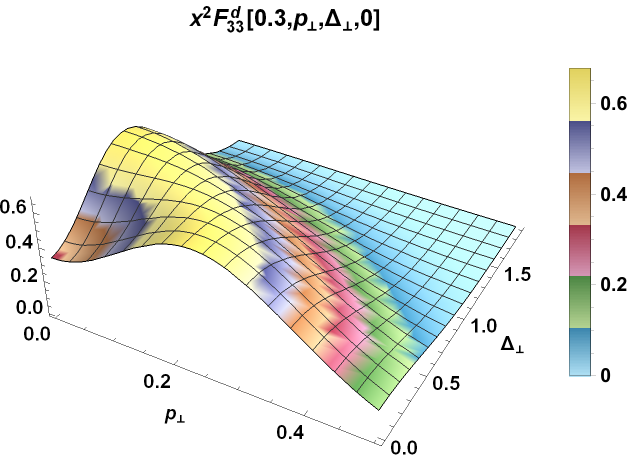}
		\hspace{0.05cm}
		(g)\includegraphics[width=7.3cm]{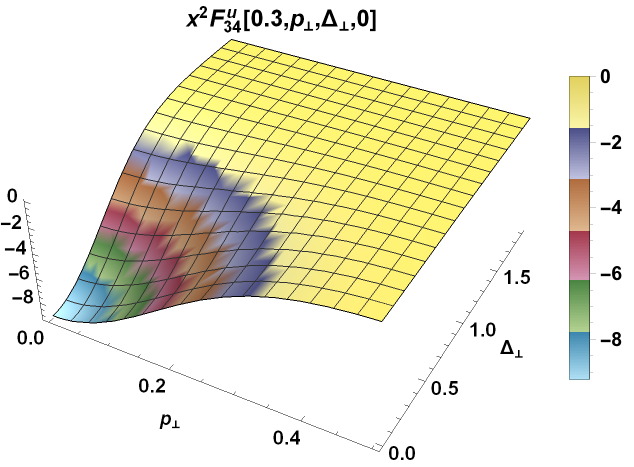}
		\hspace{0.05cm}
		(h)\includegraphics[width=7.3cm]{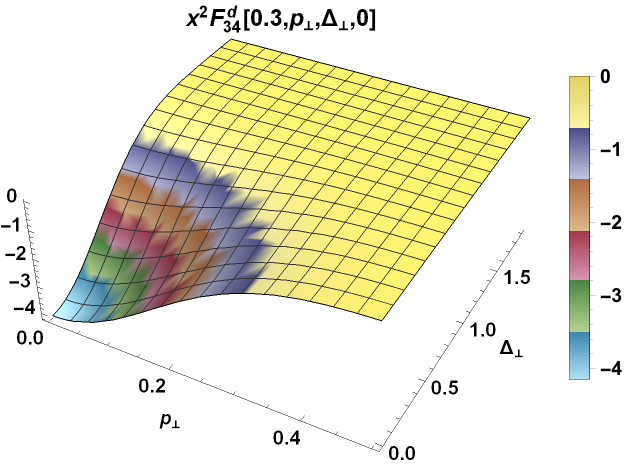}
		\hspace{0.05cm}\\
	\end{minipage}
	\caption{\label{fig3dPDF} (Color online) The twist-4 GTMDs 		
		$x^2 F_{3,1}^{\nu}(x, p_{\perp},\Delta_{\perp},\theta)$,
		$x^2 F_{3,2}^{\nu}(x, p_{\perp},\Delta_{\perp},\theta)$,
		$x^2 F_{3,3}^{\nu}(x, p_{\perp},\Delta_{\perp},\theta)$ and
		$x^2 F_{3,4}^{\nu}(x, p_{\perp},\Delta_{\perp},\theta)$
		plotted with respect to ${ p_\perp}$ and ${{ \Delta_\perp}}$ for $x= 0.3$ for ${\bfp} \parallel {\Dp}$. The left and right column correspond to $u$ and $d$ quarks sequentially.
	}
\end{figure*}

\begin{figure*}
	\centering
	\begin{minipage}[c]{0.98\textwidth}
		(a)\includegraphics[width=7.3cm]{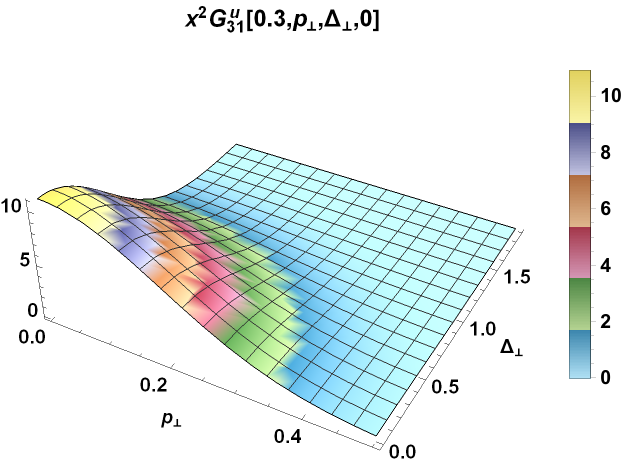}
		\hspace{0.05cm}
		(b)\includegraphics[width=7.3cm]{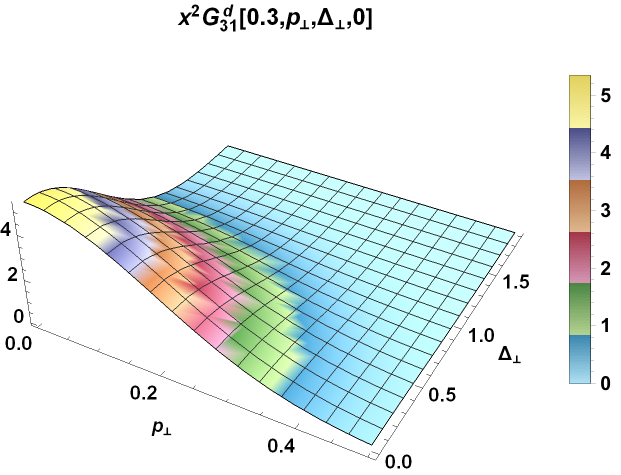}
		\hspace{0.05cm}
		(c)\includegraphics[width=7.3cm]{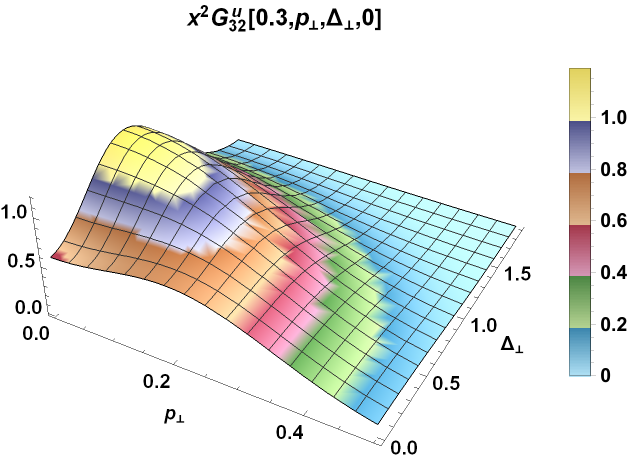}
		\hspace{0.05cm}
		(d)\includegraphics[width=7.3cm]{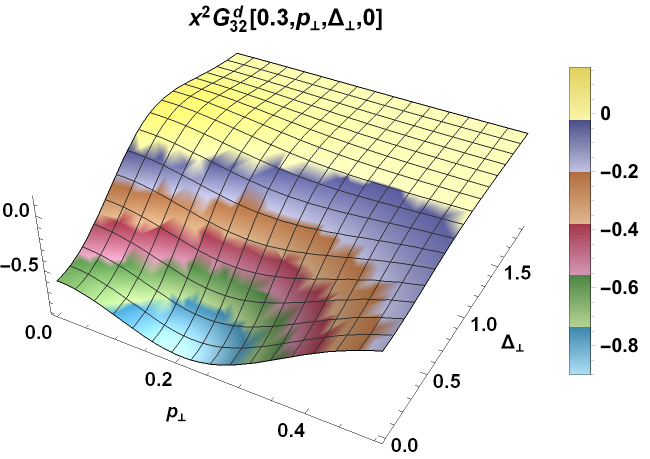}
		\hspace{0.05cm}
		(e)\includegraphics[width=7.3cm]{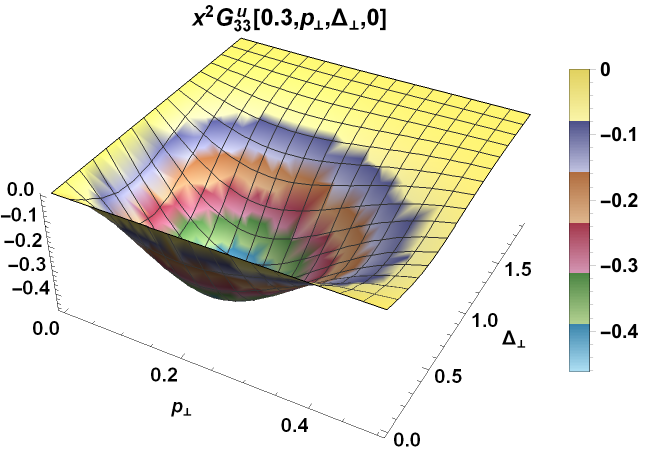}
		\hspace{0.05cm}
		(f)\includegraphics[width=7.3cm]{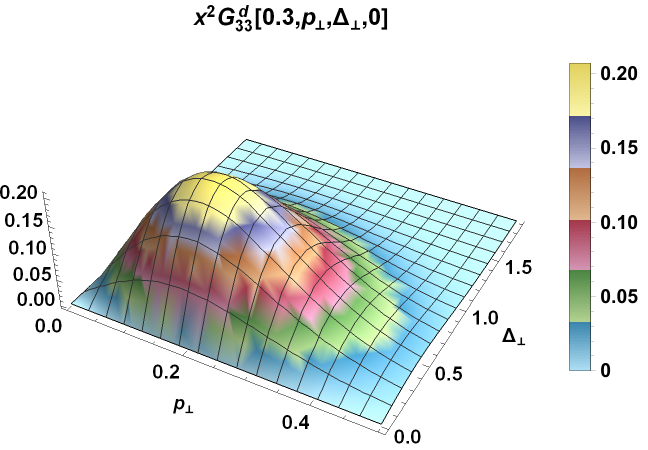}
		\hspace{0.05cm}
		(g)\includegraphics[width=7.3cm]{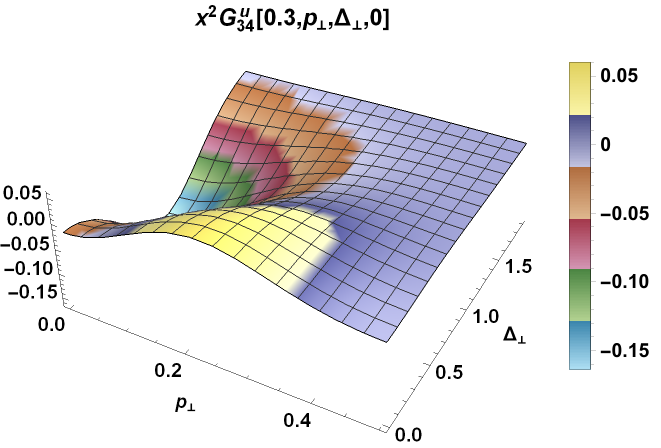}
		\hspace{0.05cm}
		(h)\includegraphics[width=7.3cm]{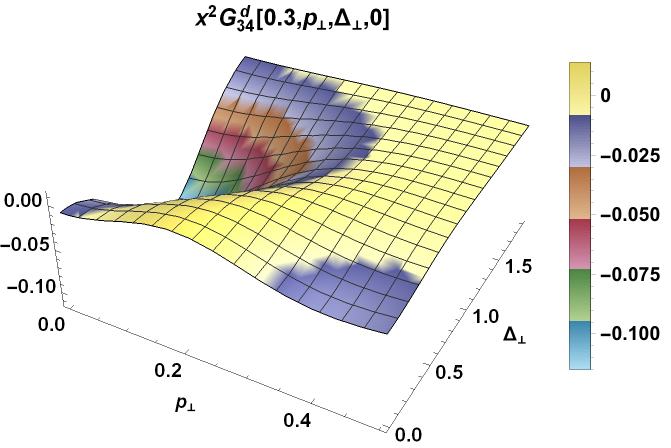}
		\hspace{0.05cm}\\
	\end{minipage}
	\caption{\label{fig3dPDG} (Color online)  The twist-4 GTMDs 		
		$x^2 G_{3,1}^{\nu}(x, p_{\perp},\Delta_{\perp},\theta)$,
		$x^2 G_{3,2}^{\nu}(x, p_{\perp},\Delta_{\perp},\theta)$,
		$x^2 G_{3,3}^{\nu}(x, p_{\perp},\Delta_{\perp},\theta)$ and
		$x^2 G_{3,4}^{\nu}(x, p_{\perp},\Delta_{\perp},\theta)$
		plotted with respect to ${ p_\perp}$ and ${{ \Delta_\perp}}$ for $x= 0.3$ for ${\bfp} \parallel {\Dp}$. The left and right column correspond to $u$ and $d$ quarks sequentially.
	}
\end{figure*}

\begin{figure*}
	\centering
	\begin{minipage}[c]{0.98\textwidth}
		(a)\includegraphics[width=7.3cm]{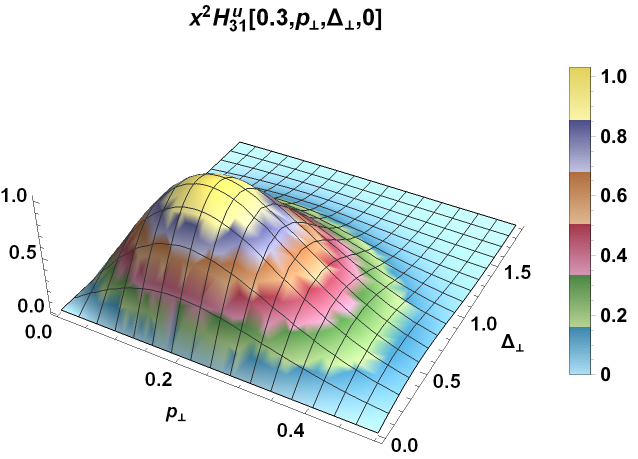}
		\hspace{0.05cm}
		(b)\includegraphics[width=7.3cm]{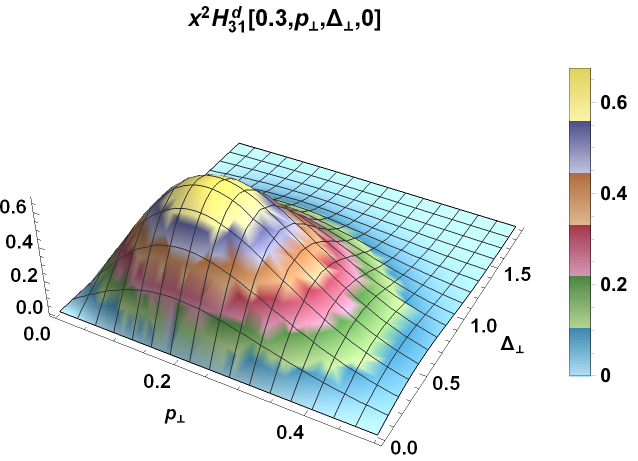}
		\hspace{0.05cm}
		(c)\includegraphics[width=7.3cm]{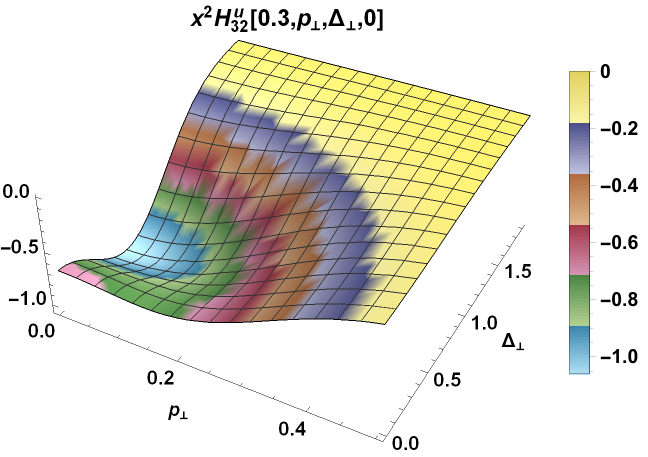}
		\hspace{0.05cm}
		(d)\includegraphics[width=7.3cm]{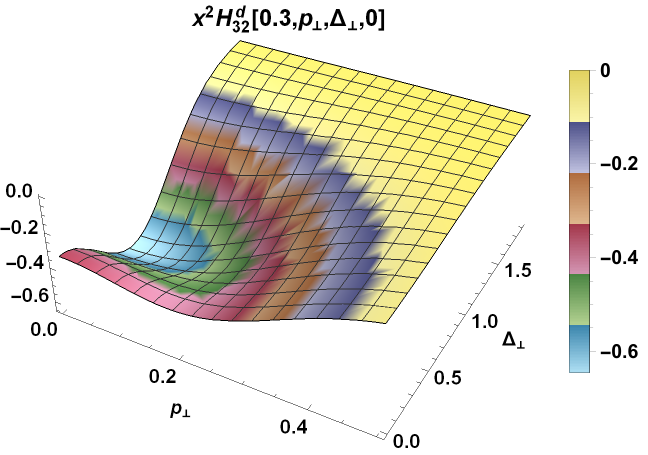}
		\hspace{0.05cm}
		(e)\includegraphics[width=7.3cm]{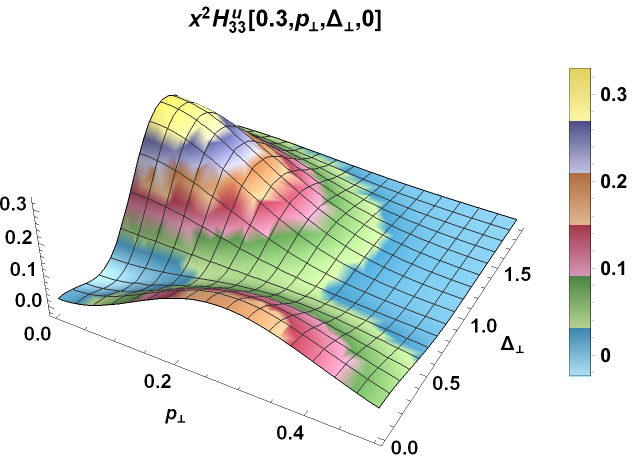}
		\hspace{0.05cm}
		(f)\includegraphics[width=7.3cm]{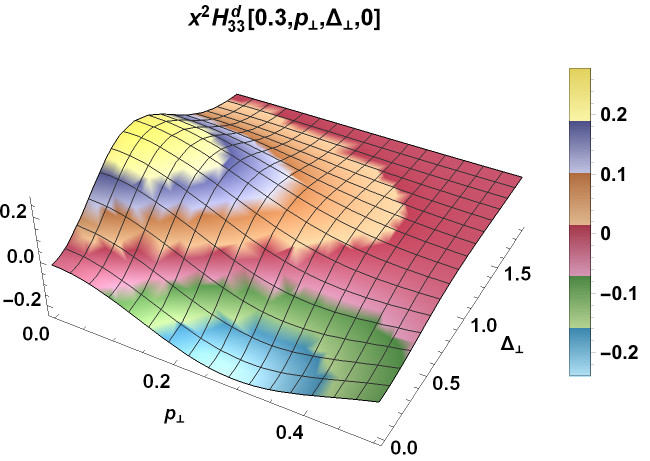}
		\hspace{0.05cm}
		(g)\includegraphics[width=7.3cm]{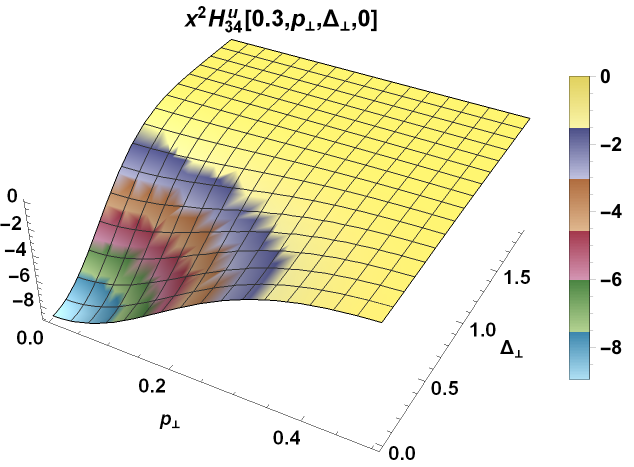}
		\hspace{0.05cm}
		(h)\includegraphics[width=7.3cm]{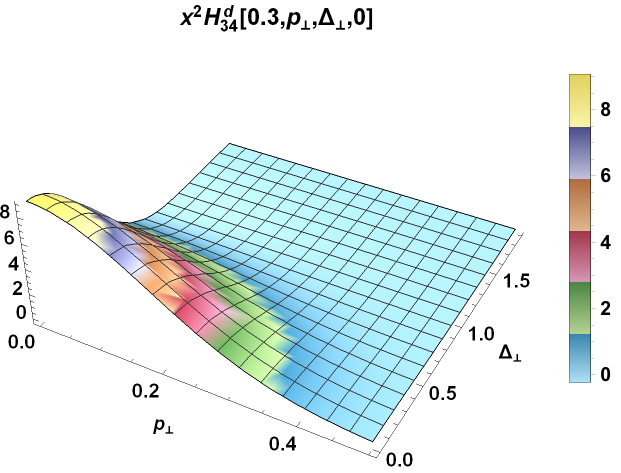}
		\hspace{0.05cm}\\
	\end{minipage}
	\caption{\label{fig3dPDH1} (Color online) The twist-4 GTMDs 		
		$x^2 H_{3,1}^{\nu}(x, p_{\perp},\Delta_{\perp},\theta)$,
		$x^2 H_{3,2}^{\nu}(x, p_{\perp},\Delta_{\perp},\theta)$,
		$x^2 H_{3,3}^{\nu}(x, p_{\perp},\Delta_{\perp},\theta)$ and
		$x^2 H_{3,4}^{\nu}(x, p_{\perp},\Delta_{\perp},\theta)$
		plotted with respect to ${ p_\perp}$ and ${{ \Delta_\perp}}$ for $x= 0.3$ for ${\bfp} \parallel {\Dp}$. The left and right column correspond to $u$ and $d$ quarks sequentially.
	}
\end{figure*}

\begin{figure*}
	\centering
	\begin{minipage}[c]{0.98\textwidth}
		(a)\includegraphics[width=7.3cm]{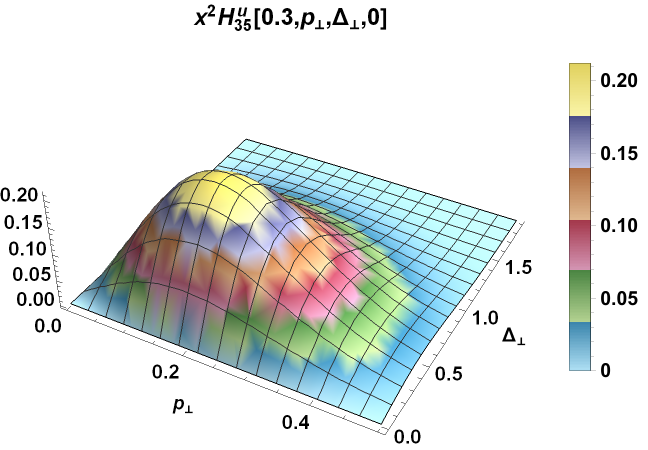}
		\hspace{0.05cm}
		(b)\includegraphics[width=7.3cm]{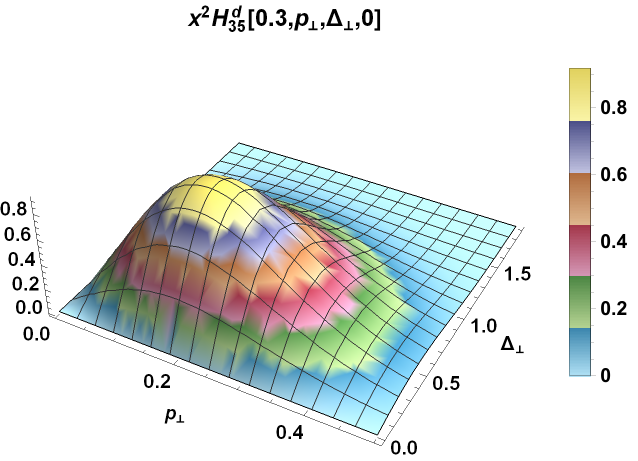}
		\hspace{0.05cm}
		(c)\includegraphics[width=7.3cm]{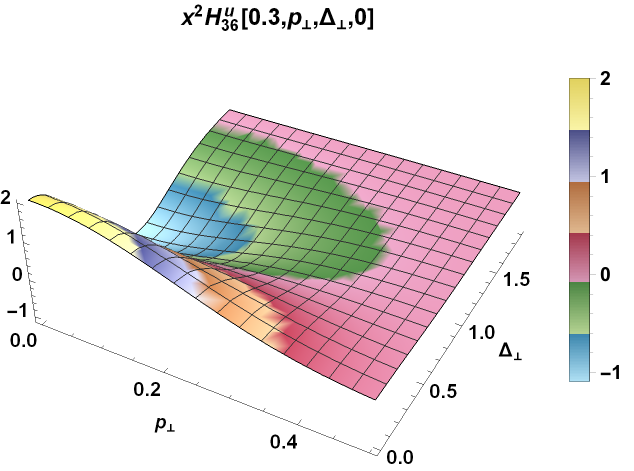}
		\hspace{0.05cm}
		(d)\includegraphics[width=7.3cm]{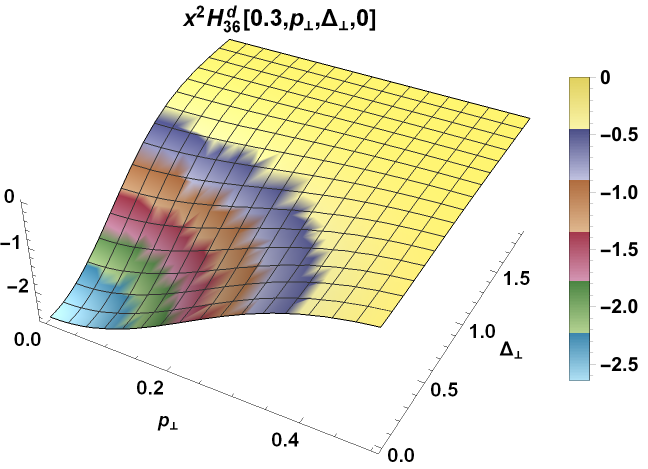}
		\hspace{0.05cm}
		(e)\includegraphics[width=7.3cm]{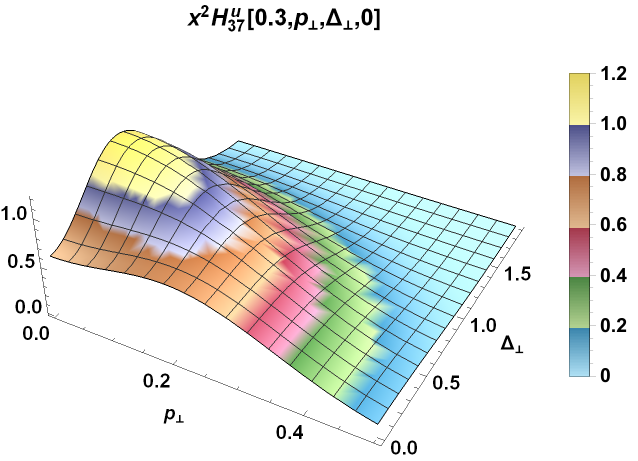}
		\hspace{0.05cm}
		(f)\includegraphics[width=7.3cm]{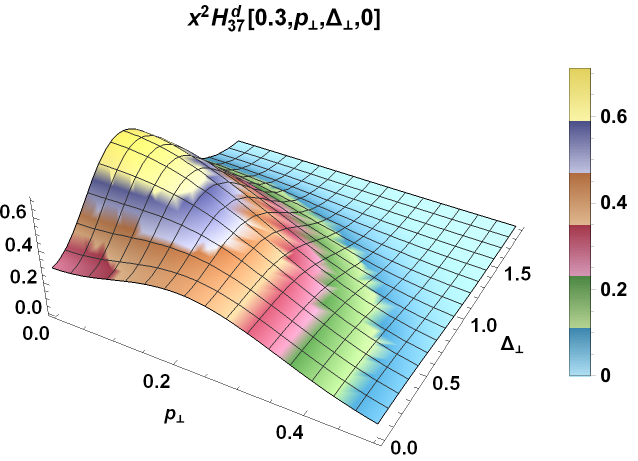}
		\hspace{0.05cm}
		(g)\includegraphics[width=7.3cm]{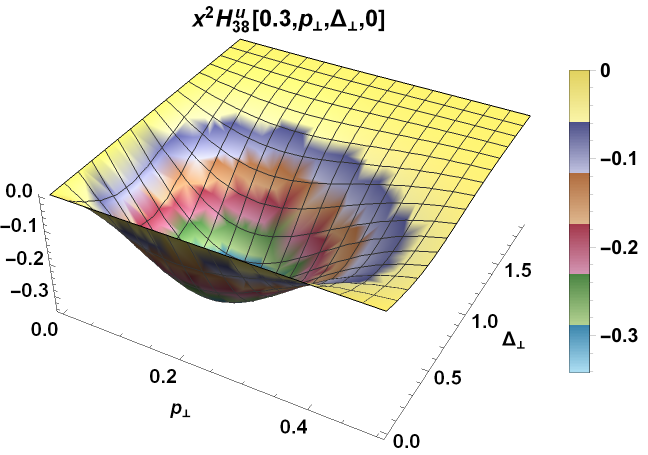}
		\hspace{0.05cm}
		(h)\includegraphics[width=7.3cm]{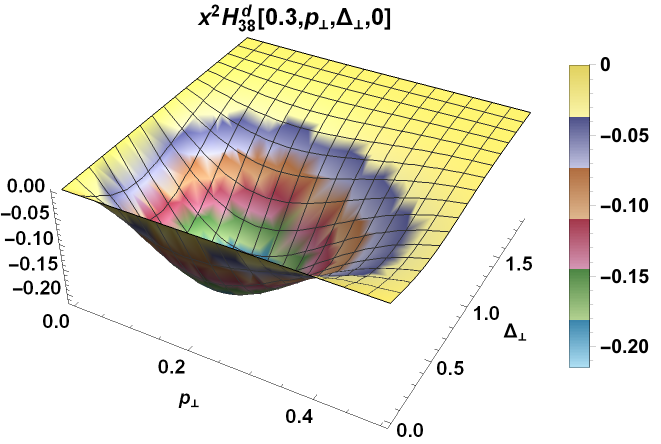}
		\hspace{0.05cm}\\
	\end{minipage}
	\caption{\label{fig3dPDH2} (Color online) The twist-4 GTMDs 		
		$x^2 H_{3,5}^{\nu}(x, p_{\perp},\Delta_{\perp},\theta)$,
		$x^2 H_{3,6}^{\nu}(x, p_{\perp},\Delta_{\perp},\theta)$,
		$x^2 H_{3,7}^{\nu}(x, p_{\perp},\Delta_{\perp},\theta)$ and
		$x^2 H_{3,8}^{\nu}(x, p_{\perp},\Delta_{\perp},\theta)$
		plotted with respect to ${ p_\perp}$ and ${{ \Delta_\perp}}$ for $x= 0.3$ for ${\bfp} \parallel {\Dp}$. The left and right column correspond to $u$ and $d$ quarks sequentially.
	}
\end{figure*}

For the study of twist-4 GTMDs with simultaneous change in the transverse momentum of quark ${ p_\perp}$ and the transverse momentum transfer between the initial and the final state of quark (or proton) ${ \Delta_\perp}$, we have plotted their 3-D variation at $x=0.3$ and $\theta=0^{\circ} $  in Figs. (\ref{fig3dPDF}), (\ref{fig3dPDG}), (\ref{fig3dPDH1}) and (\ref{fig3dPDH2}). We have chosen this value of $x$, because generally when quark carries $33\%$ of the proton longitudinal momentum, the GTMDs have maximum amplitude as evident from Figs. (\ref{fig3dXPF}) to  (\ref{fig3dXDH2}).
\subsection{Transverse-momentum dependent form factors (TMFFs)}\label{sstmff}
\begin{figure*}
	\centering
	\begin{minipage}[c]{0.98\textwidth}
		(a)\includegraphics[width=7.3cm]{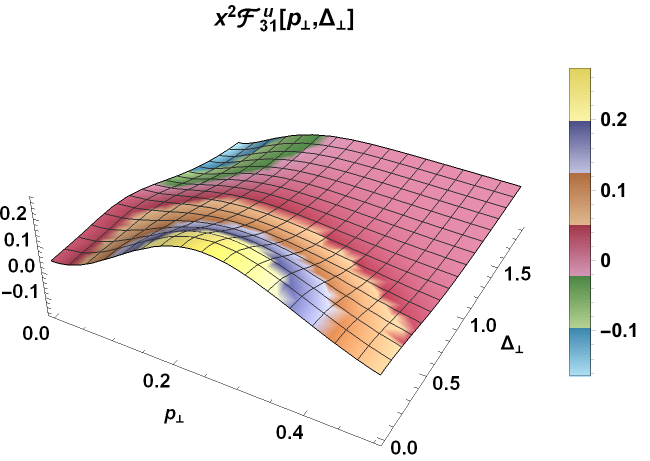}
		\hspace{0.05cm}
		(b)\includegraphics[width=7.3cm]{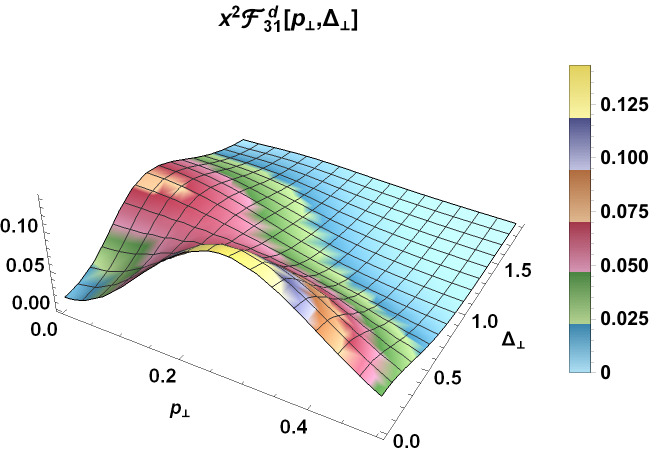}
		\hspace{0.05cm}
		(c)\includegraphics[width=7.3cm]{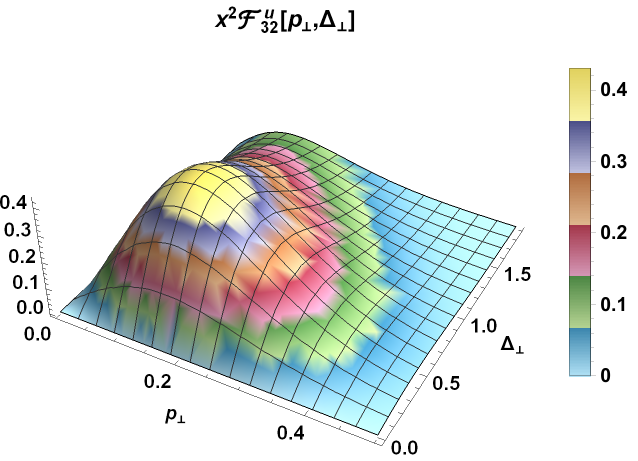}
		\hspace{0.05cm}
		(d)\includegraphics[width=7.3cm]{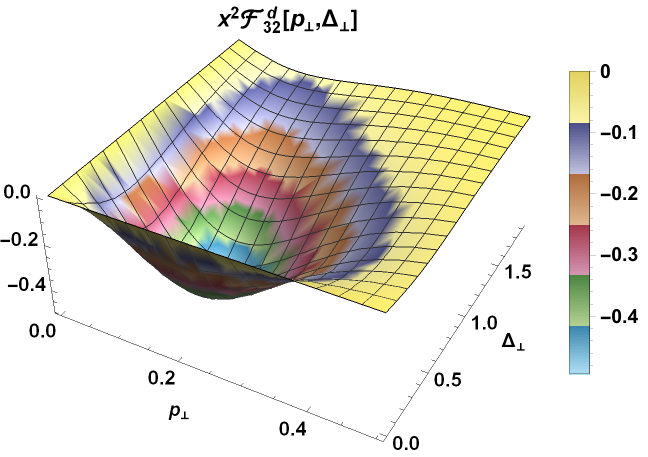}
		\hspace{0.05cm}
		(e)\includegraphics[width=7.3cm]{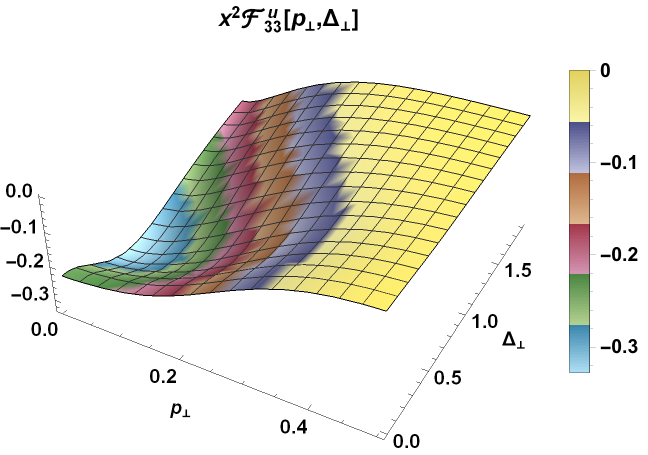}
		\hspace{0.05cm}
		(f)\includegraphics[width=7.3cm]{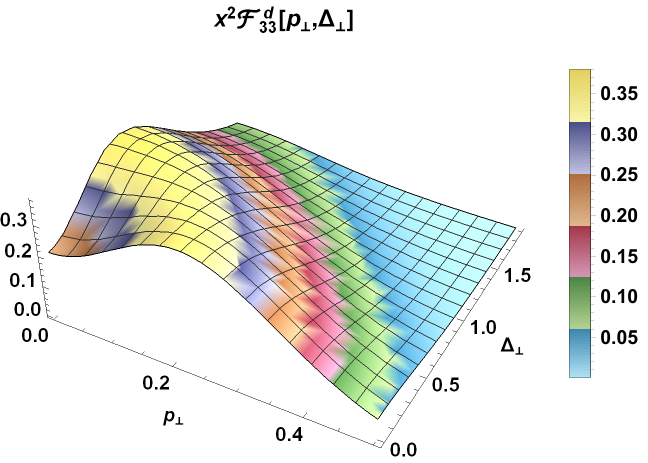}
		\hspace{0.05cm}
		(g)\includegraphics[width=7.3cm]{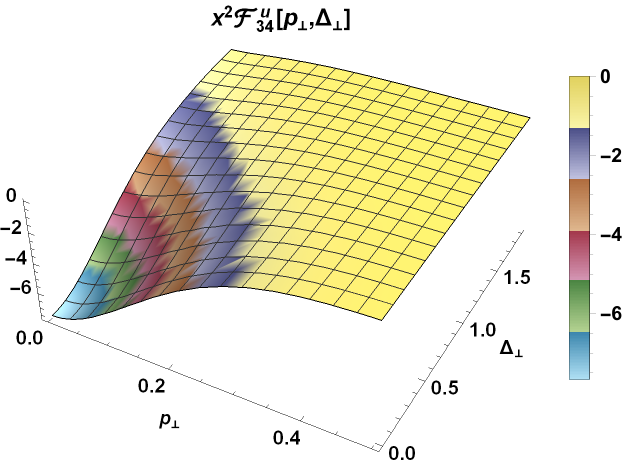}
		\hspace{0.05cm}
		(h)\includegraphics[width=7.3cm]{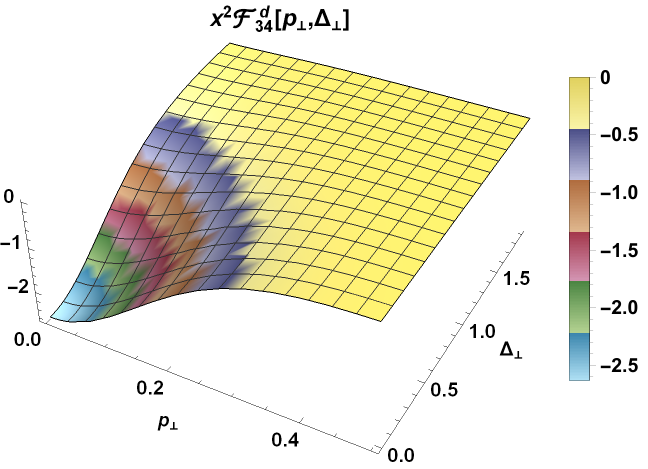}
		\hspace{0.05cm}\\
	\end{minipage}
	\caption{\label{fig3dTMFFF} (Color online) The twist-4 TMFFs 		
		$x^2 F_{3,1}^{\nu}(p_{\perp},\Delta_{\perp})$,
		$x^2 F_{3,2}^{\nu}(p_{\perp},\Delta_{\perp})$,
		$x^2 F_{3,3}^{\nu}(p_{\perp},\Delta_{\perp})$ and
		$x^2 F_{3,4}^{\nu}(p_{\perp},\Delta_{\perp})$ plotted with respect to ${ p_\perp}$ and ${{ \Delta_\perp}}$ for ${\bfp} \parallel {\Dp}$. The left and right column correspond to $u$ and $d$ quarks sequentially.
	}
\end{figure*}

\begin{figure*}
	\centering
	\begin{minipage}[c]{0.98\textwidth}
		(a)\includegraphics[width=7.3cm]{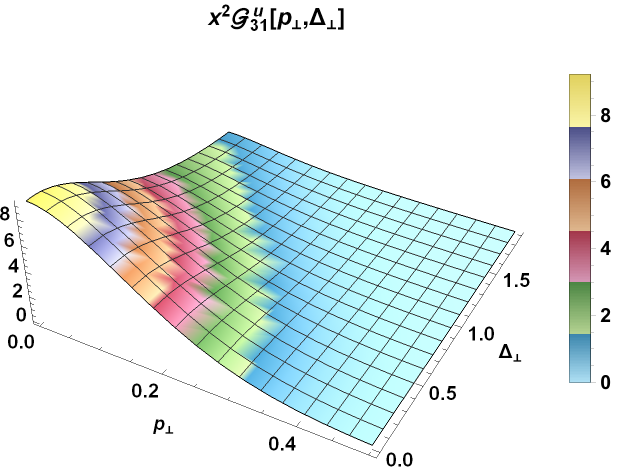}
		\hspace{0.05cm}
		(b)\includegraphics[width=7.3cm]{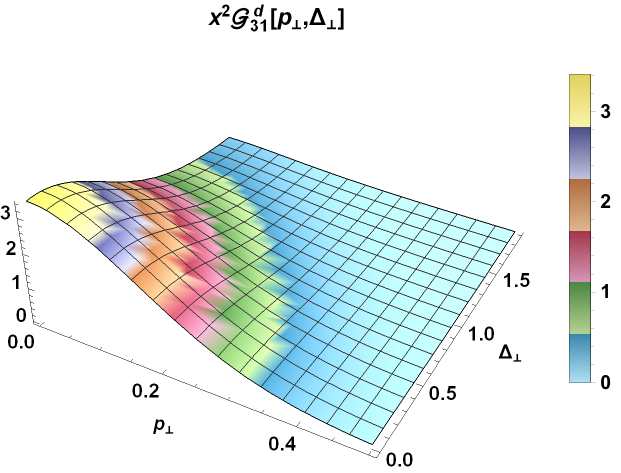}
		\hspace{0.05cm}
		(c)\includegraphics[width=7.3cm]{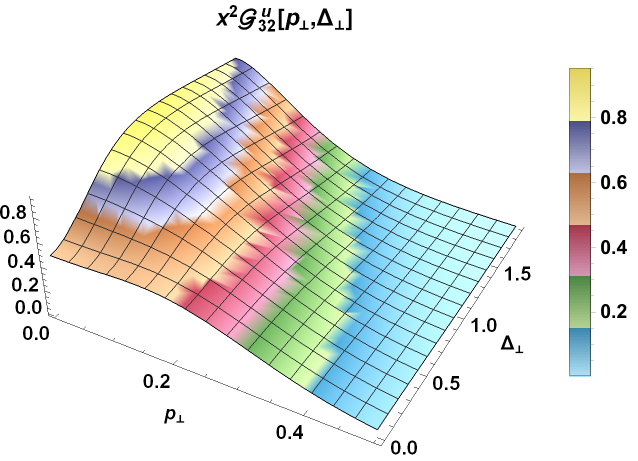}
		\hspace{0.05cm}
		(d)\includegraphics[width=7.3cm]{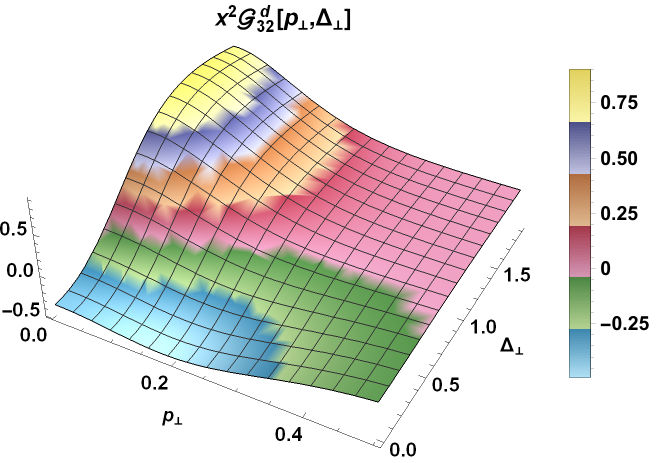}
		\hspace{0.05cm}
		(e)\includegraphics[width=7.3cm]{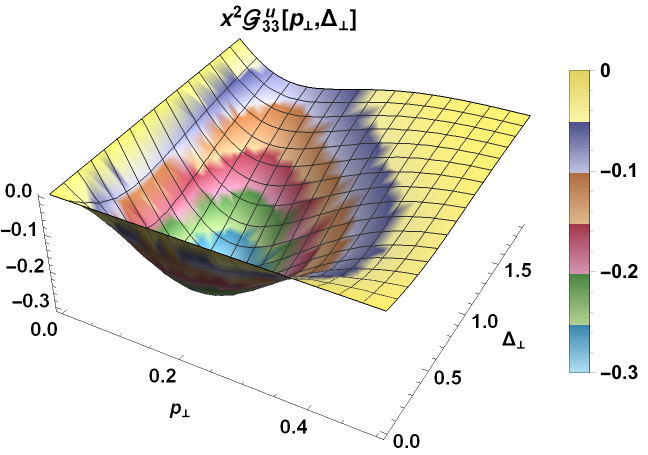}
		\hspace{0.05cm}
		(f)\includegraphics[width=7.3cm]{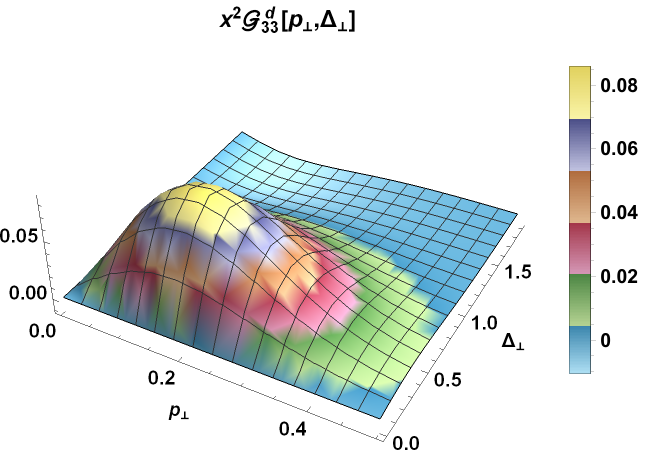}
		\hspace{0.05cm}
		(g)\includegraphics[width=7.3cm]{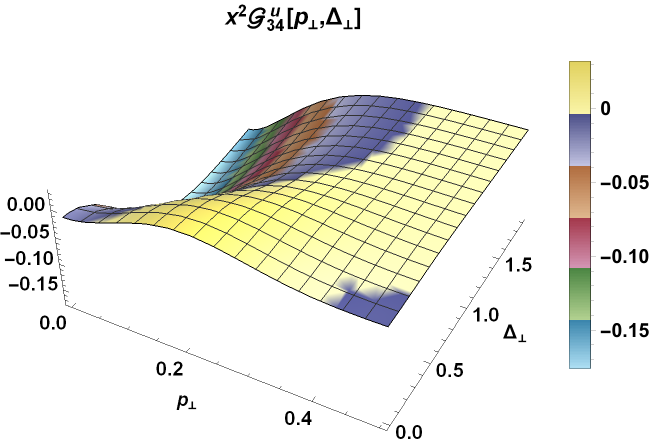}
		\hspace{0.05cm}
		(h)\includegraphics[width=7.3cm]{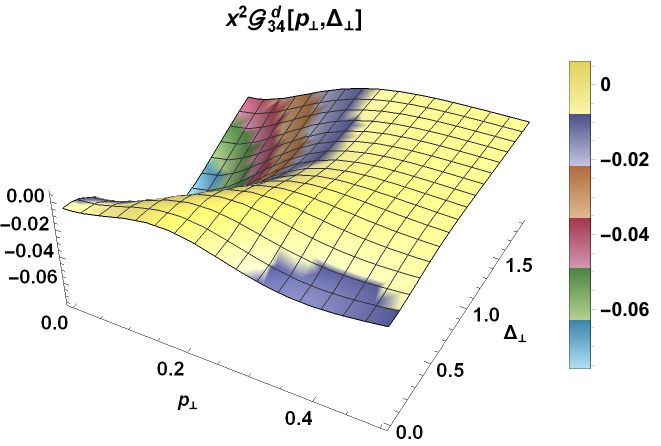}
		\hspace{0.05cm}\\
	\end{minipage}
	\caption{\label{fig3dTMFFG} (Color online)  The twist-4 TMFFs 		
		$x^2 G_{3,1}^{\nu}(p_{\perp},\Delta_{\perp})$,
		$x^2 G_{3,2}^{\nu}(p_{\perp},\Delta_{\perp})$,
		$x^2 G_{3,3}^{\nu}(p_{\perp},\Delta_{\perp})$ and
		$x^2 G_{3,4}^{\nu}(p_{\perp},\Delta_{\perp})$ plotted with respect to ${ p_\perp}$ and ${{ \Delta_\perp}}$ for ${\bfp} \parallel {\Dp}$. The left and right column correspond to $u$ and $d$ quarks sequentially.
	}
\end{figure*}

\begin{figure*}
	\centering
	\begin{minipage}[c]{0.98\textwidth}
		(a)\includegraphics[width=7.3cm]{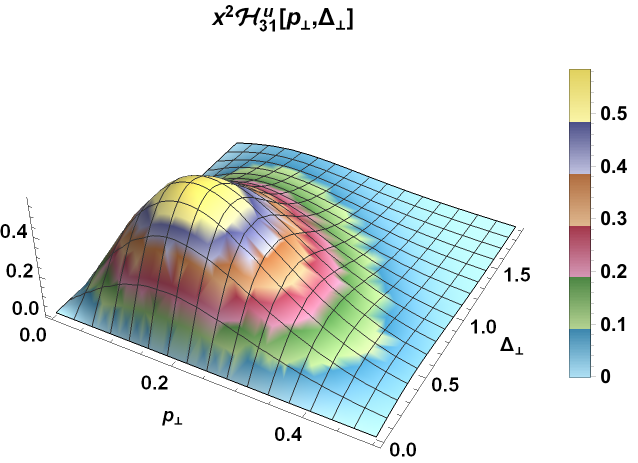}
		\hspace{0.05cm}
		(b)\includegraphics[width=7.3cm]{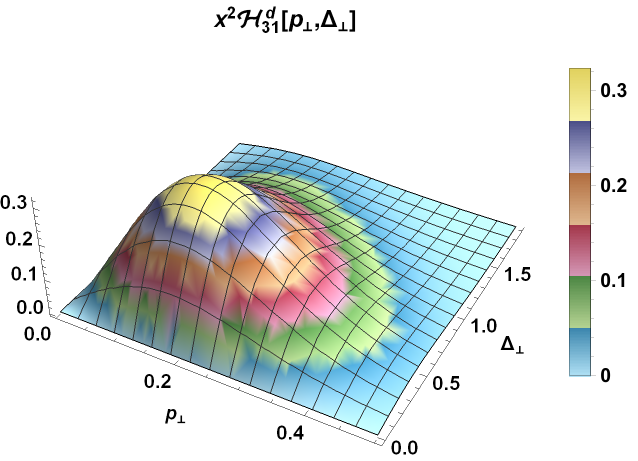}
		\hspace{0.05cm}
		(c)\includegraphics[width=7.3cm]{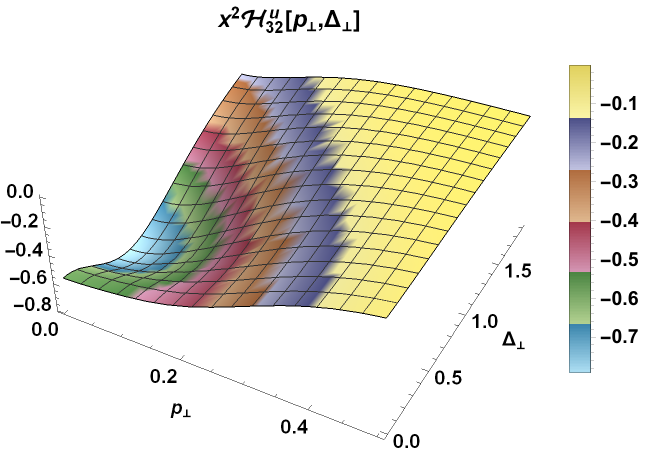}
		\hspace{0.05cm}
		(d)\includegraphics[width=7.3cm]{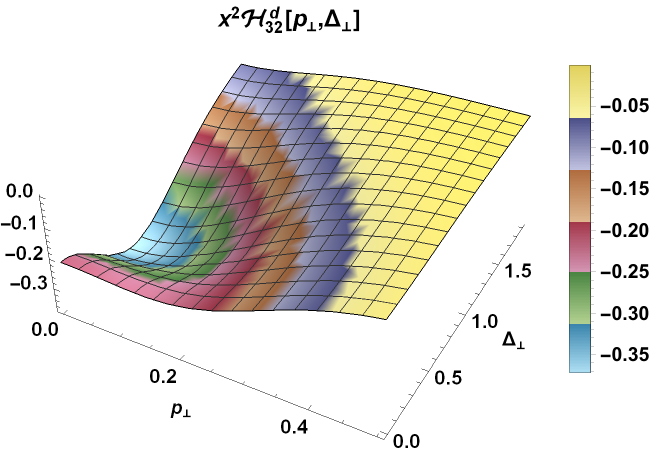}
		\hspace{0.05cm}
		(e)\includegraphics[width=7.3cm]{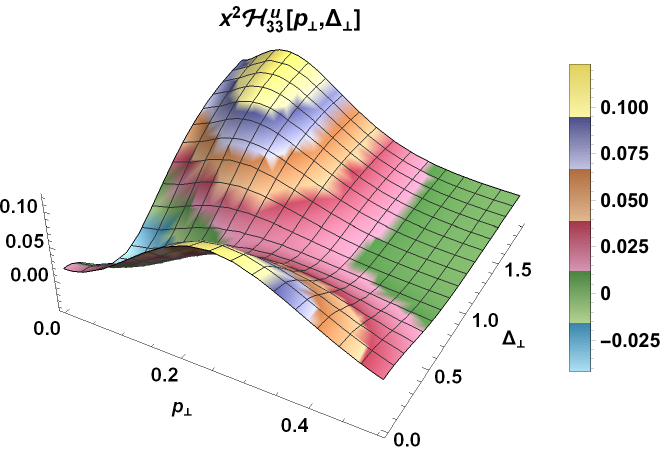}
		\hspace{0.05cm}
		(f)\includegraphics[width=7.3cm]{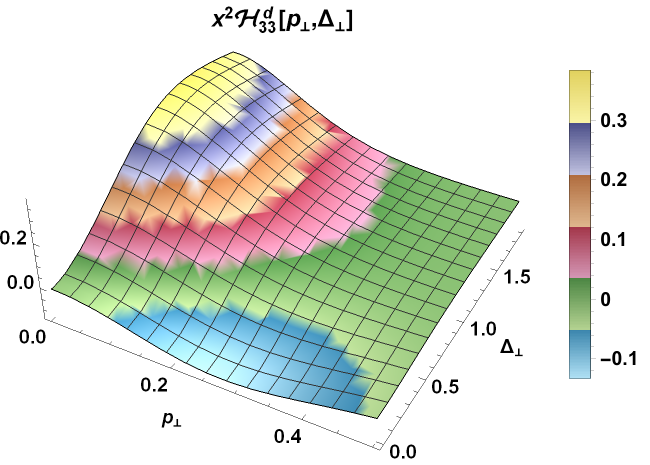}
		\hspace{0.05cm}
		(g)\includegraphics[width=7.3cm]{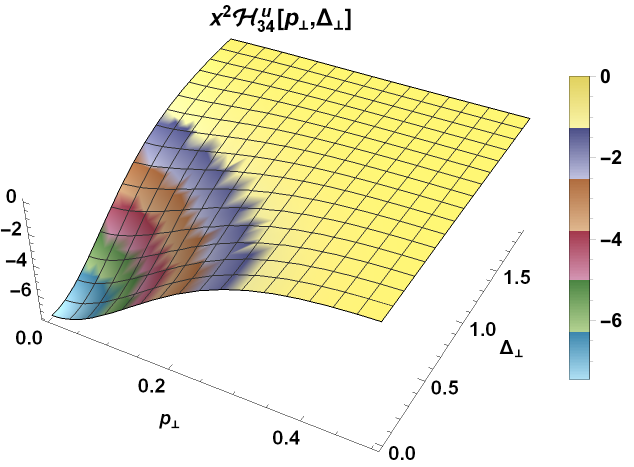}
		\hspace{0.05cm}
		(h)\includegraphics[width=7.3cm]{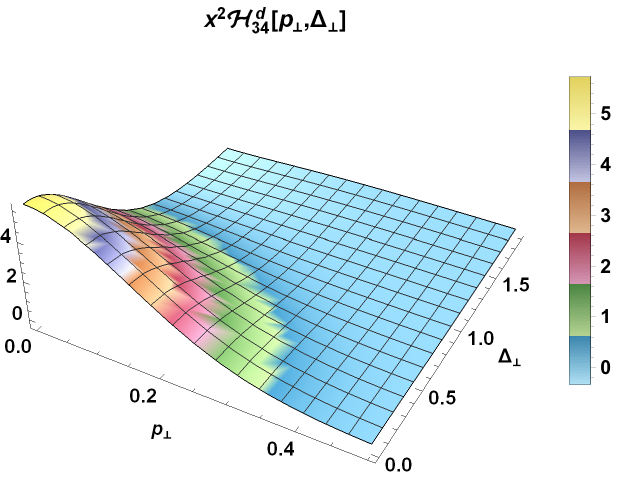}
		\hspace{0.05cm}\\
	\end{minipage}
	\caption{\label{fig3dTMFFH1} (Color online) The twist-4 TMFFs 		
		$x^2 H_{3,1}^{\nu}(p_{\perp},\Delta_{\perp})$,
		$x^2 H_{3,2}^{\nu}(p_{\perp},\Delta_{\perp})$,
		$x^2 H_{3,3}^{\nu}(p_{\perp},\Delta_{\perp})$ and
		$x^2 H_{3,4}^{\nu}(p_{\perp},\Delta_{\perp})$ plotted with respect to ${ p_\perp}$ and ${{ \Delta_\perp}}$ for ${\bfp} \parallel {\Dp}$. The left and right column correspond to $u$ and $d$ quarks sequentially.
	}
\end{figure*}

\begin{figure*}
	\centering
	\begin{minipage}[c]{0.98\textwidth}
		(a)\includegraphics[width=7.3cm]{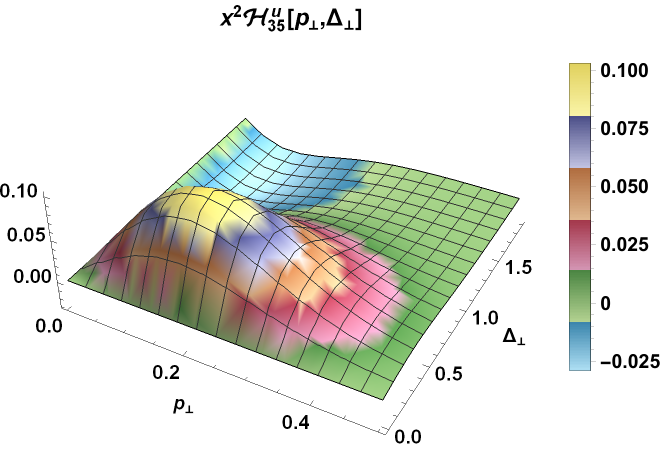}
		\hspace{0.05cm}
		(b)\includegraphics[width=7.3cm]{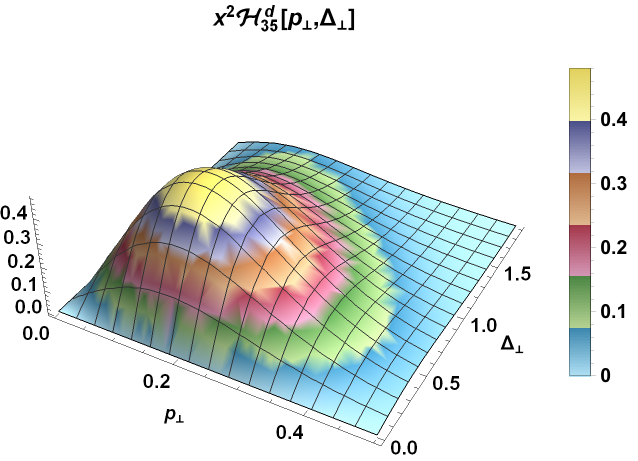}
		\hspace{0.05cm}
		(c)\includegraphics[width=7.3cm]{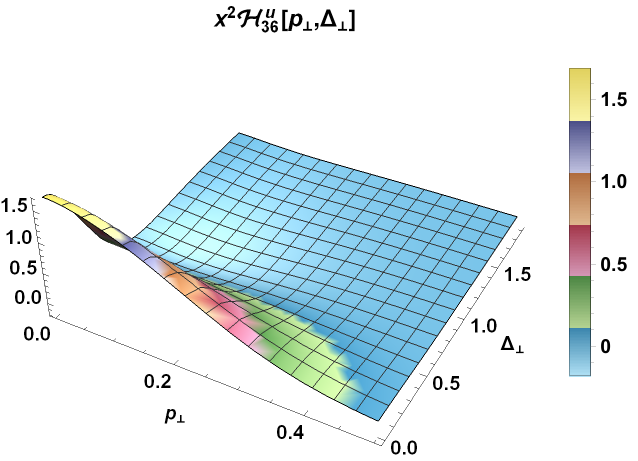}
		\hspace{0.05cm}
		(d)\includegraphics[width=7.3cm]{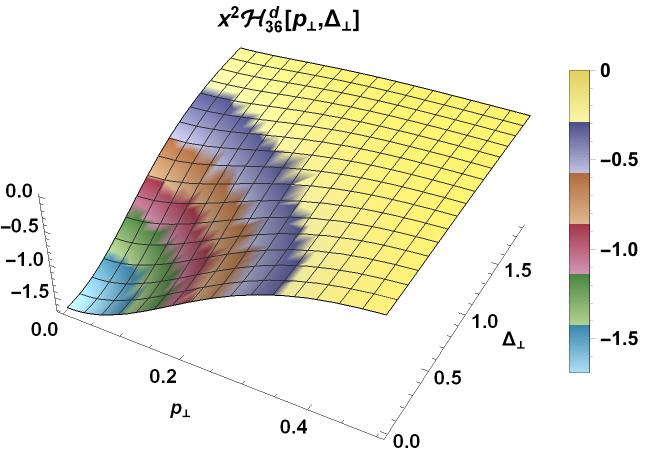}
		\hspace{0.05cm}
		(e)\includegraphics[width=7.3cm]{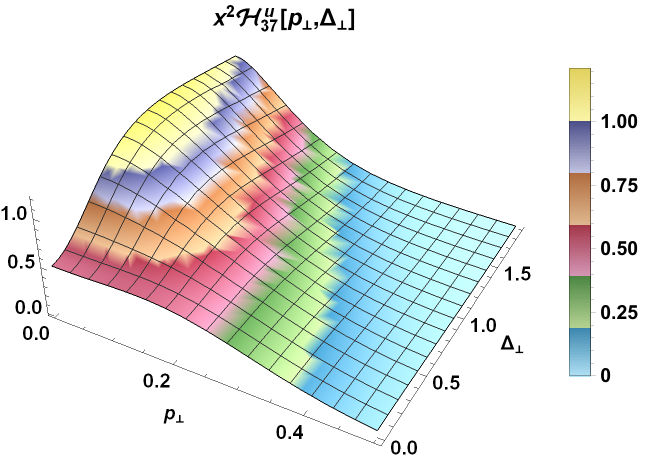}
		\hspace{0.05cm}
		(f)\includegraphics[width=7.3cm]{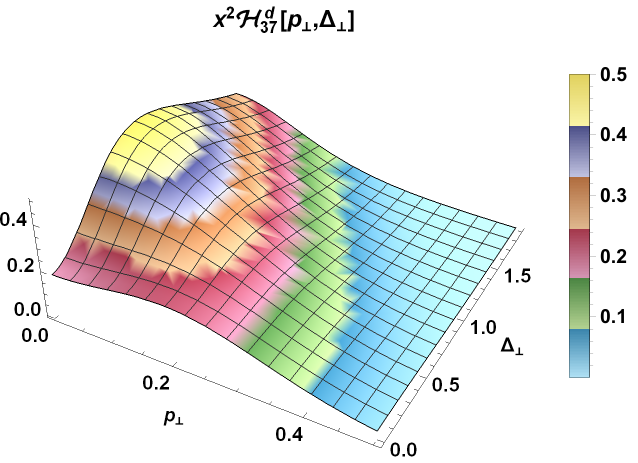}
		\hspace{0.05cm}
		(g)\includegraphics[width=7.3cm]{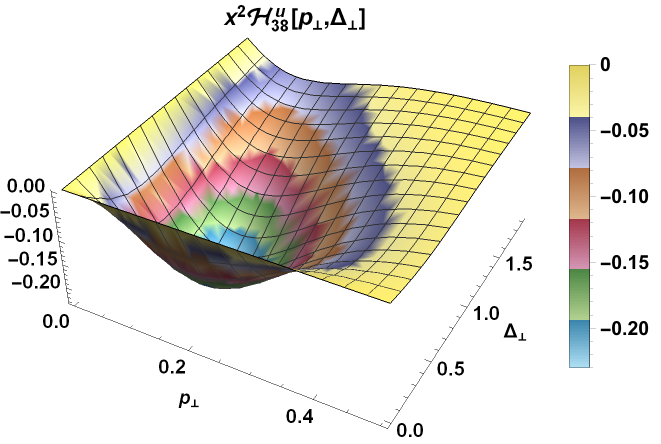}
		\hspace{0.05cm}
		(h)\includegraphics[width=7.3cm]{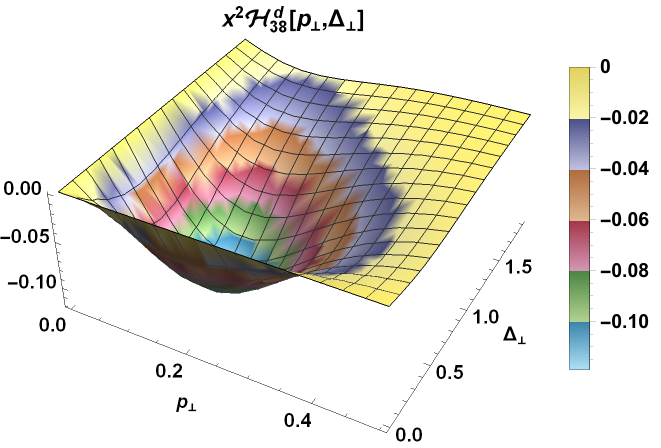}
		\hspace{0.05cm}\\
	\end{minipage}
	\caption{\label{fig3dTMFFH2} (Color online) The twist-4 TMFFs 		
		$x^2 H_{3,5}^{\nu}(p_{\perp},\Delta_{\perp})$,
		$x^2 H_{3,6}^{\nu}(p_{\perp},\Delta_{\perp})$,
		$x^2 H_{3,7}^{\nu}(p_{\perp},\Delta_{\perp})$ and
		$x^2 H_{3,8}^{\nu}(p_{\perp},\Delta_{\perp})$ plotted with respect to ${ p_\perp}$ and ${{ \Delta_\perp}}$ for ${\bfp} \parallel {\Dp}$. The left and right column correspond to $u$ and $d$ quarks sequentially.
	}
\end{figure*}

 Transverse-momentum dependent form factors (TMFFs) have been obtained from GTMDs via the integration over the longitudinal momentum fraction $x$ as it has been observed that the variation of GTMDs for a fixed longitudinal momentum fraction is similar to the case when $x$ is being integrated. In  Figs. (\ref{fig3dTMFFF}), (\ref{fig3dTMFFG}), (\ref{fig3dTMFFH1}) and (\ref{fig3dTMFFH2}), TMFFs have been plotted with respect to the transverse momentum of quark ${ p_\perp}$ and the transverse momentum transferred ${ \Delta_\perp}$.  It has been observed in all these plots that, the amplitude tends to $0$ when the value of transverse momentum transfer ${ \Delta_\perp}$ is equal to or greater than $1.5~\mathrm{GeV}$ suggesting that, for any polarization possibility of twist-4 TMFFs at the model scale, this amount of momentum transfer is not possible at all. TMFFs $F_{3,1}$, $F_{3,2}$, $F_{3,3}$ and $F_{3,4}$ corresponding to the Dirac matrix structure $\Gamma=\gamma^-$, have been plotted in Fig. (\ref{fig3dTMFFF}). In Fig. \ref{fig3dTMFFF} (a) to \ref{fig3dTMFFF} (d), the TMFFs $F_{3,1}^{\nu}$ and $F_{3,2}^{\nu}$ have been plotted for $u$ and $d$ quarks. The distribution $F_{3,1}^{\nu}$ exists only when the kinetic energy of quark is high (low) and the momentum transferred between the initial and the final state is low (high). Neglecting the amplitude, the behavior of $F_{3,2}^{\nu}$ is found to be antisymmetric over the quark flavor change. The TMFF $F_{3,3}^{\nu}$ has been plotted in Fig. \ref{fig3dTMFFF} (e) and \ref{fig3dTMFFF} (f) for $u$ and $d$ quarks respectively. Except the region around ${ \Delta_\perp}=0.5~\mathrm{GeV}$, the TMFF increases with the increase in ${ p_\perp}$ for $u$ quarks. For $d$ quark, as we move diagonally in ${ p_\perp}-{ \Delta_\perp}$ plane, the TMFF $F_{3,3}^{d}$ first increases and then decreases to meet the horizontal plane. TMFF $F_{3,4}^{\nu}$ has been plotted in Fig. \ref{fig3dTMFFF} (g) and \ref{fig3dTMFFF} (h) for $u$ and $d$ quarks respectively. The TMFF varies by a similar fashion for both ${ p_\perp}$ and ${ \Delta_\perp}$.
%
%---------------G x Dp PLOT---------------------%

For the study of TMFFs $G_{3,1}$, $G_{3,2}$, $G_{3,3}$ and $G_{3,4}$, the Dirac matrix structure $\Gamma=\gamma^-\gamma^5$ is chosen. These have been plotted with respect to ${ p_\perp}$ and ${ \Delta_\perp}$ in Fig. (\ref{fig3dTMFFG}), while $x$ and $\theta$ are fixed at $0.3$ and $0^{\circ}$ respectively. In Fig. \ref{fig3dTMFFG} (a) and \ref{fig3dTMFFG} (b), the TMFF $G_{3,1}^{\nu}$ has been plotted which is positive over the entire range of ${ p_\perp}$ and ${ \Delta_\perp}$ and its behavior is same for $u$ as well as $d$ quarks with variation only in the amplitude. In this figure, as ${ p_\perp}$ (or ${ \Delta_\perp}$) increases, the amplitude of TMFF decreases to meet the horizontal plane. 
TMFF $G_{3,2}^{\nu}$ has been plotted w.r.t ${ p_\perp}$ and ${ \Delta_\perp}$ for $u$ and $d$ quarks in Fig. \ref{fig3dTMFFG} (c) and \ref{fig3dTMFFG} (d) respectively. Due to contribution equations of different dependencies, the plot is part-wise symmetric over the quark flavor. In Fig. \ref{fig3dTMFFG} (e) and \ref{fig3dTMFFG} (f), TMFF $G_{3,3}^{\nu}$ has been plotted w.r.t ${ p_\perp}$ and ${ \Delta_\perp}$ for $u$ and $d$ quarks respectively. For both quark flavors, as we move away from the critical point, the magnitude of the TMFF decreases to meet the horizontal plane. Ignoring few exceptions, as we move away from minima the magnitude of TMFF  $G_{3,4}^{\nu}$ increases to meet the horizontal plane for both $u$ and $d$ quarks as shown in Fig. \ref{fig3dTMFFG} (g) to \ref{fig3dTMFFG} (h) respectively. This shows the existence of TMFF only at low quark momentum and large momentum transfer.
%---------------H x Dp PLOT---------------------%
%
%

By considering the Dirac matrix structure $\Gamma=i\sigma^{+-}\gamma_5$, TMFFs $H_{3,1}$, $H_{3,2}$, $H_{3,3}$, $H_{3,4}$, $H_{3,5}$, $H_{3,6}$, $H_{3,7}$ and $H_{3,8}$ have been plotted in Figs. (\ref{fig3dTMFFH1}) and (\ref{fig3dTMFFH2}). In Fig. \ref{fig3dTMFFH1} (a) to \ref{fig3dTMFFH1} (d), the TMFFs $H_{3,1}^{\nu}$ and $H_{3,2}^{\nu}$ have been plotted. These TMFFs do not change their sign on changing the quark flavor between $u$ and $d$ quarks. Keeping the amplitude aside, the general trend of its variation is same in both cases such that as we move away from the critical point the magnitude of the TMFF decreases. In Fig. \ref{fig3dTMFFH1} (e) and \ref{fig3dTMFFH1} (f), the TMFF $H_{3,3}^{\nu}$ is plotted. In this case, no quark flavor symmetry exists. The TMFF $H_{3,4}^{\nu}$ has been plotted in Fig. \ref{fig3dTMFFH1} (g) and \ref{fig3dTMFFH1} (h). Here, it has been observed that as we move away from point $(0,0)$, the value of the TMFF increases (decreases) to reach the horizontal plane for $u$ ($d$) quarks. This signifies that the equations of TMFF have the same structure dependence on transverse momentum and its transfer. TMFF $H_{3,5}^{\nu}$ has been plotted for $u$ and $d$ quarks in Fig. \ref{fig3dTMFFH2} (a) and \ref{fig3dTMFFH2} (b) respectively. Due to the presence of conflicting factors, one extra critical point is observed for $u$ quark at large momentum transfer ${ \Delta_\perp}$, although the possibility associated with it is quite low. In Fig. \ref{fig3dTMFFH2} (c) and \ref{fig3dTMFFH2} (d), the TMFF $H_{3,6}^{\nu}$ is plotted and contrary to its expressions, it shows sign of symmetry symmetry over the quark flavor. In Fig. \ref{fig3dTMFFH2} (e) to \ref{fig3dTMFFH2} (h) the TMFFs $H_{3,7}^{\nu}$ and $H_{3,8}^{\nu}$ have been plotted for both the quark flavors. Keeping magnitude aside, for both TMFFs the plot of $u$ quarks is similar to that of $d$ quarks.

%$$$$$$$$$$$$$$$$$$$$$$$$$$ TILL HERE %$$$$$$$$$$$$$$$$$$$$$$$$$$


\section{Conclusion}\label{seccon}
We have presented in this work the analysis of proton GTMDs at twist-4 within the paradigm of LFQDM. We have exploited the GTMD correlator obtained from  the off-diagonal, fully unintegrated quark-quark GPCFs via integration over the light-cone energy. By deciphering the quark-quark GTMD correlator for proton at twist-4 Dirac matrix structure, we have managed to express the correlation in the overlap form of LFWFs. After solving the parametrization equations at this twist, we have successfully expressed the GTMDs in form of explicit equations in the AdS/QCD inspired LFWFs $\varphi^{(\nu)}_{i}$. We have obtained our result for both the possibilities of the diquark being a scalar or a vector and also for the case of smacked quark's flavor being a $u$ or a $d$ quark. The analysis of this multidimensional GTMD function is achieved via 3-D plots  considering two variables at a time while keeping the others fixed or integrated. At the limit of no momentum transfer $\Delta=0$, we verify our expressions with the already published twist-4 TMD results. Even though the distinction of GTMDs on the basis of polarization of proton is possible, it becomes difficult to categorize twist-4 GTMDs on the basis of different quark polarization states as can be done for the twist-2 GTMDs \cite{Lorce16}. The spinor products arising from the Dirac matrix structure $\Gamma= \gamma^-,  \gamma^-\gamma_5$ and $\sigma^{i-}\gamma_5$ decompose different quark polarizations but we cannot compute the Wigner distributions at this twist since they require the polarization specifications of the struck quark as well as the nucleon.
\par In conclusion, the contributions of twist-4 GTMDs and its associated distributions (TMDs, GPDs, PDFs) have been the focus of research in DDY, DY, DIS, DVCS studies at HERMES and those conducted in J-Lab. As an extension of this work, we plan to study  twist-3 and twist-4 GPDs as well as the behavior of GTMDs at the GPDs limit in near future.
%====================================================
\section{Acknowledgement}
H.D. would like to thank the Science and Engineering Research Board, Department of Science and Technology, Government of India through the grant (Ref No.TAR/2021/000157) under TARE scheme for financial support.

%\appendix
%
%\section{}\label{App}

%___________________________________________
%Bib copied above

%___________________________________________

%
% ****** End of file apssamp.tex ******
\end{document}